 \documentclass[preprint,12pt]{elsarticle}%review,
\usepackage{amssymb,amsmath,latexsym}
\usepackage[dvips]{color}
\usepackage[headings]{fullpage}
\usepackage{epic,epsfig,graphicx}
\usepackage{enumerate}
\usepackage{multirow}
\usepackage{setspace}
\usepackage{keyval}
\usepackage{float}
\usepackage{amssymb}
 \newtheorem{thm}{Theorem}[section]

 \newtheorem{lemma}{Lemma}[section]

 \newtheorem{rem}{Remark}[section]
  \newtheorem{assumption}{Assumption}[section]

 \numberwithin{equation}{section}
\journal{ArXiv}

\begin{document}
\begin{frontmatter}
\title{A comparative stochastic and deterministic study of a class of epidemic dynamic models for malaria: exploring the impacts of noise on eradication and persistence of disease }
%{A Scale-Structured Network Stochastic Epidemic dynamic model with varying Incubation Period}
\author{Divine Wanduku }
\address{Department of Mathematical Sciences,
Georgia Southern University, 65 Georgia Ave, Room 3042, Statesboro,
Georgia, 30460, U.S.A. E-mail:dwanduku@georgiasouthern.edu;wandukudivine@yahoo.com\footnote{Corresponding author. Tel: +14073009605.
} }%fax: +18139746720.
\begin{abstract}
 A comparative stochastic and deterministic study of a family of SEIRS epidemic dynamic models for malaria is presented.  The family type is determined by the qualitative behavior of the nonlinear incidence rates of the disease. Furthermore, the malaria models exhibit three random delays:- two of the delays represent the incubation periods of the disease inside the vector and human hosts, whereas the third delay is the period of effective natural immunity against the disease. The stochastic malaria models are improved by including the random environmental fluctuations in the disease transmission and natural death rates of humans. Insights about the effects of the delays and the noises on the malaria dynamics are gained via  comparative analyses of the family of stochastic and deterministic models, and further critical examination of the significance of the intensities of the white noises in the system on (1) the existence and stability of the equilibria, and also on (2) the eradication and persistence of malaria in the human population. The basic reproduction numbers and other threshold values for malaria in the stochastic and deterministic settings are determined and compared for the cases of constant or random delays in the system. Numerical simulation results are presented.
\end{abstract}

\end{frontmatter}
\section{Introduction\label{ch1.sec0}}
Despite all efforts to reduce the global burden of malaria, the WHO estimates released in December $2016$ exhibit that 212 million cases of the disease occurred in $2015$ resulting in about 429 thousand deaths. Furthermore, the highest mortality rates occurred in the sub-Saharan African countries  where about $90\%$ of the global malaria cases occurred and led to about $75\%$ of the total world's malaria deaths.  Moreover, more than two third of the global malaria related deaths were children in the age group under the age five years old. In addition, despite the fact that malaria is curable and preventable, and despite all other advances to control and contain the disease, the world at large is still far from  complete safety from the health and economic menace exhibited by  the disease.

Indeed, WHO reports that in $2015$, nearly half of the world's population was at risk of malaria, and the disease was actively and continuously transmitted in about $91$ countries in the world. Moreover, the most vulnerable populations include infants and children under the age of five, pregnant women, patients with HIV/AIDS and non-immune migrants, visitors and travellers to malaria endemic zones\cite{WHO,CDC}. These facts and observations sound a loud call for understanding, cooperation, national solidarity, social and scientific investigation in the fight to eradicate or ameliorate the burdens of malaria.

 Malaria is a vector-borne disease  caused by protozoa (a micro-parasitic organism) of the genus \textit{Plasmodium}. There are several different species of the parasite that cause disease in humans namely: \textit{P. falciparum, P. viviax, P. ovale} and \textit{P. malariae}. However, the species that causes the most severe and fatal disease is the \textit{P. falciparum}.  Malaria is transmitted between humans by the infectious bite of a female mosquito of the genus \textit{Anopheles}. Similar to other mosquito-borne parasitic diseases such \textit{lymphatic filariasis}, the complete life cycle of the  malaria  plasmodium entails two-hosts: (1) the female anopheles mosquito vector, and (2) the susceptible or infectious human being. The stage of the parasite infective to humans is called sporozoite, while the stage of the parasite infective to mosquitoes is called gametocyte.

 Indeed, as an infected female anopheles mosquito persistently quests and successfully bites a human being to obtain a blood meal, she injects sporozoites through the salivary glands into the blood stream of the susceptible or infected person. Inside the exposed or infectious person, the sporoziotes are thought to develop in the liver into schizont which contain numerous merozoites\footnote{Schizonts and merozoites are intermediary developmental stages of the parasite.}. The mature schizonts rupture releasing the meroziotes into the bloodstream. The meroziotes infect red blood cells, and within the red blood  they either (1.) develop to form additional schizont in the blood stream that continue to infect the human body,  or (2.) they develop to form a sexual stage of the plasmodium called gametocyte. The stages of maturation of the plasmodium from the sporozoite through the schizont stage within the human body is called the \textit{exo-erythrocytic cycle}. Moreover, the total duration of the \textit{exo-erythrocytic cycle} is estimated at between 7-30 days depending on the species of plasmodium, with the exceptions of the plasmodia- \textit{P. vivax} and \textit{P. ovale} that may be delayed for as long as 1 to 2 years.

 The gametocyte stage of the plasmodium (also referred to as the sexual stage of the malaria parasite) which is infectious to susceptible or infectious female mosquitoes is ingested by the female mosquito when she successfully takes a blood meal from an infectious human being. Within the mosquito, the gametocyte develops into female and male gametes which undergo fertilization and develop into sporozoites which can infect humans.  The stages of development from the gametocyte to infectious sporozoites within the mosquito is called the  \textit{sporogonic cycle}. It is estimated that the duration of  the \textit{sporogonic cycle} is over 2 to 3 weeks \cite{malaria,WHO,CDC}.  The delay between infection of the mosquito and maturation of the sporozoites suggests that the mosquito must survive a minimum of the 2 to 3 weeks to be able to transmit malaria. These facts are important in deriving a mathematical model to represent the dynamics of malaria.

 %%%%%%%%%%%%%%%%%%%%%%%%%%%%%%%%%%%%%%%%%%%%%%%%%%%%%%%%%%
 %%%%%%%%%%%%%%%%%%%%%%%%%%%%%%%%%%%%%%%%%%%%%%%%%%%%%%%%%%%%%%
 In the general class of infectious diseases, vector-borne diseases such as malaria and dengue fever exhibit several unique biological characteristics. For instance, as observed in the description about the life cycle of the malaria parasite above, the incubation of the disease requires two hosts - the vector and human hosts,  which may be either  directly involved in a full life cycle of the infectious agent consisting of two separate and independent segments of sub-life cycles that are completed separately in the two hosts or directly involved in two separate and independent half-life cycles of the infectious agent in the hosts. Therefore, there exists a total latent time lapse of disease incubation which extends over the two segments of delayed incubation times namely:- (1) the incubation period of the infectious agent ( or the half-life cycle) in the vector, and (2) the incubation period of the infectious agent (or the other half-life cycle) in the human being. For example,  the dengue fever virus transmitted primarily by the \textit{Aedes aegypti and Aedes albopictus} mosquitos undergoes two delay incubation periods:- (1) about 8-12 days incubation period inside the female mosquito vector, which starts immediately after the ingestion of a dengue fever virus infected blood meal, that has been successfully taken from a dengue fever infectious human being via a mosquito bite, and (2) another delay incubation period of about 2-7 days in the human being when the hosting female infectious vector  acquires another blood meal  from a susceptible human being, whereby the virus is successfully transferred  from the infectious mosquito to the susceptible person\cite{WHO,CDC}.

 Malaria confers natural immunity after recovery from the disease. The strength and effectiveness of the natural immunity against the disease depends primarily on the frequency of  exposure to the parasites and other biological factors such as age, pregnancy, and genetic nature of red blood cells of people with malaria.
  The naturally acquired immunity against malaria, especially in areas where malaria is highly endemic such as sub-sahara African,  varies across age groups and people with various biological characteristics etc. For example, newborns, pregnant women and visitors from areas with little or no malaria history exhibit low immunity levels against the malaria parasites, while adults who have suffered repeated attacks tend to exhibit higher levels of protective natural immunity against the  occurrence of severe disease. Other adults with history of malaria exposure are asymptomatic to subsequent malaria attacks. Furthermore, other biological characteristics related to the nature of red blood cells such as sickle cell trait, and Duffy blood group negativity etc. are also noted to confer long lasting protective resistance against certain species of the malaria parasite. For example, people with sickle cell trait are relatively more protected against \textit{p. falciparum} malaria, while people who are Duffy negative show strong resistance against \textit{P. vivax} malaria\cite{CDC,lars,denise}.
 %%%%%%%%%%%%%%%%%%%%%%%%%%%%%%%%%%%%%%%%%%%%%%%%%%%%%%%%%

Various types of compartmental mathematical epidemic dynamic models have been proposed and utilized  to investigate the dynamics of infectious diseases. For instance, dengue fever and measles are studied in \cite{eric,sya,pang}. Furthermore, several different authors have proposed various epidemic dynamic models for malaria beginning  with Ross\cite{ross} who studied mosquito control, Macdonald\cite{macdonald} who addressed superinfection, a combined dynamics of mosquitoes and humans investigated in \cite{ngwa-shu}, the naturally acquired immunity by continuing exposure to malaria explored in \cite{hyun,may} and several other studies such as \cite{kazeem,gungala,anita} which are based on the mosquito biting habit. There are also studies  which have instead focused on the malaria parasite as the agent of disease transmission such as \cite{tabo}.

 In general, the compartmental mathematical epidemic dynamic models are largely classified as SIS, SIR, SIRS, SEIRS,  and  SEIR etc. epidemic dynamic models depending on the compartments of the disease classes directly involved in the general disease dynamics\cite{qun,qunliu, nguyen,joaq, sena,wanduku-fundamental,Wanduku-2017, zhica}. Several studies devote interest to SEIRS and SEIR
 models\cite{joaq,sena,cesar,sen,zhica} which allow the inclusion of the compartment of individuals who are exposed to the disease, $E$, that is, infected but noninfectious individuals. This natural inclusion of the exposed class of individuals allows for more insight about the disease dynamics during the incubation stage of the disease. For example, the existence of periodic solutions are investigated in the SEIRS epidemic study\cite{joaq,zhica}. In addition, the effects of seasonal changes on the disease dynamics are investigated in the SEIRS epidemic study in \cite{zheng}.
 %Some vector-borne diseases form a natural fit for representation by SEIRS epidemic dynamic models because of the incubation period of the disease in the human being.

 Many epidemic dynamic models are modified and  improved in reality by including the time delays that occur in the disease dynamics. Generally, two distinct classes of delays are studied namely:-disease latency and immunity delay. The disease latency has been represented as the infected but noninfectious period of disease incubation and also as the period of infectiousness which nonetheless is studied as a delay in the dynamics of the disease. The immunity delay represents the period of effective naturally acquired immunity against the disease after exposure and successful recovery from infection.  Whereas, some authors  study diseases and disease scenarios under the realistic assumption of one form of these two classes of delays in the disease dynamics\cite{Wanduku-2017,wanduku-delay,kyrychko,qun}, other authors study one or more forms of the classes of delays  represented as two separate  delay times\cite{zhica,cooke-driessche,shuj,Sampath}. The occurrence of delays in the disease dynamics  may influence the dynamics of the disease in many important  ways. For instance, in \cite{zhica}, the presence of delays in the epidemic dynamic system creates periodic solutions. In \cite{cooke, baretta-takeuchi1}, the occurrence of a delay in the vector-borne disease dynamics  destabilizes the equilibrium population state of the system.

 Stochastic epidemic dynamic models more realistically represent epidemic dynamic processes because they include the randomness which naturally occurs during a disease outbreak, owing to the presence of constant random environmental fluctuations in the disease dynamics. The presence of stochastic white noise process in the epidemic dynamic system may directly impact the density of the system or indirectly influence other driving parameters of the system such as the disease transmission, natural death, birth and disease related death rates etc. In \cite{Wanduku-2017,wanduku-fundamental,wanduku-delay},  the stochastic white noise process represents the random fluctuations in the disease transmission process. In \cite{qun}, the white noise process represents the variability in the natural death of the population. In \cite{Baretta-kolmanovskii},  the white noise process represents the random fluctuations in the system which deviate the state of the system from the equilibrium state, that is, the  white noise process is proportional to the difference between the state and equilibrium of the system.
  %%%%%%%%%%%%%%%%%%%
  A stochastic white noise process driven system generally exhibits  more complex behavior in the disease dynamics. For instance, the presence of stochastic white noise process in the disease dynamics  may destabilize a disease free steady state population by exhibiting high intensity values or high standard deviation values which generally displace the population from a disease free state. In some cases, the presence of white noise may  lead to massive oscillations of the state of the system depending on the intensity value of the random fluctuations, which can decrease the population size over time  and  lead to the extinction of the population.   For example, in \cite{qun,Wanduku-2017,wanduku-fundamental,wanduku-delay,zhuhu,yanli}, the occurrence of stochastic noise in the system destabilizes the disease free steady population state. In \cite{qun}, the disease free steady state fails to exist when the intensity value of  the white noise process from the natural death process of the susceptible population is positive.

 The interaction between susceptible, $S$, and infectious individuals, $I$, during the disease transmission process of an infectious disease sometimes exhibits  more complex behavior than a simple representation  by the frequently used bilinear incidence rate or force of infection given as $\beta S(t)I(t-T)$ for vector-borne diseases, or $\beta S(t)I(t)$ for infectious diseases that involve direct human-to-human disease transmission, where $\beta$ is the effective contact rate, and $T$ is the incubation  period for the vector-borne disease. More complex behaviors such as the psychological or crowding effects stemming from behavioral change of susceptible individuals when the infectious population increases significantly over time  exist for certain types of infectious diseases and disease scenarios,  where the contact between the susceptible and infectious classes are regulated, and consequently  prevent unboundedness in the disease transmission rate, or exhibit other nonlinear behaviors for the disease transmission rate. For instance, in  \cite{yakui,xiao,huo,kyrychko,qun,muroya,liu,capasso-serio,capasso,hethcote,koro} several different functional forms for the force of infection or incidence rate are used to represent the nonlinear behavior that occurs during the disease transmission process. In \cite{yakui,xiao,capasso-serio,huo} the authors consider a Holling Type II functional form, $\beta S(t)G(I(t))=\frac{\beta S(t)I(t)}{1+\alpha I(t)}$, that saturates for large values of $I$. In \cite{muroya,xiao, capasso}, a bounded  Holling Type II function,  $\beta S(t)G(I(t))=\frac{\beta S(t)I^{p}(t)}{1+\alpha I^{p}(t)}, p\geq 0$,  is used to represent the force of infection of the disease. In \cite{hethcote,koro}, the nonlinear behavior of the incidence rate is represented by the general functional form, $\beta S(t)G(I(t))=\beta S^{p}(t)I^{q}(t), p,q\geq 0$. In addition,  the authors in \cite{yakui,huo, capasso-serio, muroya,capasso, qun} studied vector-borne diseases with several different functional forms for the nonlinear incidence rates of the disease.
 %This paper utilizes  a general nonlinear incidence rate of the form $\beta S(t)G(I(t-T))$  to  represent the force of infection for a family of vector-borne disease dynamic models.
%%%%%can-chen
%Liu et al\cite{liu} considered a nonlinear incidence rate which incorporates behavioral changes. Capasso and Serio\cite{capasso-serio} considered a nonlinear saturated incidence rate. Xiao and %Ruan\cite{xiao} considered a bounded nonlinear incidence rate. Cappasso\cite{capasso} and Huo et al\cite{} considered a nonlinear saturated incidence rate with delay.

Cooke\cite{cooke} presented a deterministic epidemic dynamic model for a vector-borne disease, where the bilinear incidence rate defined as $\beta S(t)I(t-T)$ represents the number of new infections occurring per unit time during the disease transmission process. It is assumed in the formulation of this incidence rate that the number of infectious vectors at time $t$ interacting and effectively transmitting infection to susceptible individuals, $S$, after $\beta$  number of effective contacts per unit time per infective is proportional to the infectious human population, $I$,  at earlier time $t-T$. The study above allows insight about the dynamics of the disease primarily in the human population while keeping tract of the influence of the vector on the dynamics via the disease transmission process. Whereas vector control aides in malaria prevention\cite{ngwa-shu,gungala}, in various events of emergency malaria crisis such as when severe disease erupts, urgent medical interference on the involved human being requires direct intervention by medical experts through the use of anti-malaria medications. This observation necessitates a thorough continuous understanding of the dynamics of malaria with major emphasis based in the human population especially in a realistic framework where the system is constantly bombarded by random environmental fluctuations. Very little or nothing about the dynamics of malaria in the human population in a more realistic white noise driven mathematical dynamic system is known. This study bridges the gap by providing a comparative stochastic and deterministic study of a class of malaria models, in an attempt to elucidate the influence of underlying random perturbations on the dynamics of the disease, in particular, on disease eradication via studying the stability of equilibria, extinction and permanence of disease via studying the asymptotic properties of the solutions of the systems.
 %The assumptions in \cite{cooke} result in an epidemic dynamic model for vector-borne diseases structurally formulated in terms and expressions involving the states and parameters from the human %population.
 %In the malaria studies, the authors  presented a different version of a deterministic vector-borne disease dynamic model with delays, where the epidemic dynamic system is structurally formulated as form of predator-prey model with separate but interconnected dynamics for the two species- the vectors and the human beings.
  %The epidemic dynamic model is a union of  two-species interacting system of ordinary differential equations, that are interconnected through the disease transmission process between the vectors and humans.

 This paper employs similar reasoning in \cite{cooke}, to derive a general class of SEIRS stochastic epidemic dynamic models with three delays for  vector-borne diseases such as malaria. The three delays are classified under the two general group types namely:- disease latency and immunity delay. Two of the delays represent the incubation period of the infectious agent (plasmodium for malaria) inside the vector  and human hosts, and the third delay represents the period of effective naturally acquired immunity against the vector-borne disease, where the natural immunity is conferred after recovery from infection. Moreover, the delays are random variables. In addition, the general vector-borne disease dynamics is  driven by stochastic white noise processes originating from the random environmental fluctuations in  the natural death and disease transmission rates in the population. The deterministic version of the epidemic dynamic model is a system of ordinary differential equations.  The stochastic version of the epidemic dynamic model is a system of Ito-Doob type stochastic differential equations.

 It should be noted that this study addresses some objectives of a sizeable ongoing project. To conserve space, a parallel detailed study about the qualitative behavior of the intensity  of the random fluctuations in the disease dynamics ( which are represented by the white noise processes in the stochastic model) in relation to the stochastic asymptotic stability of the steady states of the  system, and with critical examination of the effects of the intensities on disease eradication from the stochastic system appears elsewhere.  In that parallel study,  various novel mathematical techniques are utilized to diagnose and elucidate the finite properties of the white noise processes in the system, critically evaluate and describe their impact on the disease dynamics.  In the current paper, the primary goal is to gain complete comparative insight about the general asymptotic properties of the deterministic and stochastic systems:- (\ref{ch1.sec0.eq3})-(\ref{ch1.sec0.eq6}) and  (\ref{ch1.sec0.eq8})-(\ref{ch1.sec0.eq11}), and with attention given to show how the occurrence of noise and delays in the disease dynamics create several interesting features of the disease dynamics in relation to the qualitative behavior of (1) the equilibria of the systems, whenever the equilibria exist,
  and (2) the trajectories of the stochastic system near potential deterministic equilibria in the system. Moreover, the interconnection between the two different types of dynamical systems (stochastic and deterministic) with respect to the asymptotic stability of the equilibria, and asymptotic behavior of the solutions of the stochastic system near potential deterministic equilibria is established.

   This work is presented as follows:- In section~\ref{ch1.sec0}, the stochastic and deterministic epidemic dynamic models for malaria are derived. In section~\ref{ch1.sec1}, the model validation results are presented for both the deterministic and stochastic systems. %In Section~\ref{ch1.sec2a}, the conditions for extinction of disease from the stochastic system is presented.
   In section~\ref{ch1.sec2}, the existence and asymptotic properties of the disease free equilibrium population in both systems are investigated. In Section~\ref{ch1.sec3}, existence and the asymptotic properties of the endemic equilibria of both systems are also investigated. %In Sections~\ref{ch1.sec3a} and \ref{ch1.sec5}  the permanence of the disease in the deterministic and stochastic systems are both studied.
   And in Section~\ref{ch1.sec4},  numerical simulation results are given to justify the results of this paper.
 %%%%%
 %%%%%%
\section{Derivation of Model}\label{ch1.sec0}
A generalized class of stochastic SEIRS delayed epidemic dynamic models for  vector-borne diseases is presented. The delays represent the incubation period of the infectious agents in the vector $T_{1}$, and in the human host $T_{2}$. The third delay represents the naturally acquired immunity period of the disease $T_{3}$, where the delays are random variables with density functions $f_{T_{1}}, t_{0}\leq T_{1}\leq h_{1}, h_{1}>0$, and $f_{T_{2}}, t_{0}\leq T_{2}\leq h_{2}, h_{2}>0$ and $f_{T_{3}}, t_{0}\leq T_{3}<\infty$. Furthermore, the joint density of $T_{1}$ and $T_{2}$ is given by $f_{T_{1},T_{2}}, t_{0}\leq T_{1}\leq h_{1} , t_{0}\leq T_{2}\leq h_{2}$. Moreover, it is assumed that the random variables $T_{1}$ and $T_{2}$ are independent (i.e. $f_{T_{1},T_{2}}=f_{T_{1}}.f_{T_{2}}, t_{0}\leq T_{1}\leq h_{1} , t_{0}\leq T_{2}\leq h_{2}$). Indeed, the independence between $T_{1}$ and $T_{2}$ is justified from the understanding that the  incubation of  the infectious agent for the vector-borne disease depends on the suitable  biological environmental requirements for incubation inside the vector and the human body which are unrelated. Furthermore, the independence between $T_{1}$ and $T_{3}$ follows from the lack of any real biological evidence to justify the connection between the incubation of  the infectious agent inside the vector and the acquired natural immunity conferred to the human being. But $T_{2}$ and $T_{3}$ may be dependent as biological evidence suggests that the naturally acquired immunity is induced by exposure to the infectious agent.

By employing similar reasoning in \cite{cooke,qun,capasso,huo}, the expected incidence rate of the disease or force of infection of the disease at time $t$ due to the disease transmission process between the infectious vectors and susceptible humans, $S(t)$, is given by the expression $\beta \int^{h_{1}}_{t_{0}}f_{T_{1}}(s) e^{-\mu s}S(t)G(I(t-s))ds$, where $\mu$ is the natural death rate of individuals in the population, and it is assumed for simplicity that the natural death rate for the vectors and human beings are the same. The probability rate, $0<e^{-\mu s}\leq 1, s\in [t_{0}, h_{1}], h_{1}>0$,  represents the survival probability rate of  exposed vectors over the incubation period, $T_{1}$, of the infectious agent inside the vectors with the length of the period given as $T_{1}=s, \forall s \in [t_{0}, h_{1}]$, where the vectors acquired infection at the earlier time $t-s$ from an infectious human via a successful infected blood meal, and  become infectious at time $t$.  Furthermore, it is assumed that the survival of the vectors over the incubation period of length  $s\in [t_{0}, h_{1}]$ is independent of the age of the vectors. In addition, $I(t-s)$, is the infectious human population at earlier time $t-s$, $G$ is a nonlinear incidence function of the disease dynamics,  and $\beta$ is the average number of effective contacts per infectious individual per unit time. Indeed, the force of infection,  $\beta \int^{h_{1}}_{t_{0}}f_{T_{1}}(s) e^{-\mu s}S(t)G(I(t-s))ds$  signifies the expected rate of new infections at time $t$ between the infectious vectors and the susceptible human population $S(t)$ at time $t$, where the infectious agent is transmitted per infectious vector per unit time at the rate $\beta$. Furthermore, it is assumed that the number of infectious vectors at time $t$ is proportional to the infectious human population at earlier time $t-s$. Moreover, it is further assumed that the interaction between the infectious vectors and  susceptible humans exhibits nonlinear behavior, for instance, psychological and overcrowding effects,  which is characterized by the nonlinear incidence  function $G$. Therefore, the force of infection given by
 \begin{equation}\label{ch1.sec0.eqn0}
   \beta \int^{h_{1}}_{t_{0}}f_{T_{1}}(s) e^{-\mu s}S(t)G(I(t-s))ds,
 \end{equation}
  represents the expected rate at which infected individuals leave the susceptible state and become exposed at time $t$.

\begin{figure}[H]
%\begin{center}
\includegraphics[height=12cm]{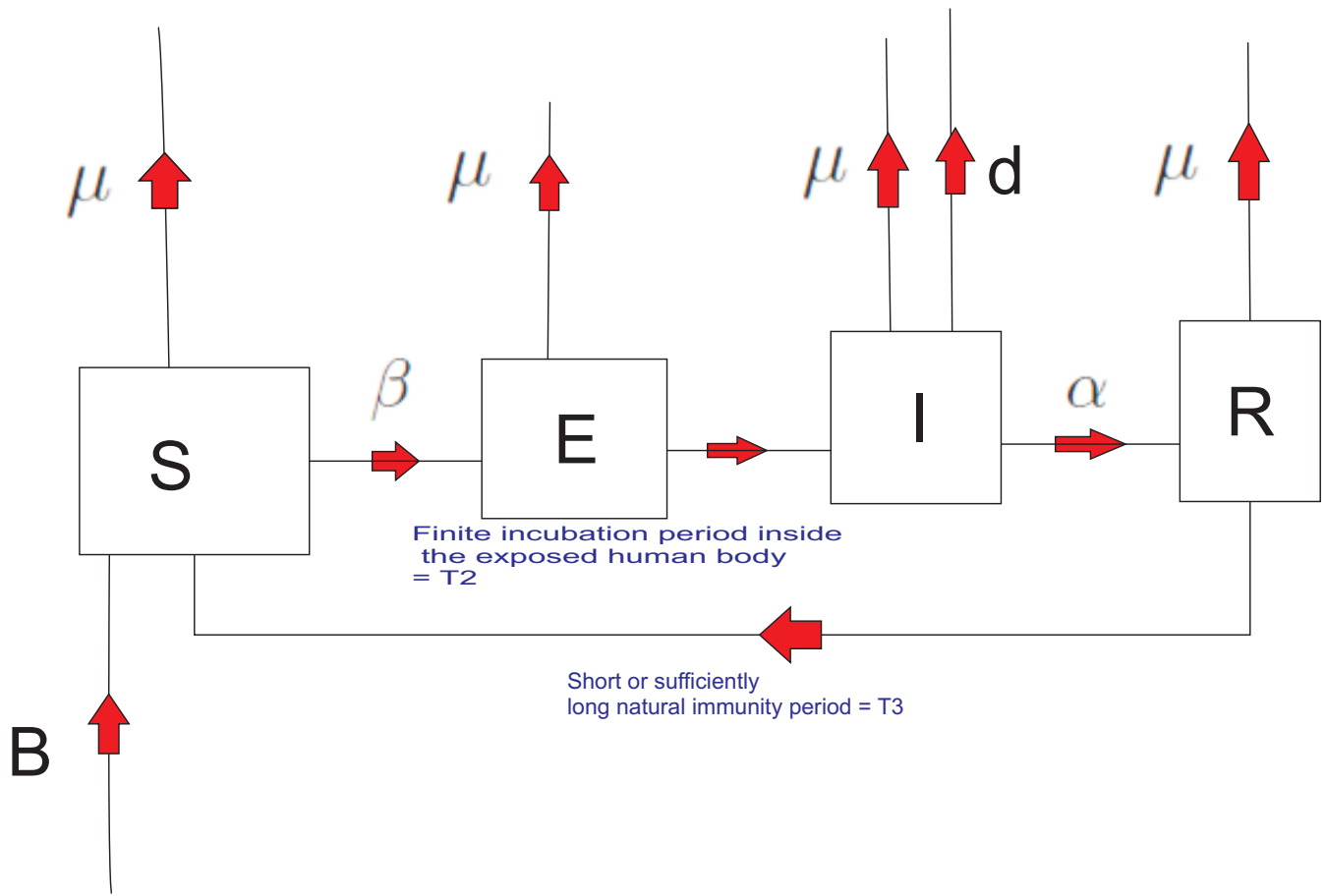}
\caption{The compartmental framework illustrates the transition rates between the states $S,E,I,R$ of the system. It also shows the incubation delay $T_{2}$ and the naturally acquired immunity $T_{3}$ periods. \label{ch1.sec4.figure 1}}
%\end{center}
\end{figure}

%%%%
%%%%
 The susceptible individuals who have acquired infection from infectious vectors but are non infectious form the exposed class $E$. The population of exposed individuals at time $t$ is denoted $E(t)$. After the incubation period, $T_{2}=u\in [t_{0}, h_{2}]$, of the infectious agent in the exposed human host, the individual becomes infectious, $I(t)$, at time $t$. Applying similar reasoning in  \cite{cooke-driessche},
 the exposed population, $E(t)$, at time $t$ can be written as follows
  \begin{equation}\label{ch1.sec0.eqn1a}
    E(t)=E(t_{0})e^{-\mu (t-t_{0})}p_{1}(t-t_{0})+\int^{t}_{t_{0}}\beta S(\xi)e^{-\mu T_{1}}G(I(\xi-T_{1}))e^{-\mu(t-\xi)}p_{1}(t-\xi)d \xi,
   \end{equation}
   where
   \begin{equation}\label{ch1.seco.eqn1b}
     p_{1}(t-\xi)=\left\{\begin{array}{l}0,t-\xi\geq T_{2},\\
 1, t-\xi< T_{2} \end{array}\right.
   \end{equation}
   represents the probability that an individual remains exposed over the time interval $[\xi, t]$.
   It is easy to see from (\ref{ch1.sec0.eqn1a}) that under the assumption that the disease has been in the population for at least a time $t>\max_{t_{0}\leq T_{1}\leq h_{1}, t_{0}\leq T_{2}\leq h_{2}} {( T_{1}+ T_{2})}$, in fact, $t>h_{1}+h_{2}$, so that all initial perturbations have died out, the expected number of exposed individuals at time $t$ is given  by
\begin{equation}\label{ch1.sec0.eqn1}
E(t)=\int_{t_{0}}^{h_{2}}f_{T_{2}}(u)\int_{t-u}^{t}\beta \int^{h_{1}}_{t_{0}} f_{T_{1}}(s) e^{-\mu s}S(v)G(I(v-s))e^{-\mu(t-u)}dsdvdu.
\end{equation}
  %%%%%
    Similarly, for the removal population, $R(t)$, at time $t$, individuals recover from the infectious state $I(t)$  at the per capita rate $\alpha$  and acquire natural immunity.  The natural immunity wanes after the varying immunity period $T_{3}=r\in [ t_{0},\infty]$, and removed individuals become susceptible again to the disease. Therefore, at time $t$, individuals leave the infectious state at the rate $\alpha I(t)$  and become part of the removal population $R(t)$. Thus, at time $t$ the removed population is given by the following equation
  \begin{equation}\label{ch1.sec0.eqn2a}
    R(t)=R(t_{0})e^{-\mu (t-t_{0})}p_{2}(t-t_{0})+\int^{t}_{t_{0}}\alpha I(\xi)e^{-\mu(t-\xi)}p_{2}(t-\xi)d \xi,
  \end{equation}
    where
    \begin{equation}\label{ch1.sec0.eqn2b}
      p_{2}(t-\xi)=\left\{\begin{array}{l}0,t-\xi\geq T_{3},\\
 1, t-\xi< T_{3} \end{array}\right.
    \end{equation}
 represents the probability that an individual remains naturally immune to the disease over the time interval $[\xi, t]$.
 But it follows from  (\ref{ch1.sec0.eqn2a}) that under the assumption that the disease has been in the population for at least a time $t> \max_{t_{0}\leq T_{1}\leq h_{1}, t_{0}\leq T_{2}\leq h_{2}, T_{3}\geq t_{0}}{(T_{1}+ T_{2}, T_{3})}\geq \max_{t_{0}\leq T_{3}}{(T_{3})}$, in fact, the disease has been in the population for sufficiently large  amount of time so that all initial perturbations have died out,  then the expected number of removal individuals at time $t$ can be written as
  %%%%%
  \begin{equation}\label{ch1.sec0.eqn2}
R(t)=\int_{t_{0}}^{\infty}f_{T_{3}}(r)\int_{t-r}^{t}\alpha I(v)e^{-\mu (t-v)}dvdr.
\end{equation}
There is also constant birth rate $B$ of susceptible individuals in the population. Furthermore, individuals die additionally due to disease related causes at the rate $d$. A compartmental framework illustrating the transition rates between the different states in the system and also showing the delays in the disease dynamics is given in Figure~\ref{ch1.sec4.figure 1}.
%%
%%%
%%%
%%%%SEIRS compartmental framework
It follows from (\ref{ch1.sec0.eqn0}), (\ref{ch1.sec0.eqn1}), (\ref{ch1.sec0.eqn2}) and the transition rates illustrated in the compartmental framework in Figure~\ref{ch1.sec4.figure 1} above that the family of SEIRS epidemic dynamic models for a vector-borne diseases in the absence of any random environmental fluctuations can be written as follows:
\begin{eqnarray}
dS(t)&=&\left[ B-\beta S(t)\int^{h_{1}}_{t_{0}}f_{T_{1}}(s) e^{-\mu s}G(I(t-s))ds - \mu S(t)+ \alpha \int_{t_{0}}^{\infty}f_{T_{3}}(r)I(t-r)e^{-\mu r}dr \right]dt,\nonumber\\
&&\label{ch1.sec0.eq3}\\
dE(t)&=& \left[ \beta S(t)\int^{h_{1}}_{t_{0}}f_{T_{1}}(s) e^{-\mu s}G(I(t-s))ds - \mu E(t)\right.\nonumber\\
&&\left.-\beta \int_{t_{0}}^{h_{2}}f_{T_{2}}(u)S(t-u)\int^{h_{1}}_{t_{0}}f_{T_{1}}(s) e^{-\mu s-\mu u}G(I(t-s-u))dsdu \right]dt,\label{ch1.sec0.eq4}\\
&&\nonumber\\
dI(t)&=& \left[\beta \int_{t_{0}}^{h_{2}}f_{T_{2}}(u)S(t-u)\int^{h_{1}}_{t_{0}}f_{T_{1}}(s) e^{-\mu s-\mu u}G(I(t-s-u))dsdu- (\mu +d+ \alpha) I(t) \right]dt,\nonumber\\
&&\label{ch1.sec0.eq5}\\
dR(t)&=&\left[ \alpha I(t) - \mu R(t)- \alpha \int_{t_{0}}^{\infty}f_{T_{3}}(r)I(t-r)e^{-\mu s}dr \right]dt,\label{ch1.sec0.eq6}
\end{eqnarray}
where the initial conditions are given in the following: Let $h= h_{1}+ h_{2}$ and define
\begin{eqnarray}
&&\left(S(t),E(t), I(t), R(t)\right)
=\left(\varphi_{1}(t),\varphi_{2}(t), \varphi_{3}(t),\varphi_{4}(t)\right), t\in (-\infty,t_{0}],\nonumber\\% t\in [t_{0}-h,t_{0}],\quad and\quad=
&&\varphi_{k}\in \mathcal{C}((-\infty,t_{0}],\mathbb{R}_{+}),\forall k=1,2,3,4, \nonumber\\
&&\varphi_{k}(t_{0})>0,\forall k=1,2,3,4,\nonumber\\
 \label{ch1.sec0.eq06a}
\end{eqnarray}
where $\mathcal{C}((-\infty,t_{0}],\mathbb{R}_{+})$ is the space of continuous functions with  the supremum norm
\begin{equation}\label{ch1.sec0.eq06b}
||\varphi||_{\infty}=\sup_{ t\leq t_{0}}{|\varphi(t)|}.
\end{equation}
%%%%%%%%%%%%%%%%%%%%%%%%%%%%%%%
%%%%%%%%%%%%%%%%%%%%%
 It is assumed that the  effects of random environmental fluctuations lead to variability in the disease transmission and natural death rates.
 For $t\geq t_{0}$, let $(\Omega, \mathfrak{F}, P)$ be a complete probability space, and $\mathfrak{F}_{t}$ be a filtration (that is, sub $\sigma$- algebra $\mathfrak{F}_{t}$ that satisfies the following: given $t_{1}\leq t_{2} \Rightarrow \mathfrak{F}_{t_{1}}\subset \mathfrak{F}_{t_{2}}; E\in \mathfrak{F}_{t}$ and $P(E)=0 \Rightarrow E\in \mathfrak{F}_{0} $ ).
% The stochastic perturbations are directly proportional to the state $X(t)=(S(t), I(t), R(t))$ of the system.
 Indeed, the variability in the disease transmission and natural death rates are represented by the white noise processes as follows:
 \begin{equation}\label{ch1.sec0.eq7}
 \mu  \rightarrow \mu  + \sigma_{i}\xi_{i}(t),\quad \xi_{i}(t) dt= dw_{i}(t),i=S,E,I,R, \quad  \beta \rightarrow \beta + \sigma_{\beta}\xi_{\beta}(t),\quad \xi_{\beta}(t)dt=dw_{\beta}(t),
 \end{equation}
 where $\xi_{i}(t)$ and $w_{i}(t)$ represent the  standard white noise and normalized Wiener processes  for the  $i^{th}$ state at time $t$, with the following properties: $w(0)=0, E(W(t))=0, var(w(t))=t$.  Furthermore,  $var(\mu(t))=\sigma^{2}_{i},i=S,E,I,R $, represents the intensity value of the environmental white noise process due to the random fluctuations in the natural death rate in the $i^{th}$ state, and $var(\beta(t))=\sigma^{2}_{\beta}$ is the intensity value of the white noise process due to the random fluctuations in the disease transmission rate.

  Indeed, the intensity values $\sigma^{2}_{i},i=S,E,I,R, \beta $ of the white noise processes:   $\tilde{\mu}(t)=\mu  + \sigma_{i}\xi_{i}(t)$ and $\tilde{\beta} (t)=\beta + \sigma_{\beta}\xi_{\beta}(t)$ representing the variability in the natural death rate, $\tilde{\mu}(t)$, and disease transmission rate, $\tilde{\beta} (t)$,  at time $t$, owing to the random fluctuations that occur during the disease transmission and natural death processes of the disease dynamics,  measures the average deviation of the random variable disease transmission, $\tilde{\beta}$,  and natural death, $\tilde{\mu}$, rates from their constant mean values - $\beta$ and $\mu$, respectively, over the infinitesimally small time interval $[t, t+dt]$. This measure reflects the force of the random fluctuations that occur during the disease outbreak at anytime,  which lead to oscillations in the natural death and disease transmission rates overtime, and consequently lead to oscillations of the  sizes of the susceptible, exposed, infectious and removal classes of the total population over time during the disease outbreak.

 Substituting (\ref{ch1.sec0.eq7}) into the deterministic system (\ref{ch1.sec0.eq3})-(\ref{ch1.sec0.eq6}) leads to the following generalized system of Ito-Doob stochastic differential equations describing the dynamics of  vector-borne diseases in the human population.
 \begin{eqnarray}
dS(t)&=&\left[ B-\beta S(t)\int^{h_{1}}_{t_{0}}f_{T_{1}}(s) e^{-\mu s}G(I(t-s))ds - \mu S(t)+ \alpha \int_{t_{0}}^{\infty}f_{T_{3}}(r)I(t-r)e^{-\mu r}dr \right]dt\nonumber\\
&&-\sigma_{S}S(t)dw_{S}(t)-\sigma_{\beta} S(t)\int^{h_{1}}_{t_{0}}f_{T_{1}}(s) e^{-\mu s}G(I(t-s))dsdw_{\beta}(t) \label{ch1.sec0.eq8}\\
dE(t)&=& \left[ \beta S(t)\int^{h_{1}}_{t_{0}}f_{T_{1}}(s) e^{-\mu s}G(I(t-s))ds - \mu E(t)\right.\nonumber\\
&&\left.-\beta \int_{t_{0}}^{h_{2}}f_{T_{2}}(u)S(t-u)\int^{h_{1}}_{t_{0}}f_{T_{1}}(s) e^{-\mu s-\mu u}G(I(t-s-u))dsdu \right]dt\nonumber\\
&&-\sigma_{E}E(t)dw_{E}(t)+\sigma_{\beta} S(t)\int^{h_{1}}_{t_{0}}f_{T_{1}}(s) e^{-\mu s}G(I(t-s))dsdw_{\beta}(t)\nonumber\\
&&-\sigma_{\beta} \int_{t_{0}}^{h_{2}}f_{T_{2}}(u)S(t-u)\int^{h_{1}}_{t_{0}}f_{T_{1}}(s) e^{-\mu s-\mu u}G(I(t-s-u))dsdudw_{\beta}(t)\label{ch1.sec0.eq9}\\
dI(t)&=& \left[\beta \int_{t_{0}}^{h_{2}}f_{T_{2}}(u)S(t-u)\int^{h_{1}}_{t_{0}}f_{T_{1}}(s) e^{-\mu s-\mu u}G(I(t-s-u))dsdu- (\mu +d+ \alpha) I(t) \right]dt\nonumber\\
&&-\sigma_{I}I(t)dw_{I}(t)+\sigma_{\beta} \int_{t_{0}}^{h_{2}}f_{T_{2}}(u)S(t-u)\int^{h_{1}}_{t_{0}}f_{T_{1}}(s) e^{-\mu s-\mu u}G(I(t-s-u))dsdudw_{\beta}(t)\nonumber\\
&&\label{ch1.sec0.eq10}\\
dR(t)&=&\left[ \alpha I(t) - \mu R(t)- \alpha \int_{t_{0}}^{\infty}f_{T_{3}}(r)I(t-r)e^{-\mu s}dr \right]dt-\sigma_{R}R(t)dw_{R}(t),\label{ch1.sec0.eq11}
\end{eqnarray}
where the initial conditions are given in the following: Let $h= h_{1}+ h_{2}$ and define
\begin{eqnarray}
&&\left(S(t),E(t), I(t), R(t)\right)
=\left(\varphi_{1}(t),\varphi_{2}(t), \varphi_{3}(t),\varphi_{4}(t)\right), t\in (-\infty,t_{0}],\nonumber\\% t\in [t_{0}-h,t_{0}],\quad and\quad=
&&\varphi_{k}\in \mathcal{C}((-\infty,t_{0}],\mathbb{R}_{+}),\forall k=1,2,3,4, \nonumber\\
&&\varphi_{k}(t_{0})>0,\forall k=1,2,3,4,\nonumber\\
 \label{ch1.sec0.eq12}
\end{eqnarray}
where $\mathcal{C}((-\infty,t_{0}],\mathbb{R}_{+})$ is the space of continuous functions with  the supremum norm
\begin{equation}\label{ch1.sec0.eq13}
||\varphi||_{\infty}=\sup_{ t\leq t_{0}}{|\varphi(t)|}.
\end{equation}
 %and $w$ is a Wierner process.
Furthermore, the random continuous functions $\varphi_{k},k=1,2,3,4$ are
$\mathfrak{F}_{0}-measurable$, or  independent of $w(t)$
for all $t\geq t_{0}$.
%%%--------------------------------------------------------------
%%----------------------------------------------------------------
%%%
%%%
%%%Discuss the deterministic system
%%%
%%%
%%%%%%%%Discuss the nonlinear incidence rate
%%%%%%%%
%%%%%%%%%

Several epidemiological studies \cite{gumel,zhica,joaq,kyrychko,qun} have been conducted involving families of SIR, SEIRS, SIS etc. epidemic dynamic  models, where the family type is determined by the class of functions satisfying different general assumptions which characterize the nonlinear character of the incidence function $G(I)$ of the disease dynamics. Some general properties of the incidence function $G$ assumed in this study include the following:
\begin{assumption}\label{ch1.sec0.assum1}
\begin{enumerate}
  \item [$A1$]$G(0)=0$.
  \item [$A2$]$G(I)$ is strictly monotonic on $[0,\infty)$.
  \item [$A3$] $G''(I)<0$ $\Leftrightarrow$ $G(I)$ is differentiable concave on $[0,\infty)$.
  \item [$A4$] $\lim_{I\rightarrow \infty}G(I)=C, 0\leq C<\infty$ $\Leftrightarrow$  $G(I)$ has a horizontal asymptote $0\leq C<\infty$.
  \item [$A5$] $G(I)\leq I, \forall I>0$ $\Leftrightarrow$ $G(I)$ is at most as large as the identity function $f:I\mapsto I$ over the positive all $I\in (0,\infty)$.
\end{enumerate}
\end{assumption}
An incidence function $G$ that satisfies  Assumption~\ref{ch1.sec0.assum1} $A1$-$A5$ can be used to describe the disease transmission process of a vector-borne disease scenario, where the disease dynamics is represented by the system (\ref{ch1.sec0.eq8})-(\ref{ch1.sec0.eq11}), and the disease transmission rate between the vectors and the human beings initially increases or decreases for small values of the infectious population size, and is bounded or steady for sufficiently large size of the infectious  individuals in the population.  It is noted that Assumption~\ref{ch1.sec0.assum1} is a generalization of some subcases of the assumptions $A1$-$A5$ investigated in \cite{gumel,zhica,kyrychko, qun}. Some examples of frequently used incidence functions in the literature that  satisfy Assumption~\ref{ch1.sec0.assum1}$A1$-$A5$ include:  $G(I(t))=\frac{I(t)}{1+\alpha I(t)}, \alpha>0$, $G(I(t))=\frac{I(t)}{1+\alpha I^{2}(t)}, \alpha>0$, $G(I(t))=I^{p}(t),0<p<1$ and $G(I)=1-e^{-aI}, a>0$.

%%%%%%%%
%%%%%%%% that characterize the nonlinear behavior of the  incidence function $G(I)$.
%%%%%%%%
Observe that (\ref{ch1.sec0.eq9}) and (\ref{ch1.sec0.eq11}), and the corresponding equations (\ref{ch1.sec0.eq4}) and (\ref{ch1.sec0.eq6}) all decouple from the other two equations in their respective systems: (\ref{ch1.sec0.eq8})-(\ref{ch1.sec0.eq11}) and (\ref{ch1.sec0.eq3})-(\ref{ch1.sec0.eq6}). Nevertheless, for convenience most of the results in this paper related to the systems (\ref{ch1.sec0.eq8})-(\ref{ch1.sec0.eq11}) and (\ref{ch1.sec0.eq3})-(\ref{ch1.sec0.eq6}) will be shown mostly for the vector $X(t)=(S(t), E(t), I(t))^{T}$. The following notations are utilized:
\begin{equation}\label{ch1.sec0.eq13b}
\left\{
  \begin{array}{lll}
    Y(t)&=&(S(t), E(t), I(t), R(t))^{T} \\
   X(t)&=&(S(t), E(t), I(t))^{T} \\
   N(t)&=&S(t)+ E(t)+ I(t)+ R(t).
  \end{array}
  \right.
\end{equation}
\section{Model Validation Results\label{ch1.sec1}}
%The following lemmas will be utilized to establish the existence results.
The analysis and results in this manuscript are exhibited for both the deterministic and stochastic systems (\ref{ch1.sec0.eq3})-(\ref{ch1.sec0.eq6}) and  (\ref{ch1.sec0.eq8})-(\ref{ch1.sec0.eq11}). These necessitate the existence and uniqueness of the solutions of the stochastic and deterministic systems.  The standard methods  utilized in the earlier studies\cite{wanduku-determ,Wanduku-2017,wanduku-delay,divine5} are applied to establish the results.
  The following Lemma describes the behavior of the positive local solutions for the systems (\ref{ch1.sec0.eq3})-(\ref{ch1.sec0.eq6}) and (\ref{ch1.sec0.eq8})-(\ref{ch1.sec0.eq11}). This result will be useful in   establishing the existence and uniqueness results for the global solutions of the deterministic and stochastic systems (\ref{ch1.sec0.eq3})-(\ref{ch1.sec0.eq6}) and (\ref{ch1.sec0.eq8})-(\ref{ch1.sec0.eq11}).
\begin{lemma}\label{ch1.sec1.lemma1}
Suppose for some $\tau_{e}>t_{0}\geq 0$ the systems (\ref{ch1.sec0.eq3})-(\ref{ch1.sec0.eq6}) and (\ref{ch1.sec0.eq8})-(\ref{ch1.sec0.eq11}) with initial conditions in (\ref{ch1.sec0.eq06a})-(\ref{ch1.sec0.eq06b}) and (\ref{ch1.sec0.eq12})-(\ref{ch1.sec0.eq13}) respectively have  unique positive solutions denoted $Y(t)\in \mathbb{R}^{4}_{+}$, for all $t\in (-\infty, \tau_{e}]$, then  if $N(t_{0})\leq \frac{B}{\mu}$, it follows that  $N(t)\leq \frac{B}{\mu}$ . In addition, the set denoted by
%of positive solutions for the system (\ref{ch1.sec0.eq3})-(\ref{ch1.sec0.eq6}) namely:
\begin{equation}\label{ch1.sec1.lemma1.eq1}
  D(\tau_{e})=\left\{Y(t)\in \mathbb{R}^{4}_{+}: N(t)=S(t)+ E(t)+ I(t)+ R(t)\leq \frac{B}{\mu}, \forall t\in (-\infty, \tau_{e}] \right\}=\bar{B}^{(-\infty, \tau_{e}]}_{\mathbb{R}^{4}_{+},}\left(0,\frac{B}{\mu}\right),
\end{equation}
is locally self-invariant with respect to the systems (\ref{ch1.sec0.eq3})-(\ref{ch1.sec0.eq6}) and (\ref{ch1.sec0.eq8})-(\ref{ch1.sec0.eq11}), where $\bar{B}^{(-\infty, \tau_{e}]}_{\mathbb{R}^{4}_{+},}\left(0,\frac{B}{\mu}\right)$ is the closed ball in $\mathbb{R}^{4}_{+}$ centered at the origin with radius $\frac{B}{\mu}$ containing the local positive solutions defined over $(-\infty, \tau_{e}]$.
\end{lemma}
Proof:\\
The proof of the result for (\ref{ch1.sec0.eq3})-(\ref{ch1.sec0.eq6}) and  (\ref{ch1.sec0.eq8})-(\ref{ch1.sec0.eq11}) are the same, hence without of loss of generality, the result will be shown only for the stochastic system (\ref{ch1.sec0.eq8})-(\ref{ch1.sec0.eq11}).
It follows directly from (\ref{ch1.sec0.eq8})-(\ref{ch1.sec0.eq11}) that
\begin{equation}\label{ch1.sec1.lemma1.eq2}
dN(t)=[B-\mu N(t)-dI(t)]dt.
\end{equation}
The result then follows easily by observing that for $Y(t)\in \mathbb{R}^{4}_{+}$, the equation (\ref{ch1.sec1.lemma1.eq2}) leads to  $N(t)\leq \frac{B}{\mu}-\frac{B}{\mu}e^{-\mu(t-t_{0})}+N(t_{0})e^{-\mu(t-t_{0})}$. And under the assumption that $N(t_{0})\leq \frac{B}{\mu}$, the result follows.
%%%%

 The following set of theorems presents the existence and uniqueness results for the global solutions  of the deterministic and stochastic systems (\ref{ch1.sec0.eq3})-(\ref{ch1.sec0.eq6}) and  (\ref{ch1.sec0.eq8})-(\ref{ch1.sec0.eq11}). First, the existence results  for the deterministic system (\ref{ch1.sec0.eq3})-(\ref{ch1.sec0.eq6}) is established. The standard technique applied in \cite{wanduku-determ} is utilized to establish the results.
\begin{thm}\label{ch1.sec1.thm1a}
  Given the initial conditions (\ref{ch1.sec0.eq06a})-(\ref{ch1.sec0.eq06b}), there exists a unique solution $Y(t)=(S(t),E(t), I(t), R(t))^{T}$ satisfying (\ref{ch1.sec0.eq3})-(\ref{ch1.sec0.eq6}), for all $t\geq t_{0}$. Moreover, the solution is nonnegative for all $t\geq t_{0}$  and also lies in $D(\infty)$.  That is, $S(t)>0,E(t)>0,  I(t)>0, R(t)>0, \forall t\geq t_{0}$  and
   \begin{equation}\label{ch1.sec1.thm1a.eq0}
\limsup_{t\rightarrow \infty} N(t)\leq S^{*}_{0}=\frac{B}{\mu},
   \end{equation}
    for $N(t)=S(t)+E(t)+ I(t)+R(t)$,  and $Y(t)\in D(\infty)=\bar{B}^{(-\infty, \infty)}_{\mathbb{R}^{4}_{+},}\left(0,\frac{B}{\mu}\right)$, where $D(\infty)$ is defined in (\ref{ch1.sec1.lemma1.eq1}). %Lemma~\ref{ch1.sec1.lemma1},
%\end{equation}
 %Then the solutions  $S^{ru}_{ia}(t,w),I^{ru}_{ia(t,w)}> 0$,
\end{thm}
Proof:\\
It is easy to see that the rate functions of the system (\ref{ch1.sec0.eq3})-(\ref{ch1.sec0.eq6}) are nonlinear, continuous in their argument variables, and satisfy the local Lipschitz condition for the given initial data (\ref{ch1.sec0.eq06a})-(\ref{ch1.sec0.eq06b}). Therefore, there exists a unique local solution $Y(t)=(S(t), E(t), I(t), R(t))^{T}$ on $t\in (-\infty,\tau_{e}]$, where $\tau_{e}> t_{0}\geq 0$. The rest of the result such as showing that the local solution is positive and extending the local solution inductively to a global positive solution follow a standard technique \cite{wanduku-determ}. Moreover,  from  Lemma~\ref{ch1.sec1.lemma1}, it follows that $Y(t)\in D(\infty)$ and (\ref{ch1.sec1.thm1a.eq0}) holds.

 The next theorem presents the existence and uniqueness results for the global solutions  of the stochastic system (\ref{ch1.sec0.eq8})-(\ref{ch1.sec0.eq11}). The standard technique applied in \cite{Wanduku-2017,wanduku-delay} is utilized to establish the results.
\begin{thm}\label{ch1.sec1.thm1}
  Given the initial conditions (\ref{ch1.sec0.eq12}) and (\ref{ch1.sec0.eq13}), there exists a unique solution process $X(t,w)=(S(t,w),E(t,w), I(t,w))^{T}$ satisfying (\ref{ch1.sec0.eq8})-(\ref{ch1.sec0.eq11}), for all $t\geq t_{0}$. Moreover, the solution process is positive for all $t\geq t_{0}$ a.s. and lies in $D(\infty)$.  That is, $S(t,w)>0,E(t,w)>0,  I(t,w)>0, \forall t\geq t_{0}$ a.s. and $X(t,w)\in D(\infty)=\bar{B}^{(-\infty, \infty)}_{\mathbb{R}^{4}_{+},}\left(0,\frac{B}{\mu}\right)$, where $D(\infty)$ is defined in Lemma~\ref{ch1.sec1.lemma1}, (\ref{ch1.sec1.lemma1.eq1}).
%\end{equation}
 %Then the solutions  $S^{ru}_{ia}(t,w),I^{ru}_{ia(t,w)}> 0$,
\end{thm}
Proof:\\
It is easy to see that the coefficients of (\ref{ch1.sec0.eq8})-(\ref{ch1.sec0.eq11}) satisfy the local Lipschitz condition for the given initial data (\ref{ch1.sec0.eq12}). Therefore there exist a unique maximal local solution $X(t,w)=(S(t,w), E(t,w), I(t,w))$ on $t\in (-\infty,\tau_{e}(w)]$, where $\tau_{e}(w)$ is the first hitting time or the explosion time\cite{mao}. The following shows that $X(t,w)\in D(\tau_{e})$ almost surely,  where $D(\tau_{e}(w))$ is defined in Lemma~\ref{ch1.sec1.lemma1} (\ref{ch1.sec1.lemma1.eq1}).
Define the following stopping time
%%%%%%%%%%%%%%%%%%%%%%%%%%%%
%%%%%%%%%%%%%%%%%%%%%%%%%%%%
\begin{equation}\label{ch1.sec1.thm1.eq1}
\left\{
\begin{array}{lll}
\tau_{+}&=&sup\{t\in (t_{0},\tau_{e}(w)): S|_{[t_{0},t]}>0,\quad E|_{[t_{0},t]}>0,\quad and\quad I|_{[t_{0},t]}>0 \},\\
\tau_{+}(t)&=&\min(t,\tau_{+}),\quad for\quad t\geq t_{0}.\\
\end{array}
\right.
\end{equation}
and lets show that $\tau_{+}(t)=\tau_{e}(w)$ a.s. Suppose on the contrary that $P(\tau_{+}(t)<\tau_{e}(w))>0$. Let $w\in \{\tau_{+}(t)<\tau_{e}(w)\}$, and $t\in [t_{0},\tau_{+}(t))$. Define
%\begin{equation}%\label{ch2.thm1.eq2}
\begin{equation}\label{ch1.sec1.thm1.eq2}
\left\{
\begin{array}{ll}
V(X(t))=V_{1}(X(t))+V_{2}(X(t))+V_{3}(X(t)),\\
V_{1}(X(t))=\ln(S(t)),\quad V_{2}(X(t))=\ln(E(t)),\quad V_{3}(X(t))=\ln(I(t)),\forall t\leq\tau_{+}(t).
\end{array}
\right.
\end{equation}
It follows from (\ref{ch1.sec1.thm1.eq2}) that
\begin{equation}\label{ch1.sec1.thm1.eq3}
  dV(X(t))=dV_{1}(X(t))+dV_{2}(X(t))+dV_{3}(X(t)),
\end{equation}
where
\begin{eqnarray}
  dV_{1}(X(t)) &=& \frac{1}{S(t)}dS(t)-\frac{1}{2}\frac{1}{S^{2}(t)}(dS(t))^{2}\nonumber \\
   &=&\left[ \frac{B}{S(t)}-\beta \int^{h_{1}}_{t_{0}}f_{T_{1}}(s) e^{-\mu s}G(I(t-s))ds - \mu + \frac{\alpha}{S(t)} \int_{t_{0}}^{\infty}f_{T_{3}}(r)I(t-r)e^{-\mu r}dr \right.\nonumber\\
   &&\left.-\frac{1}{2}\sigma^{2}_{S}-\frac{1}{2}\sigma^{2}_{\beta}\left(\int^{h_{1}}_{t_{0}}f_{T_{1}}(s) e^{-\mu s}G(I(t-s))ds\right)^{2}\right]dt\nonumber\\
&&-\sigma_{S}dw_{S}(t)-\sigma_{\beta} \int^{h_{1}}_{t_{0}}f_{T_{1}}(s) e^{-\mu s}G(I(t-s))dsdw_{\beta}(t), \label{ch1.sec1.thm1.eq4}
\end{eqnarray}
%%%%%%%%%%%%%%%%%%%%%%%%%%%
%%%%%%%%%%%%%%%%%%%%%%%%
\begin{eqnarray}
  dV_{2}(X(t)) &=& \frac{1}{E(t)}dE(t)-\frac{1}{2}\frac{1}{E^{2}(t)}(dE(t))^{2} \nonumber\\
  &=& \left[ \beta \frac{S(t)}{E(t)}\int^{h_{1}}_{t_{0}}f_{T_{1}}(s) e^{-\mu s}G(I(t-s))ds - \mu \right.\nonumber\\
&&\left.-\beta\frac{1}{E(t)} \int_{t_{0}}^{h_{2}}f_{T_{2}}(u)S(t-u)\int^{h_{1}}_{t_{0}}f_{T_{1}}(s) e^{-\mu s-\mu u}G(I(t-s-u))dsdu \right.\nonumber\\
&&\left.-\frac{1}{2}\sigma^{2}_{E}-\frac{1}{2}\sigma^{2}_{\beta}\frac{S^{2}(t)}{E^{2}(t)}\left(\int^{h_{1}}_{t_{0}}f_{T_{1}}(s) e^{-\mu s}G(I(t-s))ds\right)^{2}\right.\nonumber\\
&&\left.-\frac{1}{2}\sigma^{2}_{\beta}\frac{1}{E^{2}(t)}\left(\int_{t_{0}}^{h_{2}}f_{T_{2}}(u)S(t-u)\int^{h_{1}}_{t_{0}}f_{T_{1}}(s) e^{-\mu s-\mu u}G(I(t-s-u))dsdu \right)^{2}\right]dt\nonumber\\
&&-\sigma_{E}dw_{E}(t)+\sigma_{\beta} \frac{S(t)}{E(t)}\int^{h_{1}}_{t_{0}}f_{T_{1}}(s) e^{-\mu s}G(I(t-s))dsdw_{\beta}(t)\nonumber\\
&&-\sigma_{\beta}\frac{1}{E(t)} \int_{t_{0}}^{h_{2}}f_{T_{2}}(u)S(t-u)\int^{h_{1}}_{t_{0}}f_{T_{1}}(s) e^{-\mu s-\mu u}G(I(t-s-u))dsdudw_{\beta}(t),\nonumber\\
\label{ch1.sec1.thm1.eq5}
\end{eqnarray}
%%%%%%%%%%%%%%%%%%%%%%%%%
%%%%%%%%%%%%%%%%%%%%%%%%%%
and
\begin{eqnarray}
  dV_{3}(X(t)) &=& \frac{1}{I(t)}dI(t)-\frac{1}{2}\frac{1}{I^{2}(t)}(dI(t))^{2}\nonumber \\
  &=& \left[\beta \frac{1}{I(t)}\int_{t_{0}}^{h_{2}}f_{T_{2}}(u)S(t-u)\int^{h_{1}}_{t_{0}}f_{T_{1}}(s) e^{-\mu s-\mu u}G(I(t-s-u))dsdu- (\mu +d+ \alpha)\right.  \nonumber\\
  &&\left.-\frac{1}{2}\sigma^{2}_{I}-\frac{1}{2}\sigma^{2}_{\beta}\left(\int_{t_{0}}^{h_{2}}f_{T_{2}}(u)S(t-u)\int^{h_{1}}_{t_{0}}f_{T_{1}}(s) e^{-\mu s-\mu u}G(I(t-s-u))dsdu \right)^{2}\right]dt\nonumber\\
&&-\sigma_{I}dw_{I}(t)+\sigma_{\beta}\frac{1}{I(t)} \int_{t_{0}}^{h_{2}}f_{T_{2}}(u)S(t-u)\int^{h_{1}}_{t_{0}}f_{T_{1}}(s) e^{-\mu s-\mu u}G(I(t-s-u))dsdudw_{\beta}(t)\nonumber\\
&&\label{ch1.sec1.thm1.eq6}
\end{eqnarray}
It follows from (\ref{ch1.sec1.thm1.eq3})-(\ref{ch1.sec1.thm1.eq6}) that for $t<\tau_{+}(t)$,
\begin{eqnarray}
  V(X(t))-V(X(t_{0})) &\geq& \int^{t}_{t_{0}}\left[-\beta \int^{h_{1}}_{t_{0}}f_{T_{1}}(s) e^{-\mu s}G(I(\xi-s))ds-\frac{1}{2}\sigma^{2}_{S}\right.\nonumber\\
   &&\left.-\frac{1}{2}\sigma^{2}_{\beta}\left(\int^{h_{1}}_{t_{0}}f_{T_{1}}(s) e^{-\mu s}G(I(\xi-s))ds\right)^{2}\right]d\xi\nonumber\\
   %%%%%%%%%%%%%%%%%%%%%%%%%%%%%%%%
   %%%%%%%%%%%%%%%%%V_2
   %%%%%%%%%%%%%%%%%%%%%
   &&+ \int_{t}^{t_{0}}\left[-\beta\frac{1}{E(\xi)} \int_{t_{0}}^{h_{2}}f_{T_{2}}(u)S(\xi-u)\int^{h_{1}}_{t_{0}}f_{T_{1}}(s) e^{-\mu s-\mu u}G(I(\xi-s-u))dsdu
\right.\nonumber\\
%&&\left. \right.\nonumber\\
&&\left.-\frac{1}{2}\sigma^{2}_{E}-\frac{1}{2}\sigma^{2}_{\beta}\frac{S^{2}(\xi)}{E^{2}(\xi)}\left(\int^{h_{1}}_{t_{0}}f_{T_{1}}(s) e^{-\mu s}G(I(\xi-s))ds\right)^{2}\right.\nonumber\\
&&\left.-\frac{1}{2}\sigma^{2}_{\beta}\frac{1}{E^{2}(\xi)}\left(\int_{t_{0}}^{h_{2}}f_{T_{2}}(u)S(\xi-u)\int^{h_{1}}_{t_{0}}f_{T_{1}}(s) e^{-\mu s-\mu u}G(I(\xi-s-u))dsdu \right)^{2}\right]d\xi\nonumber\\
%%%%%%%%%%%%%%%%%%%%%%%%%%%%%%%%
   %%%%%%%%%%%%%%%%%%%%%%%%%%%%%%%%% V_3
   %%%%%%%%%%%%%%%%%%%%%%%%%%%%%%%%%%%%%%%%%%%%%%%%%%%%%%%%%%%%%
   &&+ \int_{t}^{t_{0}}\left[- (3\mu +d+ \alpha)-\frac{1}{2}\sigma^{2}_{I}\right.  \nonumber\\
  &&\left.-\frac{1}{2}\sigma^{2}_{\beta}\left(\int_{t_{0}}^{h_{2}}f_{T_{2}}(u)S(\xi-u)\int^{h_{1}}_{t_{0}}f_{T_{1}}(s) e^{-\mu s-\mu u}G(I(\xi-s-u))dsdu \right)^{2}\right]d\xi\nonumber\\
   %%%%%%%%%%%%%%%%%%%%%%%%%%%%%%%%%%%%%%%%%%%%%%
   %%%%%%%%%%%%%%%%%%%%%%%%%%%%%%%%%%%%%%%%%%%%%%%%%%%%%%%%
   %%%Diffussion v_1
   %%%%%%%%%%%%%%%%%%%%%%%%%%%%%%%%%%%%%%%%%%%%%%%%%%%%%%%
   %%%%%%%%%%%%%%%%%%%%%%%%%%%%%%%%%%%%%%%%%%%%%%%%%%%%%%
&&+\int_{t}^{t_{0}}\left[-\sigma_{S}dw_{S}(\xi)-\sigma_{\beta} \int^{h_{1}}_{t_{0}}f_{T_{1}}(s) e^{-\mu s}G(I(\xi-s))dsdw_{\beta}(\xi)\right]\nonumber \\
%%%%%%%%%%%%%%%%%%%%%%%%%%%%%%%%%%%%%%
%%%%%%%%%%%%%%%%%%%%%%%%%%%%%%%%%%%%%%%%V_2
%%%%%%%%%%%%%%%%%%%%%%%%%%%%%%%%%%%%%%%
  &&+\int_{t}^{t_{0}}\left[-\sigma_{E}dw_{E}(\xi)+\sigma_{\beta} \frac{S(\xi)}{E(\xi)}\int^{h_{1}}_{t_{0}}f_{T_{1}}(s) e^{-\mu s}G(I(\xi-s))dsdw_{\beta}(\xi)\right]\nonumber\\
&&-\int_{t}^{t_{0}}\left[\sigma_{\beta}\frac{1}{E(\xi)} \int_{t_{0}}^{h_{2}}f_{T_{2}}(u)S(\xi-u)\int^{h_{1}}_{t_{0}}f_{T_{1}}(s) e^{-\mu s-\mu u}G(I(\xi-s-u))dsdudw_{\beta}(\xi)\right]\nonumber\\
%%%%%%%%%%%%%%%%%%%%
%%%%%%%%%%%%%%%%%%%%%V_3
&&+\int_{t_{0}}^{t}\left[-\sigma_{I}dw_{I}(\xi)\right.\nonumber\\
&&\left.+\sigma_{\beta}\frac{1}{I(\xi)} \int_{t_{0}}^{h_{2}}f_{T_{2}}(u)S(\xi-u)\int^{h_{1}}_{t_{0}}f_{T_{1}}(s) e^{-\mu s-\mu u}G(I(\xi-s-u))dsdudw_{\beta}(\xi)\right].\nonumber\\
&&\label{ch1.sec1.thm1.eq7}
%%%%%%%%%%%%%%%%%%%%%
\end{eqnarray}
Taking the limit on (\ref{ch1.sec1.thm1.eq7}) as $t\rightarrow \tau_{+}(t)$, it follows from (\ref{ch1.sec1.thm1.eq1})-(\ref{ch1.sec1.thm1.eq2}) that the left-hand side $V(X(t))-V(X(t_{0}))\leq -\infty$. This contradicts the finiteness of the right-handside of the inequality (\ref{ch1.sec1.thm1.eq7}). Hence $\tau_{+}(t)=\tau_{e}(w)$ a.s., that is, $X(t,w)\in D(\tau_{e})$.

The following shows that $\tau_{e}(w)=\infty$. Let $k>0$ be a positive integer such that $||\vec{\varphi}||_{1}\leq k$, where $\vec{\varphi}=\left(\varphi_{1}(t),\varphi_{2}(t), \varphi_{3}(t)\right), t\in (-\infty,t_{0}]$ defined in (\ref{ch1.sec0.eq12}), and $||.||_{1}$ is the $p-sum$ norm defined on $\mathbb{R}^{3}$, when $p=1$. Define the stopping time
\begin{equation}\label{ch1.sec1.thm1.eq8}
\left\{
\begin{array}{ll}
\tau_{k}=sup\{t\in [t_{0},\tau_{e}): ||X(s)||_{1}=S(s)+E(s)+I(s)\leq k, s\in[t_{0},t] \}\\
\tau_{k}(t)=\min(t,\tau_{k}).
\end{array}
\right.
\end{equation}
%where,
%\begin{equation}\label{ch1.sec1.thm1.eq9}
%||X(s)||_{1}=S(s)+E(s)+I(s).
%\end{equation}
It is easy to see that as $k\rightarrow \infty$, $\tau_{k}$ increases. Set $\lim_{k\rightarrow \infty}\tau_{k}(t)=\tau_{\infty}$. Then it follows that $\tau_{\infty}\leq \tau_{e}$ a.s.
 We show in the following that: (1.) $\tau_{e}=\tau_{\infty}\quad a.s.\Leftrightarrow P(\tau_{e}\neq \tau_{\infty})=0$, (2.)  $\tau_{\infty}=\infty\quad a.s.\Leftrightarrow P(\tau_{\infty}=\infty)=1$.

Suppose on the contrary that $P(\tau_{\infty}<\tau_{e})>0$. Let $w\in \{\tau_{\infty}<\tau_{e}\}$ and $t\leq \tau_{\infty}$.
 Define
\begin{equation}\label{ch1.sec1.thm1.eq9}
\left\{
\begin{array}{ll}
\hat{V}_{1}(X(t))=e^{\mu t}(S(t)+E(t)+I(t)),\\
\forall t\leq\tau_{k}(t).
\end{array}
\right.
\end{equation}
The Ito-Doob differential $d\hat{V}_{1}$ of (\ref{ch1.sec1.thm1.eq9}) with respect to the system (\ref{ch1.sec0.eq8})-(\ref{ch1.sec0.eq11}) is given as follows:
\begin{eqnarray}
 d\hat{V}_{1} &=& \mu e^{\mu t}(S(t)+E(t)+I(t)) dt + e^{\mu t}(dS(t)+dE(t)+dI(t))  \\
   &=& e^{\mu t}\left[B+\alpha \int_{t_{0}}^{\infty}f_{T_{3}}(r)I(t-r)e^{-\mu r}dr-(\alpha + d)I(t)\right]dt\nonumber\\
   &&-\sigma_{S}e^{\mu t}S(t)dw_{S}(t)-\sigma_{E}e^{\mu t}E(t)dw_{E}(t)-\sigma_{I}e^{\mu t}I(t)dw_{I}(t)\label{ch1.sec1.thm1.eq10}
\end{eqnarray}
%%%%%%%%%%%%%%%%%%%%%%%%%%%%%%%%%%%%%%%%%%%%%%%%%%%%%
Integrating (\ref{ch1.sec1.thm1.eq9}) over the interval $[t_{0}, \tau]$, and applying some algebraic manipulations and  simplifications  it follows that
\begin{eqnarray}
  V_{1}(X(\tau)) &=& V_{1}(X(t_{0}))+\frac{B}{\mu}\left(e^{\mu \tau}-e^{\mu t_{0}}\right)\nonumber\\
  &&+\int_{t_{0}}^{\infty}f_{T_{3}}(r)e^{-\mu r}\left(\int_{t_{0}-r}^{t_{0}}\alpha I(\xi)d\xi-\int_{\tau-r}^{\tau}\alpha I(\xi)d\xi\right)dr-\int_{t_{0}}^{\tau}d I(\xi)d\xi \nonumber\\
  &&+\int^{\tau}_{t_{0}}\left[-\sigma_{S}e^{\mu \xi}S(\xi)dw_{S}(\xi)-\sigma_{E}e^{\mu \xi}E(\xi)dw_{E}(\xi)-\sigma_{I}e^{\mu \xi}I(\xi)dw_{I}(\xi)\right]\label{ch1.sec1.thm1.eq11}
\end{eqnarray}
Removing negative terms from (\ref{ch1.sec1.thm1.eq11}), it implies from (\ref{ch1.sec0.eq12}) that
\begin{eqnarray}
  V_{1}(X(\tau)) &\leq& V_{1}(X(t_{0}))+\frac{B}{\mu}e^{\mu \tau}\nonumber\\
  &&+\int_{t_{0}}^{\infty}f_{T_{3}}(r)e^{-\mu r}\left(\int_{t_{0}-r}^{t_{0}}\alpha \varphi_{3}(\xi)d\xi\right)dr \nonumber\\
  &&+\int^{\tau}_{t_{0}}\left[-\sigma_{S}e^{\mu \xi}S(\xi)dw_{S}(\xi)-\sigma_{E}e^{\mu \xi}E(\xi)dw_{E}(\xi)-\sigma_{I}e^{\mu \xi}I(\xi)dw_{I}(\xi)\right]\label{ch1.sec1.thm1.eq12}
\end{eqnarray}
%%%%%%%%%%%%%%%%%%%%%%%%%%%%%%%%%%%%%%%%%%%%%%%%%%%%%
But from (\ref{ch1.sec1.thm1.eq9}) it is easy to see that for $\forall t\leq\tau_{k}(t)$,
\begin{equation}\label{ch1.sec1.thm1.eq12a}
  ||X(t)||_{1}=S(t)+E(t)+I(t)\leq V(X(t)).
\end{equation}
 Thus setting $\tau=\tau_{k}(t)$, then it follows from
(\ref{ch1.sec1.thm1.eq8}), (\ref{ch1.sec1.thm1.eq12}) and  (\ref{ch1.sec1.thm1.eq12a}) that
\begin{equation}\label{ch1.sec1.thm1.eq13}
  k=||X(\tau_{k}(t))||_{1}\leq V_{1}(X(\tau_{k}(t)))
\end{equation}
%%%%%%%%%%%%%%%%%%%%%%%%%%%%%%%%%%%%%%%%%%%%%%%%%%%%%%%%
Taking the limit on (\ref{ch1.sec1.thm1.eq13}) as $k\rightarrow \infty$ leads to a contradiction because the left-hand-side of the inequality (\ref{ch1.sec1.thm1.eq13}) is infinite, but following the right-hand-side  from (\ref{ch1.sec1.thm1.eq12}) leads to a finite value. Hence $\tau_{e}=\tau_{\infty}$ a.s. The following shows that $\tau_{e}=\tau_{\infty}=\infty$ a.s.
 %for all $k$ satisfying $||\varphi^{00}_{00}||_{1}\leq k$

  Let $\ w\in \{\tau_{e}<\infty\}$. It follows from (\ref{ch1.sec1.thm1.eq11})-(\ref{ch1.sec1.thm1.eq12}) that
  \begin{eqnarray}
  I_{\{\tau_{e}<\infty\}}V_{1}(X(\tau)) &\leq& I_{\{\tau_{e}<\infty\}}V_{1}(X(t_{0}))+I_{\{\tau_{e}<\infty\}}\frac{B}{\mu}e^{\mu \tau}\nonumber\\
  &&+I_{\{\tau_{e}<\infty\}}\int_{t_{0}}^{\infty}f_{T_{3}}(r)e^{-\mu r}\left(\int_{t_{0}-r}^{t_{0}}\alpha \varphi_{3}(\xi)d\xi\right)dr\nonumber\\
  &&+I_{\{\tau_{e}<\infty\}}\int^{\tau}_{t_{0}}\left[-\sigma_{S}e^{\mu \xi}S(\xi)dw_{S}(\xi)-\sigma_{E}e^{\mu \xi}E(\xi)dw_{E}(\xi)-\sigma_{I}e^{\mu \xi}I(\xi)dw_{I}(\xi)\right].
  \nonumber\\
  \label{ch1.sec1.thm1.eq14}
\end{eqnarray}
Suppose $\tau=\tau_{k}(t)\wedge T$, where $ T>0$ is arbitrary, then taking the expected value of (\ref{ch1.sec1.thm1.eq14}) follows that
\begin{equation}\label{ch1.sec1.thm1.eq14a}
  E(I_{\{\tau_{e}<\infty\}}V_{1}(X(\tau_{k}(t)\wedge T))) \leq V_{1}(X(t_{0}))+\frac{B}{\mu}e^{\mu T}
\end{equation}
But from (\ref{ch1.sec1.thm1.eq12a}) it is easy to see that
\begin{equation}\label{ch1.sec1.thm1.eq15}
 I_{\{\tau_{e}<\infty,\tau_{k}(t)\leq T\}}||X(\tau_{k}(t))||_{1}\leq I_{\{\tau_{e}<\infty\}}V_{1}(X(\tau_{k}(t)\wedge T))
\end{equation}
%%%%%%%%%%%%%%%%%%%%%%%%%%%%%%%%%%%%%%%%%%%%%%%%%%%%%%%%
%%%%%%%%%%%%%%%%%%%%%%%%%%%%%%%%%%%%%%%%%%%%%%%%%%%%%%%
It follows from (\ref{ch1.sec1.thm1.eq14})-(\ref{ch1.sec1.thm1.eq15}) and
   (\ref{ch1.sec1.thm1.eq8}) that
 \begin{eqnarray}
 P(\{\tau_{e}<\infty,\tau_{k}(t)\leq T\})k&=&E\left[I_{\{\tau_{e}<\infty,\tau_{k}(t)\leq T\}}||X(\tau_{k}(t))||_{1}\right]\nonumber\\
 &\leq& E\left[I_{\{\tau_{e}<\infty\}}V_{1}(X(\tau_{k}(t)\wedge T))\right]\nonumber\\
 &\leq& V_{1}(X(t_{0}))+\frac{B}{\mu}e^{\mu T}.
%&&+\sum_{r=1}^{M}\sum_{i=1}^{n_{r}}\sum_{q=1}^{M}\sum_{l=1}^{n_{q}}\int_{0}^{\infty}f^{rq}_{il}(t)\left[\varrho^{q}_{l}\int^{t_{0}}_{-t}\varphi^{rq}_{il2}(s)
%e^{\delta^{q}_{l}s}ds\right]dt\nonumber\\
\label{ch1.sec1.thm1.eq16}
 \end{eqnarray}
  It follows immediately from (\ref{ch1.sec1.thm1.eq16}) that
 $P(\{\tau_{e}<\infty,\tau_{\infty}\leq T\})\rightarrow 0$ as $k\rightarrow \infty$. Furthermore, since $T<\infty$ is arbitrary, we conclude that $P(\{\tau_{e}<\infty,\tau_{\infty}< \infty\})= 0$.
Finally,  by the total probability principle,
 \begin{eqnarray}
 P(\{\tau_{e}<\infty\})&=&P(\{\tau_{e}<\infty,\tau_{\infty}=\infty\})+P(\{\tau_{e}<\infty,\tau_{\infty}<\infty\})\nonumber\\
 &\leq&P(\{\tau_{e}\neq\tau_{\infty}\})+P(\{\tau_{e}<\infty,\tau_{\infty}<\infty\})\nonumber\\
 &=&0.\label{ch1.sec1.thm1.eq17}
 \end{eqnarray}
 Thus from (\ref{ch1.sec1.thm1.eq17}), $\tau_{e}=\tau_{\infty}=\infty$ a.s.. In addition, $X(t)\in D(\infty)$.
%%%%%%%%%%%%%%%%%%%%%%%%%%%%%%%%%%%%%%%%%%%%%%%%%%%%%%%%%%%%%%%%%%%%%%%%%%%%%%%%%%%%%%%%%%%%%%%%%%%
%%%%%%%%%%%%%%%%%%%%%%%%%%%%%%%%%%%%%%%%%%%%%%%%%%%%%%%
%%%%%%%%%%%%%%%%%%%%%%%%%%%%%%%%%%%%%%%%%%%%%%%%%%%%%%
%%%%%%%%%%%%%%%%%%%%%%%%
%%%%%
\begin{rem}\label{ch1.sec0.remark1}
\item[1.] Theorem~\ref{ch1.sec1.thm1} and Lemma~\ref{ch1.sec1.lemma1} signify that the stochastic system (\ref{ch1.sec0.eq8})-(\ref{ch1.sec0.eq11}) has a unique global positive solution process  $Y(t)\in \mathbb{R}^{4}_{+}$,  for all $t\in (-\infty, \infty)$. Furthermore, from Lemma~\ref{ch1.sec1.lemma1} it follows that a positive solution of the system that starts in the closed ball centered at the origin with a radius of $\frac{B}{\mu}$, given by $D(\infty)=\bar{B}^{(-\infty, \infty)}_{\mathbb{R}^{4}_{+},}\left(0,\frac{B}{\mu}\right)$, will continue to oscillate in the closed ball for all time $t\geq t_{0}$. Hence, the set $D(\infty)=\bar{B}^{(-\infty, \infty)}_{\mathbb{R}^{4}_{+},}\left(0,\frac{B}{\mu}\right)$ is a positive self-invariant set for the stochastic system (\ref{ch1.sec0.eq8})-(\ref{ch1.sec0.eq11}).
  %%%%%
\item [2.] Theorem~\ref{ch1.sec1.thm1a} and Lemma~\ref{ch1.sec1.lemma1} also signify that the deterministic system (\ref{ch1.sec0.eq3})-(\ref{ch1.sec0.eq6}) has a unique global positive solution denoted by $Y(t)\in \mathbb{R}^{4}_{+}$,  for all $t\in (-\infty, \infty)$. Furthermore, from Lemma~\ref{ch1.sec1.lemma1} it follows that any positive solution of the deterministic system that starts in the closed ball centered at the origin with a radius of $\frac{B}{\mu}$, given by  $D(\infty)=\bar{B}^{(-\infty, \infty)}_{\mathbb{R}^{4}_{+},}\left(0,\frac{B}{\mu}\right)$, grows and becomes bounded as signified by (\ref{ch1.sec1.thm1a.eq0}), within the closed ball for all time $t\geq t_{0}$. Hence, the set $D(\infty)=\bar{B}^{(-\infty, \infty)}_{\mathbb{R}^{4}_{+},}\left(0,\frac{B}{\mu}\right)$ is a positive self-invariant set for the deterministic system (\ref{ch1.sec0.eq3})-(\ref{ch1.sec0.eq6}).
\end{rem}
%%%%%%%%%%%%%%%%
\section{Existence and Asymptotic Behavior of Disease Free Equilibrium \label{ch1.sec2}}
In this section, the existence and the general asymptotic properties of the deterministic and stochastic systems: (\ref{ch1.sec0.eq3})-(\ref{ch1.sec0.eq6}) and (\ref{ch1.sec0.eq8})-(\ref{ch1.sec0.eq11}), respectively with respect to the  disease free equilibrium of the systems  are investigated. Generally, the equilibria for a deterministic or stochastic system  are obtained by solving a system of  algebraic equations obtained via setting the rate functions of the deterministic system, or the  drift and diffusion parts of the stochastic system to zero.  In the case of a disease free equilibrium,  the additional condition that $ E= I = R = 0$ is utilized in the event when there is no disease in the population. Let the equilibria of the two delayed systems (\ref{ch1.sec0.eq3})-(\ref{ch1.sec0.eq6}) and  (\ref{ch1.sec0.eq8})-(\ref{ch1.sec0.eq11}) be denoted generally by $E=(S^{*}, E^{*}, I^{*})$.

Note that the existence of a disease free steady state solution for the stochastic system is determined by the intensity values of the white noise processes representing the random fluctuations in the disease transmission and natural death rates, that is, $\sigma_{i}, i=S, E, I, \beta$.   For easy reference, the following result characterizes the existence of the disease-free steady solution of the systems:  (\ref{ch1.sec0.eq3})-(\ref{ch1.sec0.eq6}) and (\ref{ch1.sec0.eq8})-(\ref{ch1.sec0.eq11}).
\begin{thm}\label{ch1.sec2.thm0}
\item[1.]  There exists a disease-free steady state solution $E_{0}=(S^{*}_{0}, 0, 0)$ for the deterministic  system (\ref{ch1.sec0.eq3})-(\ref{ch1.sec0.eq6}), where $S^{*}_{0}=\frac{B}{\mu}$.
    \item[2.] When $\sigma_{i}\geq 0, i= E, I, \beta$ and $\sigma_{S}=0$, there exists a disease-free steady state solution $E_{0}=(S^{*}_{0}, 0, 0)$, for the stochastic  system (\ref{ch1.sec0.eq8})-(\ref{ch1.sec0.eq11}), where $S^{*}_{0}=\frac{B}{\mu}$.
    \item[3.] When $\sigma_{i}\geq 0, i= E, I, \beta$ and $\sigma_{S}>0$,  the system (\ref{ch1.sec0.eq8})-(\ref{ch1.sec0.eq11}) does not have a disease-free steady state solution.
\end{thm}
 Proof:\\
The results follow immediately by applying the method of finding the equilibria of the system described above.
\begin{rem}
Theorem~\ref{ch1.sec2.thm0}[1.] signifies that the deterministic system (\ref{ch1.sec0.eq3})-(\ref{ch1.sec0.eq6}) always has a disease free equilibrium given by $E_{0}$.  Theorem~\ref{ch1.sec2.thm0}[2.] and Theorem~\ref{ch1.sec2.thm0}[3.] signify that regardless of the intensity values $\sigma_{i}\geq 0, i= E, I, \beta$ of the white processes due to random fluctuations in the natural death rates of the exposed, infectious and removal populations, and also from the disease transmission rate, there exists a steady state disease-free population $E_{0}$, which is exactly the same as that of the deterministic system, provided the intensity value of the white noise process due to the random fluctuations in the natural death rate of the susceptible population is equal to zero. That is, $\sigma_{S}=0$

These observations suggest that the source- disease transmission or natural death rates, and also the intensity of the random fluctuations in the system represented by the white noise processes in the stochastic system (\ref{ch1.sec0.eq8})-(\ref{ch1.sec0.eq11}) have bearings on the asymptotic properties of the stochastic system (\ref{ch1.sec0.eq8})-(\ref{ch1.sec0.eq11}) with respect to the  disease free steady state population $E_{0}$. A detailed qualitative study of the behavior of the intensity values $\sigma_{i}, i= S, E, I, \beta$  of the white noise processes in the system in relation to the stochastic asymptotic stability of the disease-free steady state population $E_{0}$, and with focus on  disease eradication from the system appears in a parallel study by the author. In this section, the primary goal is to gain complete comparative insight about the general asymptotic properties of the deterministic and stochastic systems (\ref{ch1.sec0.eq3})-(\ref{ch1.sec0.eq6}) and  (\ref{ch1.sec0.eq8})-(\ref{ch1.sec0.eq11}) including but without limitation to (1) the importance of the delays in the system,  and (2) the asymptotic properties of the two systems with respect to the disease free equilibrium of the systems.
\end{rem}
 %One can see that the disease free equilibrium state for the system (\ref{ch1.sec0.eq3})-(\ref{ch1.sec0.eq5}) is given by $E_{0}=(S^{*}_{0}, 0, 0)$ where $S^{*}_{0}=\frac{B}{\mu}$.
%%%%%%%The stochastic system (\ref{ch1.sec0.eq8})-(\ref{ch1.sec0.eq11})  has the same disease free equilibrium $E_{0}$ in the absence of the noise due to the natural death $\mu$, that is, when %%%%%$\sigma_{i}=0,i=S,E,I,R $. When $\sigma_{i}\neq 0,i=S,E,I,R $, the system (\ref{ch1.sec0.eq8})-(\ref{ch1.sec0.eq11}) does not have a disease free equilibrium.
In the following, the asymptotic stability  of the disease free equilibrium, $E_{0}$, of the deterministic system (\ref{ch1.sec0.eq3})-(\ref{ch1.sec0.eq6}) and the stochastic system (\ref{ch1.sec0.eq8})-(\ref{ch1.sec0.eq11}), whenever $\sigma_{S}=0$  are investigated.  The  deterministic and stochastic versions of the Lyapunov functional techniques \cite{wanduku-determ,Wanduku-2017,wanduku-delay} are utilized to establish the stability results. In order to study the qualitative properties of the systems: (\ref{ch1.sec0.eq3})-(\ref{ch1.sec0.eq6}) and  (\ref{ch1.sec0.eq8})-(\ref{ch1.sec0.eq11}) with respect to the equilibrium state $E_{0}=(S^{*}_{0},0,0), S^{*}_{0}=\frac{B}{\mu} $, first the following transformation of the variables of the systems  which shifts the equilibrium states of the systems  to the origin is used:
\begin{equation}\label{ch1.sec2.eq1a}
\left\{
\begin{array}{lll}
U(t)&=&S(t)-S^{*}_{0}\\
V(t)&=&E(t)\\
W(t)&=&I(t).
\end{array}
\right.
\end{equation}
By employing the  transformation in (\ref{ch1.sec2.eq1a}) to the system (\ref{ch1.sec0.eq8})-(\ref{ch1.sec0.eq10}), the following system is obtained:
%%%%%%%%%%%%%%%%%%%%%%%%%%%%%%%%%%%
%%%%%%%%%%%%%%%%%%%%%%%%%%%%%%%%%%%
 \begin{eqnarray}
dU(t)&=&\left[ -\beta U(t)\int^{h_{1}}_{t_{0}}f_{T_{1}}(s) e^{-\mu s}G(W(t-s))ds - \mu U(t)+ \alpha \int_{t_{0}}^{\infty}f_{T_{3}}(r)W(t-r)e^{-\mu r}dr \right]dt\nonumber\\
&&-\sigma_{S}(S^{*}_{0}+U(t))dw_{S}(t)-\sigma_{\beta} (S^{*}_{0}+U(t))\int^{h_{1}}_{t_{0}}f_{T_{1}}(s) e^{-\mu s}G(W(t-s))dsdw_{\beta}(t) \label{ch1.sec2.eq1}\\
dV(t)&=& \left[ \beta (S^{*}_{0}+U(t))\int^{h_{1}}_{t_{0}}f_{T_{1}}(s) e^{-\mu s}G(W(t-s))ds - \mu V(t)\right.\nonumber\\
&&\left.-\beta \int_{t_{0}}^{h_{2}}f_{T_{2}}(u)(S^{*}_{0}+U(t-u))\int^{h_{1}}_{t_{0}}f_{T_{1}}(s) e^{-\mu s-\mu u}G(W(t-s-u))dsdu \right]dt\nonumber\\
&&-\sigma_{E}V(t)dw_{E}(t)+\sigma_{\beta} (S^{*}_{0}+U(t))\int^{h_{1}}_{t_{0}}f_{T_{1}}(s) e^{-\mu s}G(W(t-s))dsdw_{\beta}(t)\nonumber\\
&&-\sigma_{\beta} \int_{t_{0}}^{h_{2}}f_{T_{2}}(u)(S^{*}_{0}+U(t-u))\int^{h_{1}}_{t_{0}}f_{T_{1}}(s) e^{-\mu s-\mu u}G(W(t-s-u))dsdudw_{\beta}(t)\label{ch1.sec2.eq2}\\
dW(t)&=& \left[\beta \int_{t_{0}}^{h_{2}}f_{T_{2}}(u)(S^{*}_{0}+U(t-u))\int^{h_{1}}_{t_{0}}f_{T_{1}}(s) e^{-\mu s-\mu u}G(W(t-s-u))dsdu- (\mu +d+ \alpha) W(t) \right]dt\nonumber\\
&&-\sigma_{I}W(t)dw_{I}(t)+\sigma_{\beta} \int_{t_{0}}^{h_{2}}f_{T_{2}}(u)(S^{*}_{0}+U(t-u))\int^{h_{1}}_{t_{0}}f_{T_{1}}(s) e^{-\mu s-\mu u}G(W(t-s-u))dsdudw_{\beta}(t)\nonumber\\
&&\label{ch1.sec2.eq3}
%\\
%dR(t)&=&\left[ \alpha I(t) - \mu R(t)- \alpha \int_{t_{0}}^{\infty}f_{T_{3}}(r)I(t-r)e^{-\mu s}dr \right]dt-\sigma_{R}R(t)dw_{R}(t),\label{ch1.sec0.eq11}
\end{eqnarray}
%%%%%%%%%%%%%%%%%%%%%%%%%%%%%%%%%%%%
%%%%%%%%%%%%%%%%%%%%%%%%%%%%%%%%%
%%
%%
The lemmas  that follow in this section will be utilized to establish the asymptotic results for the system (\ref{ch1.sec0.eq8})-(\ref{ch1.sec0.eq11}) with respect to the steady state solution $E_{0}$.
%%%%
%The stochastic asymptotic and mean square stability of the disease free equilibrium $E_{0}$ of the system  (\ref{ch1.sec0.eq8})-(\ref{ch1.sec0.eq10}) is shown to coincide with the stability of %the equilibrium $E_{0}$ for the deterministic system (\ref{ch1.sec0.eq3})-(\ref{ch1.sec0.eq5})  only when the system is perturbed by fluctuations in the disease transmission rate $\beta$. When %the system (\ref{ch1.sec0.eq3})-(\ref{ch1.sec0.eq5}) is perturbed by the noise due to the natural death $\mu$, the stochastic system (\ref{ch1.sec0.eq8})-(\ref{ch1.sec0.eq10}) loses the disease %equilibrium $E_{0}$, but however, continues to oscillate near $E_{0}$.
 %%%%
 Recall the following lemma in the earlier study [\cite{wanduku-fundamental}, Lemma~4.1].
\begin{lemma}\label{ch1.sec2.lemma2a-1}
 Let   $V_{1}\in\mathcal{C}^{2, 1}(\mathbb{R}^{3}\times \mathbb{R}_{+}, \mathbb{R}_{+})$,  defined by
\begin{eqnarray}
V_{1}(x,t)&=&(S(t)-S^{*}+E(t))^{2}+c(E(t))^{2}+(I(t))^{2}\\
x(t)&=&(S(t)-S^{*},E(t),I(t))^{T},\label{ch2.sec2.thm2a.eq2}
\end{eqnarray}
where  $ c$ is a positive constant. There exists two increasing positive real valued functions $\phi_{1}$, and $\phi_{2}$, such that $V_{1}$ satisfies the inequality
 \begin{eqnarray}
\phi_{1}(||x||)&\leq& V_{1}(x,(t))
\leq \phi_{2}(||x||).\label{ch2.sec2.thm2a.eq3}
\end{eqnarray}
%%%%%%%%%%%%%
\end{lemma}
Proof:\\
The result follows directly from Lemma~4.1 in \cite{wanduku-fundamental}.
%%%%
%%%%
\begin{lemma}\label{ch1.sec2.lemma2a-2}
Let the hypothesis of Theorem~\ref{ch1.sec1.thm1}  be satisfied.
The differential operator\cite{wanduku-fundamental,wanduku-determ} applied to the Lyapunov function $V_{1}$ in (\ref{ch2.sec2.thm2a.eq2}) with
respect to the system  of stochastic differential equation (\ref{ch1.sec0.eq8})-(\ref{ch1.sec0.eq11}) is given by
\begin{eqnarray}
 &&dV_{1}=LV_{1}dt-2\sigma_{S}(U(t)+V(t))(S^{*}_{0}+U(t))dw_{S}(t)\nonumber\\
 &&-2\sigma_{E}(U(t)V(t)+(c+1)V^{2}(t))dw_{E}(t)-2\sigma_{I}W^{2}(t))dw_{I}(t)\nonumber\\
 &&-2c\sigma_{\beta}(S^{*}_{0}+U(t))V(t)\int_{t_{0}}^{h_{1}}f_{T_{1}}(s)e^{-\mu s}G(W(t-s))dsdw_{\beta}\nonumber\\
 &&-2\sigma_{E}[U(t)+(c+1)V(t)+W(t)]\times\nonumber\\
 &&\times\int_{t_{0}}^{h_{2}}\int_{t_{0}}^{h_{1}}f_{T_{2}}(u)f_{T_{1}}(s)e^{-\mu (s+u)}(S^{*}_{0}+U(t-u))G(W(t-s-u))dsdu dw_{\beta}(t)\label{ch2.sec2.thm2a.eq4}
\end{eqnarray}
where for some positive valued function $\tilde{K}(\mu)$ that depends on $\mu$, the drift part $LV_{1}$ of $dV_{1}$ in (\ref{ch2.sec2.thm2a.eq4}),  satisfies the inequality
\begin{eqnarray}
  LV_{1}(x,t) &\leq&(2\beta S^{*}_{0}+\beta +\alpha + 2\frac{\mu}{\tilde{K}(\mu)^{2}} -2\mu ) U^{2}(t)\nonumber\\
  &&+\left[2\mu \tilde{K}(\mu)^{2} + \alpha + \beta (2S^{*}_{0}+1 ) + c\beta (3S^{*}_{0}+1) -2(1+c)\mu \right]V^{2}(t)\nonumber\\
  &&+2[\beta S^{*}_{0}-(\mu+d+ \alpha)]W^{2}(t) \nonumber \\
   &&+2\alpha \int_{t_{0}}^{\infty}f_{T_{3}}(r)e^{-2\mu r} W^{2}(t-r)dr  \nonumber\\
   &&+[2\beta S^{*}_{0}\left(1+c\right)+ {\sigma}^{2}_{\beta}(S^{*}_{0})^{2}(4c+2(1-c)^{2})]\int_{t_{0}}^{h_{1}}f_{T_{1}}(s)e^{-2\mu s}G^{2}(W(t-s))ds\nonumber\\
   &&+\left[\beta S^{*}_{0}(4+c)+\beta (S^{*}_{0})^{2}(2+c)+{\sigma}^{2}_{\beta}(S^{*}_{0})^{2}(4c+10)\right]\times\nonumber\\
   &&\times\int_{t_{0}}^{h_{2}}\int_{t_{0}}^{h_{1}}f_{T_{2}}(u)f_{T_{1}}(s)e^{-2\mu (s+u)}G^{2}(W(t-s-u))dsdu\nonumber\\
   %%%%%%%%%%%%%%%%%%%%%%%%----G
   &&+{\sigma}^2_{S}\left(S^{*}_{0}+U(t)\right)^{2}+{\sigma}^2_{E}(c+1)V^{2}(t)+{\sigma}^2_{I}W^{2}(t),\label{ch2.sec2.thm2a.eq5}
\end{eqnarray}
%%%%%%%%%%%%%%%%%%%%%%%%%%%%%%%%%%%%%%%%%%%%%%%%%%%%%%%%%%%%%%%%%%%%%%%%%%%%%
\end{lemma}
%%%%%STochastic stability of disease free equilibrium
%%%%%%%%%%%%%%%%%%%%%%%%%%%%%%%%%%%%%%%%%%%%%%%%%%%%%%%%%%%%%%
%%%%%%%%%%%%%%%%%%%%%%%%%%%%%%%%%%%%%%%%%%%%%%%%%%%%%%%%%%%%%%%%%%%%
Proof:\\
%%%%%%%%%%%%%%%%%%%%%%%%%%%%%%%%%%%%%%%%
 The computation of the drift part $LV$\cite{wanduku-determ,wanduku-delay} of the differential operator $dV$   applied to the Lyapunov function $V_{1}$ in (\ref{ch2.sec2.thm2a.eq2}) with
respect to the system  of stochastic differential equation (\ref{ch1.sec0.eq8})-(\ref{ch1.sec0.eq11}) gives the following:
\begin{eqnarray}
  LV_{1}(x,t) &=&
  %%%%%%%%%%%%%%%%%%%%%%%%---A
  -4\mu U(t)V(t)-2\mu U^{2}(t)-2(1+c)\mu V^{2}(t)-2(\mu+d+ \alpha)W^{2}(t) \nonumber \\
   &&+2\alpha( U(t)+V(t))\int_{t_{0}}^{\infty}f_{T_{3}}(r)e^{-\mu r} W(t-r)dr  \nonumber\\
   %%%%%%%%%%%%%%%%%%%%--B
   &&+2\beta \left[S^{*}_{0} U(t)+ (1+c)S^{*}_{0} V(t) + cV(t) U(t)\right]\int_{t_{0}}^{h_{1}}f_{T_{1}}(s)e^{-\mu s}G(W(t-s))ds\nonumber\\
   %%%%%%%%%%%%%%%%%%%%%%%%%%%%%%%--C
   &&-2\beta \left[ U(t)+(1+c)V(t)-W(t)\right]\times\nonumber\\
   &&\times\int_{t_{0}}^{h_{2}}\int_{t_{0}}^{h_{1}}f_{T_{2}}(u)f_{T_{1}}(s)e^{-\mu (s+u)}(S^{*}_{0}+U(t-u))G(W(t-s-u))dsdu
   \nonumber\\
   %%%%%%%%%%%%%%%%%%%%%%%%----D
   && +{\sigma}^{2}_{\beta}c\left(S^{*}_{0}+U(t)\right)^{2}\left(\int_{t_{0}}^{h_{1}}f_{T_{1}}(s)e^{-\mu s}G(W(t-s))ds\right)^{2}\nonumber\\
   %%%%%%%%%%%%%%%%%%%%%%%%----E
   &&+{\sigma}^2_{\beta}(c+2)\left(\int_{t_{0}}^{h_{2}}\int_{t_{0}}^{h_{1}}f_{T_{2}}(u)f_{T_{1}}(s)e^{-\mu (s+u)}(S^{*}_{0}+U(t-u))G(W(t-s-u))dsdu \right)^{2}\nonumber\\
   %%%%%%%%%%%%%%%%%%%%%%%%----F
   &&+{\sigma}^2_{\beta}(1-c) \left(S^{*}_{0}+U(t)\right)\left(\int_{t_{0}}^{h_{1}}f_{T_{1}}(s)e^{-\mu s}G(W(t-s))ds\right)\nonumber\\
     &&\times\left(\int_{t_{0}}^{h_{2}}\int_{t_{0}}^{h_{1}}f_{T_{2}}(u)f_{T_{1}}(s)e^{-\mu (s+u)}(S^{*}_{0}+U(t-u))G(W(t-s-u))dsdu \right)\nonumber\\
   %%%%%%%%%%%%%%%%%%%%%%%%----G
   &&+{\sigma}^2_{S}\left(S^{*}_{0}+U(t)\right)^{2}+{\sigma}^2_{E}(c+1)V^{2}(t)+{\sigma}^2_{E}W^{2}(t).\label{ch2.sec2.thm2.proof.eq1a}
\end{eqnarray}
%%%%%%%%%%%%%%%%%%%%%%%%%%%%%%%%%%%
Applying Theorem~\ref{ch1.sec1.thm1}, $Cauchy-Swartz$, $H\ddot{o}lder$ inequalities,  and the following algebraic inequality
\begin{equation}\label{ch2.sec2.thm2.proof.eq2a}
2ab\leq \frac{a^{2}}{g(c)}+b^{2}g(c)
\end{equation}
where $a,b,c\in \mathbb{R}$,  and the function $g$ is such that $g(c)> 0$,  to estimate the terms with integral signs  in  (\ref{ch2.sec2.thm2.proof.eq1a}), one can see the following:
%%%%%%%%%%%%%%%%%%%%%%%%%%%%%%%%%%%
\begin{equation}\label{ch2.sec2.thm2.proof.eq3a}
  2\alpha( U(t)+V(t))\int_{t_{0}}^{\infty}f_{T_{3}}(r)e^{-\mu r} W(t-r)dr\leq \alpha U^{2}(t)+\alpha V^{2}(t)+2\alpha \int_{t_{0}}^{\infty}f_{T_{3}}(r)e^{-2\mu r} W^{2}(t-r)dr.   \\
  \end{equation}
%%%%%%%%%%%%%%%%%%%%%%%%%%%%%%%%%%%%%
\begin{eqnarray}
  &&2\beta \left[S^{*}_{0} U(t)+ (1+c)S^{*}_{0} V(t) + cV(t) U(t)\right]\int_{t_{0}}^{h_{1}}f_{T_{1}}(s)e^{-\mu s}G(W(t-s))ds\nonumber\\
  &&\leq \beta S^{*}_{0}U^{2}(t)+\beta S^{*}_{0}\left(1+2c \right)V^{2}(t) + 2\beta S^{*}_{0} \left(1+ c \right)\int_{t_{0}}^{h_{1}}f_{T_{1}}(s)e^{-2\mu s}G^{2}(W(t-s))ds\nonumber\\
  &&\label{ch2.sec2.thm2.proof.eq4a}
\end{eqnarray}
%%%%%%%%%%%%%%%%%%%%%%%%%%%%%%%%%
\begin{eqnarray}
 &&-2\beta \left[ U(t)+(1+c)V(t)-W(t)\right]\times\nonumber\\
   &&\times\int_{t_{0}}^{h_{2}}\int_{t_{0}}^{h_{1}}f_{T_{2}}(u)f_{T_{1}}(s)e^{-\mu (s+u)}(S^{*}_{0}+U(t-u))G(W(t-s-u))dsdu\nonumber\\
   && \leq  \beta (S^{*}_{0}+1) U^{2}(t)+(1+c) \beta (S^{*}_{0}+1) V^{2}(t)+2\beta S^{*}_{0} W^{2}(t)\nonumber\\
  &&+\left[\beta S^{*}_{0} (4+c) + \beta (S^{*}_{0})^{2}(2+c)\right]\int_{t_{0}}^{h_{2}}\int_{t_{0}}^{h_{1}}f_{T_{2}}(u)f_{T_{1}}(s)e^{-2\mu (s+u)}G^{2}(W(t-s-u))dsdu.\nonumber\\
  &&\label{ch2.sec2.thm2.proof.eq5ab}
\end{eqnarray}
%%%%%%%%%%%%%%%%%%%%%%%%%%%%%%%%%
\begin{eqnarray}
{\sigma}^{2}_{\beta}c\left(S^{*}_{0}+U(t)\right)^{2}\left(\int_{t_{0}}^{h_{1}}f_{T_{1}}(s)e^{-\mu s}G(W(t-s))ds\right)^{2}\leq4c{\sigma}^{2}_{\beta}(S^{*}_{0})^{2}\int_{t_{0}}^{h_{1}}f_{T_{1}}(s)e^{-2\mu s}G^{2}(W(t-s))ds\nonumber\\
&&\label{ch2.sec2.thm2.proof.eq5a}
\end{eqnarray}
\begin{eqnarray}
  &&{\sigma}^2_{\beta}(c+2)\left(\int_{t_{0}}^{h_{2}}\int_{t_{0}}^{h_{1}}f_{T_{2}}(u)f_{T_{1}}(s)e^{-\mu (s+u)}(S^{*}_{0}+U(t-u))G(W(t-s-u))dsdu \right)^{2}\nonumber\\
&&\leq 4(c+2){\sigma}^{2}_{\beta}(S^{*}_{0})^{2}\int_{t_{0}}^{h_{2}}\int_{t_{0}}^{h_{1}}f_{T_{2}}(u)f_{T_{1}}(s)e^{-2\mu (s+u)}G^{2}(W(t-s-u))dsdu.\nonumber\\
&&\label{ch2.sec2.thm2.proof.eq5b}
\end{eqnarray}
\begin{eqnarray}
  &&{\sigma}^2_{\beta}(1-c) \left(S^{*}_{0}+U(t)\right)\left(\int_{t_{0}}^{h_{1}}f_{T_{1}}(s)e^{-\mu s}G(W(t-s))ds\right)\nonumber\\
     &&\times\left(\int_{t_{0}}^{h_{2}}\int_{t_{0}}^{h_{1}}f_{T_{2}}(u)f_{T_{1}}(s)e^{-\mu (s+u)}(S^{*}_{0}+U(t-u))G(W(t-s-u))dsdu \right)\nonumber\\
   %%%%%%%%%%%%%%%%%%%%%%%%----G
   &&\leq 2{\sigma}^2_{\beta}(1-c)^{2}(S^{*}_{0})^{2}\int_{t_{0}}^{h_{1}}f_{T_{1}}(s)e^{-2\mu s}G^{2}(W(t-s))ds \nonumber\\
   %%%%%%%%%%%%%%%%%%%%%%%%----G
   &&+ 2{\sigma}^{2}_{\beta}(S^{*}_{0})^{2}\int_{t_{0}}^{h_{2}}\int_{t_{0}}^{h_{1}}f_{T_{2}}(u)f_{T_{1}}(s)e^{-2\mu (s+u)}G^{2}(W(t-s-u))dsdu.\nonumber\\
&&\label{ch2.sec2.thm2.proof.eq5c}
\end{eqnarray}
%%%%%%%%%%%%%%%%%%%%
%%%%%%%%%%%%%%%%
%%%%%%%%%%%%%%%%%
The result (\ref{ch2.sec2.thm2a.eq5}) follows from (\ref{ch2.sec2.thm2.proof.eq3a})-(\ref{ch2.sec2.thm2.proof.eq5c}) and the inequality (\ref{ch2.sec2.thm2.proof.eq2a}) that (\ref{ch2.sec2.thm2.proof.eq1a}) becomes
\begin{eqnarray}
  LV_{1}(x,t) &\leq&(2\beta S^{*}_{0}+\beta +\alpha + 2\frac{\mu}{\tilde{K}(\mu)^{2}} -2\mu ) U^{2}(t)\nonumber\\
  &&+\left[2\mu \tilde{K}(\mu)^{2} + \alpha + \beta (2S^{*}_{0}+1 ) + c\beta (3S^{*}_{0}+1) -2(1+c)\mu \right]V^{2}(t)\nonumber\\
  &&+2[\beta S^{*}_{0}-(\mu+d+ \alpha)]W^{2}(t) \nonumber \\
   &&+2\alpha \int_{t_{0}}^{\infty}f_{T_{3}}(r)e^{-2\mu r} W^{2}(t-r)dr  \nonumber\\
   &&+[2\beta S^{*}_{0}\left(1+c\right)+ {\sigma}^{2}_{\beta}(S^{*}_{0})^{2}(4c+2(1-c)^{2})]\int_{t_{0}}^{h_{1}}f_{T_{1}}(s)e^{-2\mu s}G^{2}(W(t-s))ds\nonumber\\
   &&+\left[\beta S^{*}_{0}(4+c)+\beta (S^{*}_{0})^{2}(2+c)+{\sigma}^{2}_{\beta}(S^{*}_{0})^{2}(4c+10)\right]\times\nonumber\\
   &&\times\int_{t_{0}}^{h_{2}}\int_{t_{0}}^{h_{1}}f_{T_{2}}(u)f_{T_{1}}(s)e^{-2\mu (s+u)}G^{2}(W(t-s-u))dsdu\nonumber\\
   %%%%%%%%%%%%%%%%%%%%%%%%----G
   &&+{\sigma}^2_{S}\left(S^{*}_{0}+U(t)\right)^{2}+{\sigma}^2_{E}(c+1)V^{2}(t)+{\sigma}^2_{E}W^{2}(t),\label{ch2.sec2.thm2.proof.eq6}
\end{eqnarray}
where $\tilde{K}(\mu)=g(\mu)$ and $ g$ is defined in  (\ref{ch2.sec2.thm2.proof.eq2a}).

%%%%%%%%%%%%%%%%%%%%%%%%%%%%%%%%%%%%%%%%%%%%%%%%%%%%%%%%%%%%%%%%%%%%%%%%
%\subsection{Existence and Stochastic Asymptotic Behavior of Disease Free Equilibrium \label{ch1.sec2.subsec1}}
The following set of lemmas characterize the stochastic asymptotic stability of the disease free equilibrium $E_{0}$ of the system (\ref{ch1.sec0.eq8})-(\ref{ch1.sec0.eq11}) when the intensity value of the white noise process due to the natural death rate $\mu$ in the susceptible population is zero, that is, when $\sigma_{S}=0 $. Lemma~\ref{ch1.sec2.lemma2a} presents the stochastic stability results for the case of constant and finite delays in the system, and Lemma~\ref{ch1.sec2.lemma2} presents stability results for arbitrary random finite and infinite delays in the system.
%The following lemma is utilized to establish the final results.
%%%%%%%%%%%%%%%%%%%%%%%%%%%%%%%%%%%%%

It is noted that the assumption  of constant delay times representing the incubation period of the disease in the vector, $T_{1}$, incubation period of the disease in the host, $T_{2}$, and immunity period of the disease in the human population, $T_{3}$  is equivalent to the special case of letting the probability density functions $f_{T_{i}}, i=1,2,3$ of the random variables $T_{1}, T_{2}$ and $T_{3}$ be the dirac-delta function. That is,
\begin{equation}\label{ch1.sec2.eq4}
f_{T_{i}}(s)=\delta(s-T_{i})=\left\{\begin{array}{l}+\infty, s=T_{i},\\
0, otherwise,
\end{array}\right.
, i=1, 2, 3.
\end{equation}
Moreover, under the assumption that $T_{1}\geq 0, T_{2}\geq 0$ and $T_{3}\geq 0$ are constant, the following expectations can be written as  $E(e^{-2\mu (T_{1}+T_{2})})=e^{-2\mu (T_{1}+T_{2})} $, $E(e^{-2\mu T_{1}})=e^{-2\mu T_{1}} $ and $E(e^{-2\mu T_{3}})=e^{-2\mu T_{3}} $.
%%%%%%%%%%%%%%%%%%%%%%%%%%%%%%%%%
%%%%%%%%%%%%%%%%%%%%%%%%%%%%%%%%%%Constant delay
\begin{lemma}\label{ch1.sec2.lemma2a}
 Let the hypotheses of  Theorem~\ref{ch1.sec1.thm1}, Theorem~\ref{ch1.sec2.thm0}[2.] and Lemma~\ref{ch1.sec2.lemma2a-2} be satisfied. Also, let  $T_{1}, T_{2}$ and $T_{3}$ be constant positive values.  There exists a Lyapunov functional
 \begin{equation}\label{ch2.sec2.thm2.eq1aa}
V=V_{1}+V_{12},
\end{equation}
where
 %$V_{1}:\mathbb{R}^{3}\times \mathbb{R}_{+}\rightarrow \mathbb{R}_{+}$,
  $V_{1}\in\mathcal{C}^{2, 1}(\mathbb{R}^{3}\times \mathbb{R}_{+}, \mathbb{R}_{+})$ is defined by (\ref{ch2.sec2.thm2a.eq2})
and $V_{12}$ is defined as follows:
%%%%%%%%%%%%%
%We construct a Lyapunov functional
%%%%%%%%%%%%%%%%%%%%%%%%%%%%%%%%%%%%%%%%%%%%%%%%%%%%%%%%%%%%%%%%%%%%%%%%%%%%%
%
%%%%%%%%%%%%%%%%%%%%%%%%%%%%%%%%%%%%%%%%%%%%%%%%%%%%%%%%%%%%%%%%%%%%%%%%%%%%
%%%%%%%%%%%%%%%%%%%%%%%%%%%
 \begin{eqnarray}
   %%%%%%%%%%%%%%%%\nonumber \\
   &&V_{12}(x,t)=2\alpha e^{-2\mu T_{3}} \int_{t-r}^{t}I^{2}(v)dv  \nonumber\\
   &&+[2\beta S^{*}_{0}\left(1+c\right)+ {\sigma}^{2}_{\beta}(S^{*}_{0})^{2}(4c+2(1-c)^{2})]e^{-2\mu T_{1}}\int^{t}_{t-s}G^{2}(I(v))dv\nonumber\\
   &&+\left[\beta S^{*}_{0}(4+c)+\beta (S^{*}_{0})^{2}(2+c)+{\sigma}^{2}_{\beta}(S^{*}_{0})^{2}(4c+10)\right]e^{-2\mu (T_{1}+T_{2})}\int^{t}_{t-(T_{1}+T_{2})}G^{2}(I(v))dv\nonumber\\
    %%%%%%%%%%%%%%%%%%%%%%%%----G
   %%%&&
   \label{{ch2.sec2.thm2.eq4aa}}
\end{eqnarray}
Furthermore, there exists threshold values $R^{*}_{1}$, $R^{*}_{0}$,  $U_{0}$ and $V_{0}$  defined as follows:
\begin{equation}\label{ch2.sec2.thm1.eq5aa}
R^{*}_{1}=\frac{\beta S^{*}_{0} \hat{K}^{*}_{1}+\alpha}{(\mu+d+\alpha)},
\end{equation}
\begin{equation}\label{ch2.sec2.thm1.eq5aaa}
R^{*}_{0}=\frac{\beta S^{*}_{0} +\frac{1}{2}\sigma^{2}_{I}}{(\mu+d+\alpha)},
\end{equation}
\begin{equation}\label{ch2.sec2.thm1.eq5ba}
U_{0}=\frac{2\beta S^{*}_{0}+\beta +\alpha + 2\frac{\mu}{\tilde{K}(\mu)^{2}}}{2\mu},
\end{equation}
 and
 \begin{equation}\label{ch2.sec2.thm1.eq5ca}
V_{0}=\frac{(2\mu \tilde{K}(\mu)^{2} + \alpha + \beta (2S^{*}_{0}+1 ) +\sigma^{2}_{E})}{2\mu},
\end{equation}
with some constant $\hat{K}^{*}_{1}>0$  that depends on $S^{*}_{0}$ and $\sigma_{\beta}$ (in fact, $\hat{K}^{*}_{1}=4+6\frac{1}{\beta}\sigma^{2}_{\beta}S^{*}_{0} $), and some positive constants $\phi$, $\psi$, and $ \varphi $, such that,
   under the assumptions that $R^{*}_{0}< 1$, $U_{0}\leq 1$, and $V_{0}\leq 1$,  and
    \begin{equation}\label{ch2.sec2.thm1.eq5ca1}
      T_{max}\geq \frac{1}{2\mu}\log{\frac{R^{*}_{1}}{1-R^{*}_{0}}},
    \end{equation}
    where
    \begin{equation}\label{ch2.sec2.thm1.eq5ca2}
      T_{max}=\max{(T_{1}+T_{2}, T_{3})},
    \end{equation}
    the drift part $LV$ of the differential operator $dV$ applied to $V$ with respect to the stochastic dynamic system (\ref{ch1.sec0.eq8})-(\ref{ch1.sec0.eq11}) satisfies the following inequality:
 \begin{equation}\label{ch2.sec2.thm1.eq6a}
  L{V}(x,t)\leq  -\left(\phi U^{2}(t)+\psi V^{2}(t)+ \varphi W^{2}(t)\right).
\end{equation}%\l
\end{lemma}
Proof:\\
By applying the translation properties of the Dirac-Delata function (\ref{ch1.sec2.eq4}), it can be seen from Lemma~\ref{ch1.sec2.lemma2a-2} that the drift part $LV$ of the differential operator $dV$ applied to the Lyapunov functional defined in (\ref{ch2.sec2.thm2.eq1aa}), (\ref{ch2.sec2.thm2a.eq2}) and (\ref{{ch2.sec2.thm2.eq4aa}}) with respect to system (\ref{ch1.sec0.eq8})-(\ref{ch1.sec0.eq11}) leads to the following:
\begin{eqnarray}%\label{ch2.sec2.thm1.proof.eq6}
  L{V}(x,t)&=&L{V}_{1}(x,t)\nonumber\\
    &&+2\alpha e^{-2\mu T_{3}} W^{2}(t)  \nonumber\\
   &&+[2\beta S^{*}_{0}\left(1+c\right)+ {\sigma}^{2}_{\beta}(S^{*}_{0})^{2}(4c+2(1-c)^{2})]e^{-2\mu T_{1}}G^{2}(W(t))\nonumber\\
   &&+\left[\beta S^{*}_{0}(4+c)+\beta (S^{*}_{0})^{2}(2+c)+{\sigma}^{2}_{\beta}(S^{*}_{0})^{2}(4c+10)\right]e^{-2\mu (T_{1}+T_{2})}G^{2}(W(t))\nonumber\\
     %%%%%%%%%%%%%%%%%%%%%%%%----G
      &&-2\alpha e^{-2\mu T_{3}} W^{2}(t-T_{3})  \nonumber\\
   &&-[2\beta S^{*}_{0}\left(1+c\right)+ {\sigma}^{2}_{\beta}(S^{*}_{0})^{2}(4c+2(1-c)^{2})]e^{-2\mu T_{1}}G^{2}(W(t-T_{1}))\nonumber\\
   &&-\left[\beta S^{*}_{0}(4+c)+\beta (S^{*}_{0})^{2}(2+c)+{\sigma}^{2}_{\beta}(S^{*}_{0})^{2}(4c+10)\right]\times\nonumber\\
   &&\times e^{-2\mu (T_{1}+T_{2})}G^{2}(W(t-T_{1}-T_{2})). \label{ch2.sec2.thm1.proof.eq7a}
   %%%%%%%%%%%%%%%%%%%%%%%%----G
\end{eqnarray}
 It follows that
under the assumptions for $\sigma_{i},i=S,E,I, \beta $ in Theorem~\ref{ch1.sec2.thm0}[2.], and for some suitable choice of the positive constant $ c$,  it is easy to see from  (\ref{ch2.sec2.thm2a.eq5}), (\ref{ch2.sec2.thm1.proof.eq7a}), the statements of Assumption~\ref{ch1.sec0.assum1}, $A5$ (i.e. $G^{2}(x)\leq x^{2}, x\geq 0$)  and some further algebraic manipulations and simplifications that
\begin{equation}\label{ch2.sec2.thm1.proof.eq8aa}
  L{V}(x,t)\leq  -\left(\phi U^{2}(t)+\psi V^{2}(t)+ \varphi W^{2}(t)\right),
\end{equation}%\label{ch2.sec2.thm1.proof.eq6}
where,
 \begin{eqnarray}
 \phi &=&2\mu (1-U_{0})\label{ch2.sec2.thm1.proof.eq8aa1}\\
 \psi &=&2\mu (1-V_{0})-2\mu c\left(1-\frac{\beta (3S^{*}_{0}+1)+\sigma^{2}_{E}}{2\mu}\right)\label{ch2.sec2.thm1.proof.eq8ba}\\
\varphi &=&2(\mu +d+ \alpha)-\left[2\beta S^{*}_{0}+\sigma^{2}_{I}+2\alpha e^{-2\mu T_{3}}+2\left(\beta S^{*}_{0}+\sigma^{2}_{\beta}( S^{*}_{0})^{2}\right)e^{-2\mu T_{3}}\right.\nonumber\\
&&\left.+\left(4\beta S^{*}_{0}+2\beta (S^{*}_{0})^{2}+10\sigma^{2}_{\beta}( S^{*}_{0})^{2}\right)e^{-2\mu (T_{1}+T_{2})}\right] -c(3\beta S^{*}_{0}+\beta (S^{*}_{0})^{2}+4\sigma^{2}_{\beta}(S^{*}_{0})^{2})\nonumber\\
&&-2c^{2}\sigma^{2}_{\beta}(S^{*}_{0})^{2},\nonumber\\
&\geq& 2(\mu +d+ \alpha)\left[1-R^{*}_{0}-R^{*}_{1}e^{-2\mu T_{max}}\right]-c(3\beta S^{*}_{0}+\beta (S^{*}_{0})^{2}+4\sigma^{2}_{\beta}(S^{*}_{0})^{2})\nonumber\\
&&-2c^{2}\sigma^{2}_{\beta}(S^{*}_{0})^{2}.\label{ch2.sec2.thm1.proof.eq8aa2}
 \end{eqnarray}
 and $R^{*}_{0}$ and $R^{*}_{1}$ are  defined in (\ref{ch2.sec2.thm1.eq5aa})-(\ref{ch2.sec2.thm1.eq5aaa}).
 %%%%%%%%%%%%%%%%%%%%%%%%%%%%%%%
 %%%%%%%%%%%%%%%%%%%%%%%%%%%%
  It is now easy to see that under the assumptions of $R^{*}_{0}$, $R^{*}_{1}$, $U_{0}$, and $V_{0}$ in the hypothesis and also for a suitable choice of the positive constant $c$ it follows that $\phi$, $\psi$, and  $\varphi$ are positive constants and (\ref{ch2.sec2.thm1.eq6a}) follows immediately.
 %%%%%%%%%%%%%%%%%%%%%%%%%%%%%%%%%%

The following result describes the stochastic asymptotic stability of the  disease free equilibrium $E_{0}$, whenever it exists and the delays in the stochastic system are constant.
\begin{thm}\label{ch1.sec2.theorem1a}
Suppose Theorem~\ref{ch1.sec2.thm0}[2.] and the hypotheses of Lemma~\ref{ch1.sec2.lemma2a-2} and Lemma~\ref{ch1.sec2.lemma2a} are satisfied, then the
  disease free equilibrium $E_{0}$ of the stochastic dynamic system (\ref{ch1.sec0.eq8})-(\ref{ch1.sec0.eq11}) is stochastically  asymptotically stable in the large in the set $D(\infty)$. Moreover, the steady state solution $E_{0}$ is  exponentially mean square stable.
\end{thm}
Proof:\\
The result follows by applying the comparison stability results\cite{mao, wanduku-delay}.  Moreover, the disease free equilibrium state is exponentially mean square stable.
%%%%%%%%%%%%%

The following result for the deterministic system (\ref{ch1.sec0.eq3})-(\ref{ch1.sec0.eq6}) will be useful to compare and obtain insight about the influence of the noise and delays in the stochastic system (\ref{ch1.sec0.eq8})-(\ref{ch1.sec0.eq11}), whenever the delays $T_{i}, i=1,2,3$ in both systems are constant.
%%%%%%%%%%%
\begin{thm}\label{ch1.sec2.lemma2a.corrolary1}
 Let the hypotheses of  Theorem~\ref{ch1.sec1.thm1}, Theorem~\ref{ch1.sec2.thm0}[1.] and Lemma~\ref{ch1.sec2.lemma2a-2} be satisfied. Also, let  $T_{1}, T_{2}$ and $T_{3}$ be constant positive values.  There exists a Lyapunov functional
 \begin{equation}\label{ch1.sec2.lemma2a.corrolary1.eq1}
V=V_{1}+V_{13},
\end{equation}
where
 %$V_{1}:\mathbb{R}^{3}\times \mathbb{R}_{+}\rightarrow \mathbb{R}_{+}$,
  $V_{1}\in\mathcal{C}^{2, 1}(\mathbb{R}^{3}\times \mathbb{R}_{+}, \mathbb{R}_{+})$ is defined by (\ref{ch2.sec2.thm2a.eq2})
and $V_{13}$ is defined as follows:
%%%%%%%%%%%%%
%We construct a Lyapunov functional
%%%%%%%%%%%%%%%%%%%%%%%%%%%%%%%%%%%%%%%%%%%%%%%%%%%%%%%%%%%%%%%%%%%%%%%%%%%%%
%
%%%%%%%%%%%%%%%%%%%%%%%%%%%%%%%%%%%%%%%%%%%%%%%%%%%%%%%%%%%%%%%%%%%%%%%%%%%%
%%%%%%%%%%%%%%%%%%%%%%%%%%%
 \begin{eqnarray}
   %%%%%%%%%%%%%%%%\nonumber \\
   &&V_{13}(x,t)=2\alpha e^{-2\mu T_{3}} \int_{t-r}^{t}I^{2}(v)dv  \nonumber\\
   &&+[2\beta S^{*}_{0}\left(1+c\right)]e^{-2\mu T_{1}}\int^{t}_{t-s}G^{2}(I(v))dv\nonumber\\
   &&+\left[\beta S^{*}_{0}(4+c)+\beta (S^{*}_{0})^{2}(2+c)\right]e^{-2\mu (T_{1}+T_{2})}\int^{t}_{t-(T_{1}+T_{2})}G^{2}(I(v))dv\nonumber\\
    %%%%%%%%%%%%%%%%%%%%%%%%----G
   %%%&&
   \label{ch1.sec2.lemma2a.corrolary1.eq2}
\end{eqnarray}
Furthermore, there exists threshold values $\hat{R}^{*}_{1}$, $\hat{R}^{*}_{0}$,  $\hat{U}_{0}$ and $\hat{V}_{0}$  defined as follows:
\begin{equation} \label{ch1.sec2.lemma2a.corrolary1.eq3}
\hat{R}^{*}_{1}=\frac{\beta S^{*}_{0} \hat{K}^{*}_{0}+\alpha}{(\mu+d+\alpha)},
\end{equation}
\begin{equation} \label{ch1.sec2.lemma2a.corrolary1.eq4}
\hat{R}^{*}_{0}=\frac{\beta S^{*}_{0} }{(\mu+d+\alpha)},
\end{equation}
\begin{equation} \label{ch1.sec2.lemma2a.corrolary1.eq5}
\hat{U}_{0}=\frac{2\beta S^{*}_{0}+\beta +\alpha + 2\frac{\mu}{\tilde{K}(\mu)^{2}}}{2\mu},
\end{equation}
 and
 \begin{equation} \label{ch1.sec2.lemma2a.corrolary1.eq6}
\hat{V}_{0}=\frac{(2\mu \tilde{K}(\mu)^{2} + \alpha + \beta (2S^{*}_{0}+1 ) )}{2\mu},
\end{equation}
with some constant $\hat{K}^{*}_{0}>0$ (in fact, $\hat{K}^{*}_{1}=4 $), and some positive constants $\phi$, $\psi$, and $ \varphi $, such that,
   under the assumptions that $\hat{R}^{*}_{0}< 1$, $\hat{U}_{0}\leq 1$, and $\hat{V}_{0}\leq 1$,  and
    \begin{equation} \label{ch1.sec2.lemma2a.corrolary1.eq7}
      T_{max}\geq \frac{1}{2\mu}\log{\frac{\hat{R}^{*}_{1}}{1-\hat{R}^{*}_{0}}},
    \end{equation}
    where
    \begin{equation} \label{ch1.sec2.lemma2a.corrolary1.eq8}
      T_{max}=\max{(T_{1}+T_{2}, T_{3})},
    \end{equation}
    the deterministic differential operator $\dot{V}$ applied to $V$ with respect to the deterministic dynamic system (\ref{ch1.sec0.eq3})-(\ref{ch1.sec0.eq6}) satisfies the following inequality:
 \begin{equation} \label{ch1.sec2.lemma2a.corrolary1.eq9}
  \dot{V}(x,t)\leq  -\min{(\phi, \psi, \varphi )}||X(t)-E_{0}||^{2},%\left(\phi U^{2}(t)+\psi V^{2}(t)+ \varphi W^{2}(t)\right).
\end{equation}%\l
where $X(t)$ is defined in (\ref{ch1.sec0.eq13b}), and $||.||$ is the natural Euclidean norm on $\mathbb{R}^{2}$.

Furthermore, the disease free steady state $E_{0}$ is globally uniformly asymptotically stable in $D(\infty)$. Moreover, it is exponentially stable.
\end{thm}
Proof:\\
The results follow directly from Lemma~\ref{ch2.sec2.thm1.eq5aa}, where for the intensities $\sigma_{i}=0, i=1,2,3$, it is easily seen that when the assumptions in the hypothesis of Theorem~\ref{ch1.sec2.lemma2a.corrolary1} are satisfied, then
\begin{equation}\label{ch1.sec2.lemma2a.corrolary1.proof.eq1}
  \dot{V}(x,t)\leq-\left(\phi U^{2}(t)+\psi V^{2}(t)+ \varphi W^{2}(t)\right),
\end{equation}%\l-\min{(\phi, \psi, \varphi )}||X(t)-E_{0}||^{2}
where $\phi$, $\psi$, and $ \varphi $ (with $\sigma_{i}=0, i=1,2,3$ ) are defined in (\ref{ch2.sec2.thm1.proof.eq8aa1})-(\ref{ch2.sec2.thm1.proof.eq8aa2}). The result in (\ref{ch1.sec2.lemma2a.corrolary1.eq9}) follows immediately from (\ref{ch1.sec2.lemma2a.corrolary1.proof.eq1}). Furthermore, the rest of the stability results follow by applying the comparison results in \cite{divine-proceeding1,divine-proceeding2,divine5}.
 %applying the translation properties of the Dirac-Delata function (\ref{ch1.sec2.eq4}), it can be seen from Lemma~\ref{ch1.sec2.lemma2a-2} that the drift part $LV$ of the differential operator %$dV$ applied to the Lyapunov functional defined in (\ref{ch2.sec2.thm2.eq1aa}), (\ref{ch2.sec2.thm2a.eq2}) and (\ref{{ch2.sec2.thm2.eq4aa}}) with respect to system %(\ref{ch1.sec0.eq8})-(\ref{ch1.sec0.eq11}) leads to the following:
%%%%%%%%%%%%%%%%%%%%%%%%%%%%%%%%%%%%%%%%
%%%%%%%%%%%%%%%%%%%%%%%%%%%%%%%%%%%%%%%%%
\begin{rem}\label{ch1.sec2.theorem1a.rem1}
\item[1.] Theorem~\ref{ch1.sec2.theorem1a} signifies that in the absence of the white noise process in the system due to the random fluctuations from the natural death rate in the susceptible population, that is, $\sigma_{S}=0$, it follows that regardless of (1.) the  variability in the natural death of the exposed, infectious, and the removal populations, that is, $\sigma^{2}_{i}\geq 0, i= E, I, R, $, or (2.) the variability in the disease transmission process, that is, $\sigma^{2}_{ \beta}\geq 0$, or  (3.) the value of the incubation delay of the disease in the vector and in the human being ($T_{1}$  and $T_{2}$ respectively), or  the value of the delay of the natural immunity period $T_{3}$, a disease free steady state $E_{0}$ for the population exists. Furthermore, the disease free steady state $E_{0}$ for the population is stochastically asymptotically stable in the large, whenever the threshold conditions: $R^{*}_{0}< 1$, $U_{0}\leq 1$, $V_{0}\leq 1$,  and $ T_{max}\geq \frac{1}{2\mu}\log{\frac{R^{*}_{1}}{1-R^{*}_{0}}}$
     are satisfied, where $R^{*}_{0}$, $R^{*}_{1}$, $U_{0}$, $V_{0}$, $T_{max}$ are defined in (\ref{ch2.sec2.thm1.eq5aa})-(\ref{ch2.sec2.thm1.eq5ca2}).

      It should also be noted that the threshold values $R^{*}_{0}$, $U_{0}$, and $V_{0}$ are explicit in terms of the parameters of the system (\ref{ch1.sec0.eq8})-(\ref{ch1.sec0.eq11}) and  also computationally attractive, whenever specific values for the parameters of the system are given. This observation  suggests that  in a disease scenario  where there are no random fluctuations in the  natural death of susceptible individuals in the population, that is, $\sigma_{S}=0$, the stochastic dynamic system (\ref{ch1.sec0.eq8})-(\ref{ch1.sec0.eq11}) exhibits a disease free steady state for the population, regardless of whether there is (1.) significant variability in  the disease transmission, that is, $\sigma^{2}_{\beta}\geq 0$, or (2.) significant variability  in the natural death of the other states- $E, I, R$, that is, $\sigma^{2}_{i}\geq 0, i= E, I, R$. Furthermore, the disease can be eradicated from the system (\ref{ch1.sec0.eq8})-(\ref{ch1.sec0.eq11}), whenever the threshold conditions $R^{*}_{0}< 1$, $U_{0}\leq 1$, $V_{0}\leq 1$,  and $ T_{max}\geq \frac{1}{2\mu}\log{\frac{R^{*}_{1}}{1-R^{*}_{0}}}$ are satisfied. In addition, it is noted that the disease eradication  is restricted by the values of the incubation and natural immunity periods in the system, that is,   $ T_{max}\geq \frac{1}{2\mu}\log{\frac{R^{*}_{1}}{1-R^{*}_{0}}}$. This result is of significance to policy decisions related to malaria eradication from the human population noting that either the total incubation phase of the plasmodium in the vector and human hosts (lasting for $T_{1}+T_{2}$ time units) or the period of effective immunity against malaria ($T_{3}$) is delayed for at least $\frac{1}{2\mu}\log{\frac{R^{*}_{1}}{1-R^{*}_{0}}}$ time units, provided the other conditions $R^{*}_{0}< 1$, $U_{0}\leq 1$, $V_{0}\leq 1$ hold.
      \item[2.]
 Since the threshold conditions, that is, $R^{*}_{0}< 1$, $U_{0}\leq 1$, $V_{0}\leq 1$,  and $ T_{max}\geq \frac{1}{2\mu}\log{\frac{R^{*}_{1}}{1-R^{*}_{0}}}$  are sufficient for the disease free steady state population, $E_{0}$, to be stochastically asymptotically stable in the large, and consequently for the eradication of the disease from the population,  the threshold value $R^{*}_{0}$ is the noise-modified basic reproduction number for the disease dynamics described by stochastic system (\ref{ch1.sec0.eq8})-(\ref{ch1.sec0.eq11}), whenever the following assumptions:- (1) that $\sigma_{i}\geq 0, i= E, I, R, \beta$, and  $\sigma_{S}=0$, and also (2) that the incubation and natural immunity delays $T_{1}, T_{2}$ and $T_{3}$ are constant.
 \item[3.] The results in Theorem~\ref{ch1.sec2.lemma2a.corrolary1} in comparison to results in Theorem~\ref{ch1.sec2.theorem1a} exhibit the combined effects of the delays and the noise from the disease transmission and natural death rates on the dynamics of the disease.

     Indeed, the threshold value $\hat{R}^{*}_{0}=\frac{\beta S^{*}_{0} }{(\mu+d+\alpha)}$ from (\ref{ch1.sec2.lemma2a.corrolary1.eq4}), represents the total number of infectious cases  that result from one malaria infectious person present in a completely disease free population with state given by $S^{*}_{0}=\frac{B}{\mu}$, over the average lifetime given by $\frac{1 }{(\mu+d+\alpha)}$ of a person who has survived from disease related death $d$,  natural death $\mu$  and recovered at rate $\alpha$ from infection. Hence,  $\hat{R}^{*}_{0}$ is the noise-free basic reproduction number of the disease, whenever the incubation periods of the malaria parasite inside the human and mosquito hosts given by $T_{i}, i=1,2$, and also the period of effective natural immunity against malaria given by $T_{3}$, are all positive constants. Furthermore, the threshold condition $\hat{R}^{*}_{0}<1$ from Theorem~\ref{ch1.sec2.lemma2a.corrolary1} is required for the disease to be eradicated from the human population.

     Comparing the  delay threshold conditions from (\ref{ch2.sec2.thm1.eq5ca1}) and (\ref{ch1.sec2.lemma2a.corrolary1.eq7}), and the other threshold values from Theorem~\ref{ch1.sec2.lemma2a.corrolary1} and  Theorem~\ref{ch1.sec2.theorem1a}, it can be seen that
     \begin{equation}\label{ch1.sec2.theorem1a.rem1.eq1}
      T_{max}\geq \frac{1}{2\mu}\log{\frac{R^{*}_{1}}{1-R^{*}_{0}}}> \frac{1}{2\mu}\log{\frac{\hat{R}^{*}_{1}}{1-\hat{R}^{*}_{0}}},
    \end{equation}
    $R^{*}_{0}>\hat{R}^{*}_{0}$, $U_{0}=\hat{U}_{0}$ and $V_{0}>\hat{V}_{0}$,
    whenever $\sigma_{i}>0, i= E, I, R, \beta$. This observation suggests clearly that the intensities $\sigma_{i}, i=S, E, I, R, \beta$ of the white noise processes in the system from the natural and disease transmission rates exhibit bearings on the stability of the disease free equilibrium $E_{0}$, and hence on disease eradication. Furthermore, in the event where the disease free equilibrium $E_{0}$ is stable for both  deterministic and stochastic systems (\ref{ch1.sec0.eq3})-(\ref{ch1.sec0.eq6}) and (\ref{ch1.sec0.eq8})-(\ref{ch1.sec0.eq11}), that is, when
       (\ref{ch1.sec2.theorem1a.rem1.eq1}) and  $1>R^{*}_{0}>\hat{R}^{*}_{0}$, $1\geq U_{0}=\hat{U}_{0}$ and $1\geq V_{0}>\hat{V}_{0}$, it can be seen easily that for higher intensity values for $\sigma_{i}>0, i= E, I, R, \beta$, the following hold:   $1> R^{*}_{0}\gg\hat{R}^{*}_{0}$, $1\geq U_{0}=\hat{U}_{0}$,  $1\geq V_{0}\gg\hat{V}_{0}$ and
      \begin{equation}\label{ch1.sec2.theorem1a.rem1.eq2}
      T_{max}\geq \frac{1}{2\mu}\log{\frac{R^{*}_{1}}{1-R^{*}_{0}}}\gg \frac{1}{2\mu}\log{\frac{\hat{R}^{*}_{1}}{1-\hat{R}^{*}_{0}}}.
    \end{equation}
    Therefore, it is evident that the presence of random fluctuations in the disease dynamics from the disease transmission and natural death rates negatively impact the disease eradication process by setting higher bounds for the threshold values that are sufficient for the stability of the disease free population steady state $E_{0}$.  In addition, since in most malaria endemic regions, the natural immunity period $T_{3}$ is often longer than the combined duration $T_{1}+T_{2}$ of incubation of the malaria parasite inside the mosquito and human hosts, so that from (\ref{ch1.sec2.lemma2a.corrolary1.eq8}),  $T_{max}=T_{3}$,  the result in (\ref{ch1.sec2.theorem1a.rem1.eq2}) shows that the occurrence of the white noise processes in the system, and especially with high intensity values, potentially sets unrealistic bounds for the natural immunity period in the human population needed for the disease to be eradicated.

    Thus, malaria eradication policies must make efforts to reduce the intensities of the random fluctuation in the disease transmission and natural death rates, perhaps via better care of the people in malaria endemic zones, and application of effective vector control strategies or measures.
\end{rem}
%%%%%%%%%%%%%%%%%%%%%%%%%%%%%%%%%%%
\begin{lemma}\label{ch1.sec2.lemma2}
 Let the hypotheses of  Theorem~\ref{ch1.sec1.thm1}, Theorem~\ref{ch1.sec2.thm0}[2.] and Lemma~\ref{ch1.sec2.lemma2a-2} be satisfied. And let $f_{T_{i}}, i=1,2,3$  be the arbitrary  density functions of the random variables $T_{1}, T_{2}$ and $T_{3}$. There exists a Lyapunov functional
 \begin{equation}\label{ch2.sec2.thm2.eq1a}
V=V_{1}+V_{22},
\end{equation}
where
   $V_{1}\in\mathcal{C}^{2, 1}(\mathbb{R}^{3}\times \mathbb{R}_{+}, \mathbb{R}_{+})$ is defined by (\ref{ch2.sec2.thm2a.eq2})
and $V_{22}$ is defined as follows:
%%%%%%%%%%%%%
%We construct a Lyapunov functional
 \begin{eqnarray}
   %%%%%%%%%%%%%%%%\nonumber \\
   &&V_{22}(x,t)=2\alpha \int_{t_{0}}^{\infty}f_{T_{3}}(r)e^{-2\mu r} \int_{t-r}^{t}I^{2}(v)dvdr  \nonumber\\
   &&+[2\beta S^{*}_{0}\left(1+c\right)+ {\sigma}^{2}_{\beta}(S^{*}_{0})^{2}(4c+2(1-c)^{2})]\int_{t_{0}}^{h_{1}}f_{T_{1}}(s)e^{-2\mu s}\int^{t}_{t-s}G^{2}(I(v))dvds\nonumber\\
   &&+\left[\beta S^{*}_{0}(4+c)+\beta (S^{*}_{0})^{2}(2+c)+{\sigma}^{2}_{\beta}(S^{*}_{0})^{2}(4c+10)\right]\times\nonumber\\
   &&\times\left[\int_{t_{0}}^{h_{2}}\int_{t_{0}}^{h_{1}}f_{T_{2}}(u)f_{T_{1}}(s)e^{-2\mu (s+u)}\int^{t}_{t-u}G^{2}(I(v-s))dvdsdu\right.\nonumber\\
   &&\left.+\int_{t_{0}}^{h_{2}}\int_{t_{0}}^{h_{1}}f_{T_{2}}(u)f_{T_{1}}(s)e^{-2\mu (s+u)}\int^{t}_{t-s}G^{2}(I(v))dvdsdu\right]\nonumber\\
   %\int_{t_{0}}^{h_{2}}\int_{t_{0}}^{h_{1}}f_{T_{2}}(u)f_{T_{1}}(s)e^{-2\mu (s+u)}\int^{t}_{t-(s+u)}G^{2}(I(v))dvdsdu\nonumber\\
   %%%%%%%%%%%%%%%%%%%%%%%%----G
   %%%&&
   \label{{ch2.sec2.thm2.eq4a}}
\end{eqnarray}
Furthermore, there exists threshold values $R_{1}$, $U_{0}$ and $V_{0}$  defined as follows:
\begin{equation}\label{ch2.sec2.thm1.eq5a}
R_{1}=\frac{\beta S^{*}_{0} \hat{K}_{1}+\alpha+\frac{1}{2}\sigma^{2}_{I}}{(\mu+d+\alpha)},
\end{equation}
\begin{equation}\label{ch2.sec2.thm1.eq5b}
U_{0}=\frac{2\beta S^{*}_{0}+\beta +\alpha + 2\frac{\mu}{\tilde{K}(\mu)^{2}}}{2\mu},
\end{equation}
 and
 \begin{equation}\label{ch2.sec2.thm1.eq5c}
V_{0}=\frac{(2\mu \tilde{K}(\mu)^{2} + \alpha + \beta (2S^{*}_{0}+1 ) +\sigma^{2}_{E})}{2\mu},
\end{equation}
with some constant $\hat{K}_{1}>0$  that depends on $S^{*}_{0}$ and $\sigma_{\beta}$ (in fact, $\hat{K}_{1}=4+ S^{*}_{0}+ 6\frac{1}{\beta}\sigma^{2}_{\beta} S^{*}_{0}$), and some positive constants $\phi$, $\psi$, and $ \varphi $, such that,
   under the assumptions that $R_{1}\leq 1$, $U_{0}\leq 1$, and $V_{0}\leq 1$,  the drift part $LV$ of the differential operator $dV$ applied to $V$ with respect to the stochastic dynamic system (\ref{ch1.sec0.eq8})-(\ref{ch1.sec0.eq11}) satisfies the following inequality:
 \begin{equation}\label{ch2.sec2.thm1.eq6}
  L{V}(x,t)\leq  -\left(\phi U^{2}(t)+\psi V^{2}(t)+ \varphi W^{2}(t)\right).
\end{equation}%\l
 %%%%%%%%%%%%%%%%%%%%
 %%%%%%%%%%%%%%%%%%%%%%%%%%%
%%%%%%%%%%%%%%%%%%%
%%%%%%%%%%%%%%%%%%%
\end{lemma}
Proof:\\
The drift part $LV$ of the differential operator $dV$ applied to the Lyapunov functional defined in (\ref{ch2.sec2.thm2.eq1a}), (\ref{ch2.sec2.thm2a.eq2}) and (\ref{{ch2.sec2.thm2.eq4a}}) with respect to system (\ref{ch1.sec0.eq8})-(\ref{ch1.sec0.eq11}) leads to the following:
\begin{eqnarray}%\label{ch2.sec2.thm1.proof.eq6}
  L{V}(x,t)&=&L{V}_{1}(x,t)\nonumber\\
  %%%%%%%%%%%%%%%%%%%%%%%%%%
   &&+2\alpha \int_{t_{0}}^{\infty}f_{T_{3}}(r)e^{-2\mu r} W^{2}(t)dr  \nonumber\\
   &&+[2\beta S^{*}_{0}\left(1+c\right)+ {\sigma}^{2}_{\beta}(S^{*}_{0})^{2}(4c+2(1-c)^{2})]\int_{t_{0}}^{h_{1}}f_{T_{1}}(s)e^{-2\mu s}G^{2}(W(t))ds\nonumber\\
   &&+\left[\beta S^{*}_{0}(4+c)+\beta (S^{*}_{0})^{2}(2+c)+{\sigma}^{2}_{\beta}(S^{*}_{0})^{2}(4c+10)\right]\times\nonumber\\
   &&\times\int_{t_{0}}^{h_{2}}\int_{t_{0}}^{h_{1}}f_{T_{2}}(u)f_{T_{1}}(s)e^{-2\mu (s+u)}G^{2}(W(t))dsdu\nonumber\\
   %%%%%%%%%%%%%%%%%%%%%%%%----G
   %%%%%%%%%%%%%%%%%%%%%%%%%%%%%
    %%%%%%%%%%%%%%%%%\nonumber \\
   &&-2\alpha \int_{t_{0}}^{\infty}f_{T_{3}}(r)e^{-2\mu r} W^{2}(t-r)dr  \nonumber\\
   &&-[2\beta S^{*}_{0}\left(1+c\right)+ {\sigma}^{2}_{\beta}(S^{*}_{0})^{2}(4c+2(1-c)^{2})]\int_{t_{0}}^{h_{1}}f_{T_{1}}(s)e^{-2\mu s}G^{2}(W(t-s))ds\nonumber\\
   &&-\left[\beta S^{*}_{0}(4+c)+\beta (S^{*}_{0})^{2}(2+c)+{\sigma}^{2}_{\beta}(S^{*}_{0})^{2}(4c+10)\right]\times\nonumber\\
   &&\times\int_{t_{0}}^{h_{2}}\int_{t_{0}}^{h_{1}}f_{T_{2}}(u)f_{T_{1}}(s)e^{-2\mu (s+u)}G^{2}(W(t-s-u))dsdu.\nonumber\\
   %%%%%%%%%%%%%%%%%%%%%%%%----G
    %%%%%%%%%%%%%%%%%%%%%%%%%%%%%%%%%%%%%%%%%%%%%%%%%%%%%%%%%%%%%%%%
  \label{ch2.sec2.thm1.proof.eq7}
\end{eqnarray}
Under the assumptions for $\sigma_{i},i=S,E,I, \beta $ in Theorem~\ref{ch1.sec2.thm0}[2.], and for some suitable choice of the positive constant $ c$,  it follows from  (\ref{ch2.sec2.thm2a.eq5}), (\ref{ch2.sec2.thm1.proof.eq7}), the statements of Assumption~\ref{ch1.sec0.assum1}, $A5$ (i.e. $G^{2}(x)\leq x^{2}, x\geq 0$)  and some further algebraic manipulations and simplifications that
\begin{equation}\label{ch2.sec2.thm1.proof.eq8}
  L{V}(x,t)\leq  -\left(\phi U^{2}(t)+\psi V^{2}(t)+ \varphi W^{2}(t)\right),
\end{equation}%\label{ch2.sec2.thm1.proof.eq6}
where,
 \begin{eqnarray}
 \phi &=&2\mu (1-U_{0})\label{ch2.sec2.thm1.proof.eq8a}\\
 \psi &=&2\mu (1-V_{0})-2\mu c\left(1-\frac{\beta (3S^{*}_{0}+1)+\sigma^{2}_{E}}{2\mu}\right)\label{ch2.sec2.thm1.proof.eq8b}\\
\varphi &=&2(\mu +d+ \alpha) (1-R_{1})-c(3\beta S^{*}_{0}+\beta (S^{*}_{0})^{2}+4\sigma^{2}_{\beta}(S^{*}_{0})^{2})-2c^{2}\sigma^{2}_{\beta}(S^{*}_{0})^{2}.\label{ch2.sec2.thm1.proof.eq8c}
 \end{eqnarray}
 where $
R_{1}$ is defined in (\ref{ch2.sec2.thm1.eq5a}).
 %%%%%%%%%%%%%%%%%%%%%%%%%%%%%%%
 %%%%%%%%%%%%%%%%%%%%%%%%%%%%
  In addition, under the assumptions of $R_{1}$, $U_{0}$, and $V_{0}$ in the hypothesis and for a suitable choice of the positive constant $c$ it follows that $\phi$, $\psi$, and  $\varphi$ are positive constants and (\ref{ch2.sec2.thm1.eq6}) follows immediately.
 %%%%%%%%%%%%%%%%%%%%%%%%%%%%%
%%%%%
The following result describes the stochastic asymptotic stability of the  disease free equilibrium $E_{0}$, whenever it exists.
\begin{thm}\label{ch1.sec2.theorem1}
Suppose Theorem~\ref{ch1.sec2.thm0}[2.] and the hypotheses of Lemma~\ref{ch1.sec2.lemma2a-1} and Lemma~\ref{ch1.sec2.lemma2} are satisfied, then the
  disease free equilibrium $E_{0}$ of the stochastic dynamic system (\ref{ch1.sec0.eq8})-(\ref{ch1.sec0.eq11}) is stochastically  asymptotically stable in the large in the set $D(\infty)$. Moreover, the steady state solution $E_{0}$ is  exponentially mean square stable.
\end{thm}
Proof\\
The result follows by applying the comparison stability results\cite{mao, wanduku-delay}.  Moreover, the disease free equilibrium state is exponentially mean square stable.
%%%%%
%%%%
\begin{rem}\label{ch1.sec2.theorem1.rem1}
\item[1.]
Theorem~\ref{ch1.sec2.theorem1} signifies that in the absence of the white noise process in the system due to the random fluctuations in the natural death rate of the susceptible population, that is, $\sigma_{S}=0$, it follows that regardless of (1.) the  variability in  the natural death of the exposed, infectious, and the removal populations, that is, $\sigma^{2}_{i}\geq 0, i= E, I, R, $, or (2.) the variability in the disease transmission rate, that is, $\sigma^{2}_{ \beta}\geq 0$, or (3.) the variability of the incubation delay of the disease in the vector and in the human being ($T_{1}$  and $T_{2}$ respectively), or (4.) the  variability of the natural immunity period $T_{3}$, a stochastically asymptotically stable in the large disease free steady state $E_{0}$ for the population exists, whenever the threshold conditions: $R_{1}\leq 1$, $U_{0}\leq 1$, and $V_{0}\leq 1$ are satisfied, where $R_{1}$, $U_{0}$, and $V_{0}$ are defined in (\ref{ch2.sec2.thm1.eq5a})-(\ref{ch2.sec2.thm1.eq5c}). It should be noted that the threshold values $R_{1}$, $U_{0}$, and $V_{0}$ are explicit in terms of the parameters of the system (\ref{ch1.sec0.eq8})-(\ref{ch1.sec0.eq11}) and  also computationally attractive, whenever specific values for the parameters of the system are given. This observation  suggests that  in a disease scenario  where there are no random fluctuations in the  natural death rate of susceptible individuals in the population, that is, $\sigma_{S}=0$, the stochastic dynamic system (\ref{ch1.sec0.eq8})-(\ref{ch1.sec0.eq11}) exhibits a disease free steady state for the population, regardless of whether there is (1.)  a significant variability in  the disease transmission rate, that is, $\sigma^{2}_{\beta}\geq 0$, or (2.) a significant variability  in the natural death rates of the other states- $E, I, R$, that is, $\sigma^{2}_{i}\geq 0, i= E, I, R$ or (3.) a significant variability in the delay periods in the system- $T_{1}, T_{2}$ and $T_{3}$. Moreover, the disease can be eradicated from the system (\ref{ch1.sec0.eq8})-(\ref{ch1.sec0.eq11}), whenever the threshold conditions $R_{1}\leq 1$, $U_{0}\leq 1$, and $V_{0}\leq 1$ are satisfied. Furthermore, unlike in the case of constant delays in the system noted in Remark~\ref{ch1.sec2.theorem1a.rem1}, the threshold conditions in Theorem~\ref{ch1.sec2.theorem1} provide stronger bases for policy decisions related to malaria eradication with no additional restrictions to the average lengths of the incubation and natural immunity periods in the system.
%%%%%%%%%%%%%%%%
  \item[2.]
 Since the threshold conditions, that is, $R_{1}\leq 1$, $U_{0}\leq 1$, and $V_{0}\leq 1$,  are sufficient for the disease free steady state population, $E_{0}$, to be stochastically asymptotically stable in the large, and consequently for the eradication of the disease from the population, it follows that the threshold value $R_{1}$ is the noise-modified basic reproduction number for the disease dynamics described by stochastic system (\ref{ch1.sec0.eq8})-(\ref{ch1.sec0.eq11}), whenever the assumptions:-  $\sigma^{2}_{i}\geq 0, i= E, I, R, \beta$ and $\sigma^{2}_{S}=0$ are satisfied.
\end{rem}
%\subsection{Existence and deterministic Asymptotic Behavior of Disease Free Equilibrium \label{ch1.sec2.subsec2}}
The following result characterizes the stability of the disease free equilibrium of the deterministic  system (\ref{ch1.sec0.eq3})-(\ref{ch1.sec0.eq6}). The deterministic version of the Lyapunov functional  technique is used to establish the results.
\begin{thm}\label{ch1.sec2.theorem1.corollary1}
 Let the hypotheses of  Theorem~\ref{ch1.sec1.thm1a}, Theorem~\ref{ch1.sec2.thm0}[1.] and Lemma~\ref{ch1.sec2.lemma2a-2} be satisfied. And let $f_{T_{i}}, i=1,2,3$  be the arbitrary  density functions of the random variables $T_{1}, T_{2}$ and $T_{3}$. There exists a Lyapunov functional
 \begin{equation}\label{ch1.sec2.theorem1.corollary1.eq1}
V=V_{1}+V_{32},
\end{equation}
where
 %$V_{1}:\mathbb{R}^{3}\times \mathbb{R}_{+}\rightarrow \mathbb{R}_{+}$,
  $V_{1}\in\mathcal{C}^{2, 1}(\mathbb{R}^{3}\times \mathbb{R}_{+}, \mathbb{R}_{+})$ is defined by (\ref{ch2.sec2.thm2a.eq2})
and $V_{32}$ is defined as follows:
%%%%%%%%%%%%%
%We construct a Lyapunov functional
%%%%%%%%%%%%%%%%%%%%%%%%%%%%%%%%%%%%%%%%%%%%%%%%%%%%%%%%%%%%%%%%%%%%%%%%%%%%%
%
%%%%%%%%%%%%%%%%%%%%%%%%%%%%%%%%%%%%%%%%%%%%%%%%%%%%%%%%%%%%%%%%%%%%%%%%%%%%
%%%%%%%%%%%%%%%%%%%%%%%%%%%
 \begin{eqnarray}
   %%%%%%%%%%%%%%%%\nonumber \\
   &&V_{32}(x,t)=2\alpha \int_{t_{0}}^{\infty}f_{T_{3}}(r)e^{-2\mu r} \int_{t-r}^{t}I^{2}(v)dvdr  \nonumber\\
   &&+[2\beta S^{*}_{0}\left(1+c\right)]\int_{t_{0}}^{h_{1}}f_{T_{1}}(s)e^{-2\mu s}\int^{t}_{t-s}G^{2}(I(v))dvds\nonumber\\%+ {\sigma}^{2}_{\beta}(S^{*}_{0})^{2}(4c+2(1-c)^{2})\
   &&+\left[\beta S^{*}_{0}(4+c)+\beta (S^{*}_{0})^{2}(2+c)\right]\times\nonumber\\%+{\sigma}^{2}_{\beta}(S^{*}_{0})^{2}(4c+10)
   &&\times\left[\int_{t_{0}}^{h_{2}}\int_{t_{0}}^{h_{1}}f_{T_{2}}(u)f_{T_{1}}(s)e^{-2\mu (s+u)}\int^{t}_{t-u}G^{2}(I(v-s))dvdsdu\right.\nonumber\\
   &&\left.+\int_{t_{0}}^{h_{2}}\int_{t_{0}}^{h_{1}}f_{T_{2}}(u)f_{T_{1}}(s)e^{-2\mu (s+u)}\int^{t}_{t-s}G^{2}(I(v))dvdsdu\right].\nonumber\\
   %%%%%%%%%%%%%%%%%%%%%%%%----G
   %%%&&
   \label{ch1.sec2.theorem1.corollary1.eq2}
\end{eqnarray}
Furthermore, define $R_{0}$, $U_{0}$ and $V_{0}$,  as follows:
\begin{equation}\label{ch1.sec2.theorem1.corollary1.eq3}
R_{0}=\frac{\beta S^{*}_{0} \hat{K}_{0}+\alpha}{(\mu+d+\alpha)},
\end{equation}
%%%%%%%%%%%
\begin{equation}\label{ch1.sec2.theorem1.corollary1.eq4}
\hat{U}_{0}=\frac{2\beta S^{*}_{0}+\beta +\alpha + 2\frac{\mu}{\tilde{K}(\mu)^{2}}}{2\mu},
\end{equation}
 and
 \begin{equation}\label{ch1.sec2.theorem1.corollary1.eq5}
\hat{V}_{0}=\frac{(2\mu \tilde{K}(\mu)^{2} + \alpha + \beta (2S^{*}_{0}+1 ) )}{2\mu},
\end{equation}
%%%%%%%%%%%%%%%%
where, $\hat{K}_{0}>0$ is a constant that depends only on $S^{*}_{0}$  (in fact, $\hat{K}_{0}=4+ S^{*}_{0} $).
 Assume that $R_{0}\leq 1$, $\hat{U}_{0}\leq 1$, and $\hat{V}_{0}\leq 1$,  then  there exist positive constants $\phi_{1}$, $\psi_{1}$, and $ \varphi_{1} $, such that the differential operator $\dot{V}$ applied to $V$ with respect to the deterministic system (\ref{ch1.sec0.eq3})-(\ref{ch1.sec0.eq6}) satisfies the following differential inequality:
 \begin{equation}\label{ch1.sec2.theorem1.corollary1.eq6}
  \dot{V}(x,t)\leq -\min{(\phi_{1}, \psi_{1}, \varphi_{1} )}||X(t)-E_{0}||^{2},% -\left(\phi_{1} U^{2}(t)+\psi_{1} V^{2}(t)+ \varphi_{1} W^{2}(t)\right).
\end{equation}%\l
  % inequality:
 %\begin{equation} \label{ch1.sec2.lemma2a.corrolary1.eq9}
 % \dot{V}(x,t)\leq  -\min{(\phi, \psi, \varphi )}||X(t)-E_{0}||^{2},%\left(\phi U^{2}(t)+\psi V^{2}(t)+ \varphi W^{2}(t)\right).
%\end{equation}%\l
where $X(t)$ is defined in (\ref{ch1.sec0.eq13b}), and $||.||$ is the natural Euclidean norm on $\mathbb{R}^{2}$.

Moreover,  under the assumptions in the hypothesis of Theorem~\ref{ch1.sec2.thm0}[1.], the disease free equilibrium $E_{0}$ of the resulting system (\ref{ch1.sec0.eq8})-(\ref{ch1.sec0.eq11})  is uniformly globally asymptotically stable in the set $D(\infty)$.
 %%%%%%%%%%%%%%%%%%%%%%%%%%%
%%%%%%%%%%%%%%%%%%%
%%%%%%%%%%%%%%%%%%%
\end{thm}
Proof: \\
 The result follows directly from the proofs of Lemma~\ref{ch1.sec2.lemma2a-2} and Lemma~\ref{ch1.sec2.lemma2} by setting $\sigma_{i}=0,i=S,E,I, \beta $, and also applying the comparison stability results in \cite{ladde,wanduku-determ}, where from (\ref{ch2.sec2.thm1.proof.eq8a})-(\ref{ch2.sec2.thm1.proof.eq8c}),  $\phi_{1}=\phi>0$, $\psi_{1}=\psi>0$, and $ \varphi_{1}=\varphi>0 $.
%%Stochastic part
\begin{rem}\label{ch1.sec2.rem1}
\item[1.]
Theorem~\ref{ch1.sec2.theorem1.corollary1} signifies that in the absence of random fluctuations in the disease dynamics leading to variability in any of the parameters of the system- natural death or disease transmission rates, the system (\ref{ch1.sec0.eq3})-(\ref{ch1.sec0.eq6}) has a disease free steady state, $E_{0}$,  which is globally uniformly  and asymptotically stable under the threshold conditions $R_{0}\leq 1$,  $\hat{U}_{0}\leq 1$ and $\hat{V}_{0}\leq 1$, where the threshold values $R_{0}$,  $\hat{U}_{0}$ and $\hat{V}_{0}$ are defined in (\ref{ch1.sec2.theorem1.corollary1.eq3})-(\ref{ch1.sec2.theorem1.corollary1.eq5}).

 It should also be noted that the threshold values $R_{0}$,  $\hat{U}_{0}$ and $\hat{V}_{0}$ are explicit in terms of the parameters of the system (\ref{ch1.sec0.eq3})-(\ref{ch1.sec0.eq6}) and  also computationally attractive, whenever specific values for the parameters of the system are given. This observation  suggests that in a disease scenario,  where there are no random fluctuations in the disease dynamics owing to  the disease transmission or natural death rates in the  dynamic system (\ref{ch1.sec0.eq3})-(\ref{ch1.sec0.eq6}), the system has a disease free steady state, $E_{0}$,  for the population.  In addition, if the  threshold values $R_{0}$,  $\hat{U}_{0}$ and $\hat{V}_{0}$ satisfy the  conditions $R_{0}\leq 1$,  $\hat{U}_{0}\leq 1$ and $\hat{V}_{0}\leq 1$, then the disease can be eradicated from the population.
%%%%%%%%%%%%%%%%%%%%%%%%%%%%%%%%%%
\item[2]
Furthermore, comparing the threshold values for Theorem~\ref{ch1.sec2.theorem1.corollary1} and  Theorem~\ref{ch1.sec2.theorem1} defined in (\ref{ch1.sec2.theorem1.corollary1.eq3})-(\ref{ch1.sec2.theorem1.corollary1.eq5}) and (\ref{ch2.sec2.thm1.eq5a})-(\ref{ch2.sec2.thm1.eq5c}) respectively, it is easy to see that $R_{0}\leq R_{1}$, $ U_{0}=\hat{U}_{0}$ and $\hat{V}_{0}\leq V_{0}$,  whenever the intensity values of the white noise processes in the system (\ref{ch1.sec0.eq8})-(\ref{ch1.sec0.eq11}) satisfy the following conditions: $\sigma_{i}>0, i=E, I, \beta$ and $\sigma_{S}=0$. It is easy to see that the intensity values $\sigma_{i}>0, i=E, I, R, \beta$  of the corresponding white noise processes owning to the disease transmission and natural death rates of the exposed, infections and removal individuals in the stochastic dynamic system (\ref{ch1.sec0.eq8})-(\ref{ch1.sec0.eq11}) exert constraints on the  threshold values:  $R_{1},  U_{0}$, and $ V_{0}$, defined in (\ref{ch2.sec2.thm1.eq5a})-(\ref{ch2.sec2.thm1.eq5c}), than  on the set of threshold values $R_{0}$,  $\hat{U}_{0}$ and $\hat{V}_{0}$ defined in (\ref{ch1.sec2.theorem1.corollary1.eq3})-(\ref{ch1.sec2.theorem1.corollary1.eq5}). Moreover,  the  threshold values- $R_{0}$,  $\hat{U}_{0}$ and $\hat{V}_{0}$  from (\ref{ch1.sec2.theorem1.corollary1.eq3})-(\ref{ch1.sec2.theorem1.corollary1.eq5}) attain the threshold  condition of "less than the bound $1$" more rapidly compared to the set of threshold values  $R_{1},  U_{0}$, and $ V_{0}$ from (\ref{ch2.sec2.thm1.eq5a})-(\ref{ch2.sec2.thm1.eq5c}), whenever the intensity values $\sigma_{i}>0, i=E, I, R,  \beta$. Consequently, the disease would be eradicated more rapidly in a disease scenario where the disease dynamics is driven by parameters that lead to the set of  threshold values $R_{0}$,  $\hat{U}_{0}$ and $\hat{V}_{0}$,  and eradicated less rapidly in a disease scenario where the driving parameters of the disease dynamics lead to the threshold values $R_{1},  U_{0}$, and $ V_{0}$ which depend on $\sigma_{i}>0, i=E, I, \beta$.

 This observation suggests that the occurrence of random fluctuations (in the disease transmission and natural death rates) with significant large intensity values in the disease dynamics exert counter positive effects against the disease eradication process. Furthermore, the intensity levels of the random fluctuations in the disease dynamics expressed in terms of the sizes of the intensity values, $\sigma_{i}, i=E, I, R,  \beta$, of the white noise processes in the stochastic system (\ref{ch1.sec0.eq8})-(\ref{ch1.sec0.eq11}), reflect the weight of the counter positive effects exerted against the disease eradication process. Therefore, good malaria eradication policies should embark on controlling the intensities of the random fluctuations of malaria transmission and natural death rates of people who are infected ($E, I$). Perhaps this can be achieved by controlling the mosquito populations, and better care of people who are infected with the malaria parasite.
%%%%%%%%%%%%%
 %%%%
 \item[3.]
 The threshold conditions, that is, $R_{0}\leq 1$, $\hat{U}_{0}\leq 1$, and $\hat{V}_{0}\leq 1$, are sufficient for the existence of a globally uniformly asymptotically stable disease free steady state population, $E_{0}$ in the deterministic system (\ref{ch1.sec0.eq3})-(\ref{ch1.sec0.eq6}). Furthermore, these conditions are sufficient for the eradication of the disease from the population. Hence, the threshold value $R_{0}$ is the noise-free basic reproduction number for the disease dynamics described by deterministic system (\ref{ch1.sec0.eq3})-(\ref{ch1.sec0.eq6}).
 \end{rem}
%Corollary exhibits the asymptotic stability results for the disease free equilibrium $E_{0}$ of the deterministic system (\ref{ch1.sec0.eq3})-(\ref{ch1.sec0.eq5}).  Recall, the system %(\ref{ch1.sec0.eq8})-(\ref{ch1.sec0.eq11}) has no disease free equilibrium when at least one of $\sigma_{i}\neq 0,i=S,E,I,R$.  The following result characterizes the asymptotic behavior of the %stochastic system (\ref{ch1.sec0.eq8})-(\ref{ch1.sec0.eq11}) in the neighborhood of the disease free equilibrium $E_{0}$ of the deterministic system (\ref{ch1.sec0.eq3})-(\ref{ch1.sec0.eq5}).
While Theorem~\ref{ch1.sec2.thm0}[3.] asserts that the stochastic system (\ref{ch1.sec0.eq8})-(\ref{ch1.sec0.eq11}) has no disease free steady state for the population, whenever the intensity value of the white noise process due to the natural death rate of the susceptible population is positive, that is, $\sigma_{S}>0$, it is necessary to characterize  the behavior of the solutions of the stochastic system near the disease free steady state $E_{0}$ of the deterministic system (\ref{ch1.sec0.eq3})-(\ref{ch1.sec0.eq6}), in order to gain insight about the extend of behavioral deviation of the stochastic system (\ref{ch1.sec0.eq8})-(\ref{ch1.sec0.eq11}) away from the deterministic system disease free equilibrium $E_{0}$, whenever the intensity value  $\sigma_{S}>0$.

The following result describes the oscillatory behavior of the  trajectories of the stochastic system (\ref{ch1.sec0.eq8})-(\ref{ch1.sec0.eq11}) in the neighborhood of the disease free equilibrium $E_{0}$ obtained in Theorem~\ref{ch1.sec2.thm0}[1.],[2.],  whenever Theorem~\ref{ch1.sec2.thm0}[3.] is satisfied. That is, whenever   the stochastic system (\ref{ch1.sec0.eq8})-(\ref{ch1.sec0.eq11}) does not have a disease free equilibrium.  This result characterizes the expected average relative distance between  solutions of the stochastic system (\ref{ch1.sec0.eq8})-(\ref{ch1.sec0.eq11}) and the disease free steady state $E_{0}$,  in an attempt to describe the average size or amplitude of the trajectories of the stochastic system (\ref{ch1.sec0.eq8})-(\ref{ch1.sec0.eq11}) relative to the position of the deterministic disease free state $E_{0}$.
\begin{thm}\label{ch1.sec2.theorem2}
Let the hypothesis of Theorem~\ref{ch1.sec2.thm0}[3.] be satisfied. And define the following threshold values:
\begin{equation}\label{ch2.sec2.thm2.eq1}
\tilde{R}_{1}=\frac{\beta S^{*}_{0} \hat{K}_{1}+\alpha+\frac{1}{2}\sigma^{2}_{I}}{(\mu+d+\alpha)},
\end{equation}
\begin{equation}\label{ch2.sec2.thm2.eq2}
\tilde{U}_{0}=\frac{2\beta S^{*}_{0}+\beta +\alpha + 2\frac{\mu}{\tilde{K}(\mu)^{2}}+\sigma^{2}_{S}}{2\mu},
\end{equation}
 and
 \begin{equation}\label{ch2.sec2.thm2.eq3}
\tilde{V}_{0}=\frac{(2\mu \tilde{K}(\mu)^{2} + \alpha + \beta (2S^{*}_{0}+1 )+\sigma^{2}_{E} )}{2\mu},
\end{equation}
with some constant $\hat{K}_{1}>0$  that depends on $S^{*}_{0}$ and $\sigma_{\beta}$ (in fact, $\hat{K}_{1}=4+ S^{*}_{0}+ 6\frac{1}{\beta}\sigma^{2}_{\beta}S^{*}_{0} $). Let $X(t)=(S(t),E(t),I(t))$ be a solution of the decoupled system from (\ref{ch1.sec0.eq8})-(\ref{ch1.sec0.eq11}) with initial conditions (\ref{ch1.sec0.eq12}). Assume that,
       $\tilde{R}_{1}\leq 1$, $\tilde{U}_{0}\leq 1$, and $\tilde{V}_{0}\leq 1$,  then  there exists a positive constant $\mathfrak{m}>0$,  such that  the following inequality holds
 \begin{equation}\label{ch2.sec2.thm2.eq4}
  \limsup_{t\rightarrow \infty}\frac{1}{t}E\int^{t}_{0}\left[ (S(v)-S^{*}_{0})^{2}+ E^{2}(v)+  I^{2}(v)\right]dv\leq \frac{3\sigma^{2}_{S}(S^{*}_{0})^{2}}{\mathfrak{m}}.
\end{equation}%\l
 %%%%%%%%%%%%%%%%%%%%%%%%%%%
\end{thm}
Proof:
Let Theorem~\ref{ch1.sec2.thm0}[3.] be satisfied.  Applying the differential operator $dV$ to $V$ defined in (\ref{ch2.sec2.thm2.eq1a}), and utilizing  (\ref{ch2.sec2.thm2a.eq4}) and (\ref{ch2.sec2.thm2a.eq5}), it is easy to see that
\begin{eqnarray}
 &&dV=LV dt-2\sigma_{S}(U(t)+V(t))(S^{*}_{0}+U(t))dw_{S}(t)\nonumber\\
 &&-2\sigma_{E}(U(t)V(t)+(c+1)V^{2}(t))dw_{E}(t)-2\sigma_{I}W^{2}(t))dw_{I}(t)\nonumber\\
 &&-2c\sigma_{\beta}(S^{*}_{0}+U(t))V(t)\int_{t_{0}}^{h_{1}}f_{T_{1}}(s)e^{-\mu s}G(W(t-s))dsdw_{\beta}\nonumber\\
 &&-2\sigma_{E}[U(t)+(c+1)V(t)+W(t)]\times\nonumber\\
 &&\times\int_{t_{0}}^{h_{2}}\int_{t_{0}}^{h_{1}}f_{T_{2}}(u)f_{T_{1}}(s)e^{-\mu (s+u)}(S^{*}_{0}+U(t-u))G(W(t-s-u))dsdu dw_{\beta}(t)\label{ch2.sec2.thm2.proof.eq1}
\end{eqnarray}
%We construct a Lyapunov functional
where for some positive constant valued function $\tilde{K}(\mu)$, the drift part of (\ref{ch2.sec2.thm2.proof.eq1}),  $LV$,  satisfies the inequality
\begin{equation}\label{ch2.sec2.thm2.proof.eq2}
  L{V}(x,t)\leq  -\left(\tilde{\phi} U^{2}(t)+\tilde{\psi} V^{2}(t)+ \tilde{\varphi} W^{2}(t)\right),
\end{equation}%\label{ch2.sec2.thm1.proof.eq6}
where
 \begin{eqnarray}
 \tilde{\phi} &=&2\mu (1-\tilde{U}_{0})\\
\tilde{ \psi} &=&2\mu (1-\tilde{V}_{0})-(2\mu+\sigma^{2}_{E}) c\left(1-\frac{\beta (3S^{*}_{0}+1)}{(2\mu+\sigma^{2}_{E})}\right)\\
\tilde{\varphi} &=&2(\mu +d+ \alpha) (1-\tilde{R}_{1})-c(3\beta S^{*}_{0}+\beta (S^{*}_{0})^{2}+4\sigma^{2}_{\beta}(S^{*}_{0})^{2})-2c^{2}\sigma^{2}_{\beta}(S^{*}_{0})^{2}.\label{ch2.sec2.thm2.proof.eq3}
 \end{eqnarray}
 Moreover, $
\tilde{R}_{1}=\frac{\beta S^{*}_{0} \hat{K}_{1}+\alpha+\frac{1}{2}\sigma^{2}_{I}}{(\mu+d+\alpha)},
$ where $\hat{K}_{1}=4+ S^{*}_{0}+ 6\frac{1}{\beta}\sigma^{2}_{\beta}$.
%%%%%%%%%%%%%%%%%%%%%%%%%%%%%%%%%%%%%%%%%%%%%%%%%%%%%%%%%%%%%%%%%%%%%%%%%%%%%
Under the assumptions of $\tilde{R}_{1}$, $\tilde{U}_{0}$, and $\tilde{V}_{0}$ in the hypothesis and for suitable choice of the positive constant $c$ it follows that $\tilde{\phi}$, $\tilde{\psi}$, and  $\tilde{\varphi}$ are positive constants. Therefore, by integrating (\ref{ch2.sec2.thm2.proof.eq1}) from 0 to $t$ on both sides and taking the expectation, it follows from (\ref{ch2.sec2.thm2.proof.eq1})-(\ref{ch2.sec2.thm2.proof.eq3}) that
 %%%%%%%%%%%%%%%%%%%%%%%%%%%%%
 \begin{eqnarray}\label{ch2.sec2.thm2.proof.eq4}
 E(V(t)-V(0))\leq -\mathfrak{m}E\int^{t}_{0}\left[ (S(v)-S^{*}_{0})^{2}+ E^{2}(v)+  I^{2}(v)\right]dv+3\sigma^{2}_{S}(S^{*}_{0})^{2}t,
\end{eqnarray}
where $V(0)$ is constant and
\begin{equation}
\mathfrak{m}=min(\tilde{\phi},\tilde{ \psi},\tilde{\varphi}).
\end{equation}
 Hence, diving both sides of (\ref{ch2.sec2.thm2.proof.eq4}) by $t$ and $\mathfrak{m}$, and taking the limit supremum as $t\rightarrow \infty$, then (\ref{ch2.sec2.thm2.eq4}) follows immediately.
 \begin{rem}\label{ch1.sec2.rem2}
 \item[1.] Theorem~\ref{ch1.sec2.theorem2} signifies that under  conditions that warrant the nonexistence of a disease free steady state for the stochastic system (\ref{ch1.sec0.eq8})-(\ref{ch1.sec0.eq11}),   the asymptotic expected average relative distance between the trajectories of the stochastic system and the disease free steady state, $E_{0}$ obtained in Theorem~\ref{ch1.sec2.thm0}[1.][2.], does not exceed a constant multiple of the intensity value, $\sigma_{S}$, of the white noise process from the natural death rate of the susceptible population, whenever the following threshold conditions $\tilde{R}_{1}\leq 1$, $\tilde{U}_{0}\leq 1$, and $\tilde{V}_{0}\leq 1$ are satisfied. That is, asymptotically, when the physical characteristics of the disease scenario allow variability in the natural death rate of susceptible individuals with intensity value $\sigma_{S}>0$, there is no disease free steady state for the stochastic system. Nevertheless, the trajectories of the stochastic system (\ref{ch1.sec0.eq8})-(\ref{ch1.sec0.eq11}) will continue to oscillate near the deterministic disease free steady state $E_{0}$ obtained in Theorem~\ref{ch1.sec2.thm0}[1.][2.], whenever the threshold conditions $\tilde{R}_{1}\leq 1$, $\tilde{U}_{0}\leq 1$, and $\tilde{V}_{0}\leq 1$ are satisfied, whence, the threshold values $\tilde{R}_{1}$, $\tilde{U}_{0}$, and $\tilde{V}_{0}$  are defined in (\ref{ch2.sec2.thm2.eq1})-(\ref{ch2.sec2.thm2.eq3}).  Furthermore, the size or amplitude of the oscillations relative to the position of the disease free state $E_{0}$ depends on the size of the intensity value, $\sigma^{2}_{S}$.
              \item[2.] As similarly remarked in Remark~\ref{ch1.sec2.rem1}[2.], comparing the threshold values from Theorem~\ref{ch1.sec2.theorem1.corollary1},  Theorem~\ref{ch1.sec2.theorem1}, and Theorem~\ref{ch1.sec2.theorem2}, that is, $R_{1}$, $U_{0}$, $V_{0}$ in  (\ref{ch2.sec2.thm1.eq5a})-(\ref{ch2.sec2.thm1.eq5c}),  $R_{0}$, $\hat{U}_{0}$, $\hat{V}_{0}$ in (\ref{ch1.sec2.theorem1.corollary1.eq3})-(\ref{ch1.sec2.theorem1.corollary1.eq5}), and $\tilde{R}_{1}$, $\tilde{U}_{0}$, $\tilde{V}_{0}$ in (\ref{ch2.sec2.thm2.eq1})-(\ref{ch2.sec2.thm2.eq3}), it is easy to see that $R_{0}\leq R_{1}=\tilde{R}_{1}$, $ U_{0}=\hat{U}_{0}\leq \tilde{U}_{0}$ and $V_{0}\leq \hat{V}_{0}=\tilde{V}_{0}$, whenever the intensity values of the white noise processes in the system (\ref{ch1.sec0.eq8})-(\ref{ch1.sec0.eq11}) satisfy the condition:  $\sigma_{i}>0, i= S, E, I, R, \beta$. It is also easy to see that the threshold value $\tilde{U}_{0}$ in (\ref{ch2.sec2.thm2.eq2}) has been further constrained by the assumption that $\sigma_{S}>0$, from the corresponding threshold value $U_{0}=\hat{U}_{0}$ in (\ref{ch2.sec2.thm1.eq5b}) and (\ref{ch1.sec2.theorem1.corollary1.eq4}). Meanwhile it was remarked in Remark~\ref{ch1.sec2.rem1}[2.] that the threshold values $R_{1}$, $U_{0}$ and $V_{0}$ from Theorem~\ref{ch1.sec2.theorem1} would attain the threshold condition of " less than the mark of 1" less rapidly compared to the set of threshold values $R_{0}$, $\hat{U}_{0}$ and $\hat{V}_{0}$ from Theorem~\ref{ch1.sec2.theorem1.corollary1}, it is easy to see that the threshold values $\tilde{R}_{1}$, $\tilde{U}_{0}$, and $\tilde{V}_{0}$ from Theorem~\ref{ch1.sec2.theorem2} would attain the threshold condition of "less than the bound 1", even less rapidly when compared to $R_{1}$, $U_{0}$ and $V_{0}$ from Theorem~\ref{ch1.sec2.theorem1}, and even much less rapidly when compared to $R_{0}$, $\hat{U}_{0}$ and $\hat{V}_{0}$ from Theorem~\ref{ch1.sec2.theorem1.corollary1}.

                  This observation suggests that the  sources (natural death or disease transmission rates) and intensity levels of random fluctuations in the disease dynamics exhibit bearings on (1.) the existence or attainment of a disease free population steady  state for the system (\ref{ch1.sec0.eq8})-(\ref{ch1.sec0.eq11}) and also on (2.) the eradication disease. A critical analysis of the importance of the source and intensity levels of the random fluctuations in the disease transmission and natural death rates in the system is a subject for another concurrent study by this author. A valuable note in this remark, in line with the objectives of the current study, and also in comparing the asymptotic behavior of the deterministic and stochastic systems (\ref{ch1.sec0.eq3})-(\ref{ch1.sec0.eq6}) and (\ref{ch1.sec0.eq8})-(\ref{ch1.sec0.eq11}) relative to the disease free steady state $E_{0}$,  and consequently to disease eradication, is the extend to which the presence of the noise in the system affects the threshold hold values: $R_{1}$, $U_{0}$, $V_{0}$ in  (\ref{ch2.sec2.thm1.eq5a})-(\ref{ch2.sec2.thm1.eq5c}),  $R_{0}$, $\hat{U}_{0}$, $\hat{V}_{0}$ in (\ref{ch1.sec2.theorem1.corollary1.eq3})-(\ref{ch1.sec2.theorem1.corollary1.eq5}), and $\tilde{R}_{1}$, $\tilde{U}_{0}$, $\tilde{V}_{0}$ in (\ref{ch2.sec2.thm2.eq1})-(\ref{ch2.sec2.thm2.eq3}), which control the asymptotic and qualitative properties of the disease free steady state $E_{0}$ in both systems.
                   %For instance, adding the new source of random fluctuations due to the natural death rate of the susceptible population which leads to the white noise process with intensity value  $\sigma_{S}$, it follows that for infinitesimally small values for $\sigma_{S}$, there exists a disease free population state, and the disease can be eradicated, whenever the conditions in Theorem~\ref{ch1.sec2.theorem1.corollary1} and  Theorem~\ref{ch1.sec2.theorem1} are satisfied. But for significant values of the intensity value $\sigma_{S}>0$, the additional source of the white noise due to random fluctuations in the natural death of the susceptible individuals leads to the nonexistence or non-attaintment of a disease free population steady state. Moreover, in such event, the solutions of the system (\ref{ch1.sec0.eq8})-(\ref{ch1.sec0.eq11}) can oscillate closely to the disease free population steady state $E_{0}$ obtained in Theorem~\ref{ch1.sec2.thm0}[1.][2.],  provided that the conditions $\tilde{R}_{1}\leq 1$, $\tilde{U}_{0}\leq 1$, and $\tilde{V}_{0}\leq 1$ are satisfied and the value of $\sigma_{S}$ is relatively small.
      %%%%
      \item[3.]Furthermore, as similarly remarked in Remark~\ref{ch1.sec2.rem1}[2.], the results of Theorem~\ref{ch1.sec2.theorem2}  suggest that in a disease scenario that exhibits random fluctuations with significant positive intensity values ($\sigma_{S}>0$) in the natural death rate of susceptible individuals and consequently does not allow the existence of a disease free steady state population, but exhibit physical characteristics which can be represented mathematically by the parameters of the system (\ref{ch1.sec0.eq8})-(\ref{ch1.sec0.eq11}), wherein the threshold values $\tilde{R}_{1}$, $\tilde{U}_{0}$, and $\tilde{V}_{0}$ in (\ref{ch2.sec2.thm2.eq1})-(\ref{ch2.sec2.thm2.eq3}), $R_{1}$, $U_{0}$, $V_{0}$ in  (\ref{ch2.sec2.thm1.eq5a})-(\ref{ch2.sec2.thm1.eq5c}) and $R_{0}$, $\hat{U}_{0}$, $\hat{V}_{0}$ in (\ref{ch1.sec2.theorem1.corollary1.eq3})-(\ref{ch1.sec2.theorem1.corollary1.eq5}),  can all be computed, and satisfy the relationship and threshold conditions, $R_{0}\leq R_{1}=\tilde{R}_{1}\leq 1$, $ U_{0}=\hat{U}_{0}\leq \tilde{U}_{0}\leq 1$ and $V_{0}\leq \hat{V}_{0}=\tilde{V}_{0}\leq 1$,  the occurrence of the white noise processes due to the random environmental fluctuations in  the other sources namely:- from (1.) the natural death rates of the exposed, infectious and removal  populations,  and also from (2.) the disease transmission rate, exert additional  counter-positive  constraints against the disease eradication process as determined by the relationship between the threshold values and the threshold conditions $R_{0}\leq R_{1}=\tilde{R}_{1}\leq 1$, $ U_{0}=\hat{U}_{0}\leq \tilde{U}_{0}\leq 1$ and $V_{0}\leq \hat{V}_{0}=\tilde{V}_{0}\leq 1$. That is, whereas the disease can be eradicated much less rapidly when the disease scenario represented by (\ref{ch1.sec0.eq8})-(\ref{ch1.sec0.eq11}) is controlled by the threshold values $R_{1}$, $U_{0}$, $V_{0}$ in  (\ref{ch2.sec2.thm1.eq5a})-(\ref{ch2.sec2.thm1.eq5c}) than when it is controlled by the threshold values $R_{0}$, $\hat{U}_{0}$, $\hat{V}_{0}$ in (\ref{ch1.sec2.theorem1.corollary1.eq3})-(\ref{ch1.sec2.theorem1.corollary1.eq5}), it follows that when the disease scenario is controlled by the threshold values $\tilde{R}_{1}$, $\tilde{U}_{0}$, and $\tilde{V}_{0}$ in (\ref{ch2.sec2.thm2.eq1})-(\ref{ch2.sec2.thm2.eq3}), the disease cannot be eradicated at all. Nevertheless, the disease population can be maintained close to the disease free population steady state $E_{0}$ obtained in Theorem~\ref{ch1.sec2.thm0}[1.][2.], whenever the  value of $\sigma_{s}$ is small and the threshold conditions $\tilde{R}_{1}\leq 1$, $\tilde{U}_{0}\leq 1$, and $\tilde{V}_{0}\leq 1$ are satisfied. Thus, good policy decisions about malaria control must be informed by this fact that, wherever malaria cannot be completely eradicated, it is important to reduce the intensity of the the random fluctuations in the natural death of the health susceptible population (perhaps via better care of this population), whenever the other threshold conditions $R_{0}\leq R_{1}=\tilde{R}_{1}\leq 1$, $ U_{0}=\hat{U}_{0}\leq \tilde{U}_{0}\leq 1$ and $V_{0}\leq \hat{V}_{0}=\tilde{V}_{0}\leq 1$ hold, in order to contain the disease, and keep the human population close to a disease free state.
      % Moreover, the occurrence of the white noise processes   in the system  (\ref{ch1.sec0.eq8})-(\ref{ch1.sec0.eq11}) lead to the oscillations in the different classes- the susceptible, exposed, infectious and removal states the population with varied sizes or amplitudes of the oscillations depending on the intensity value of the white noise process due to the random fluctuations in the  natural death process of the susceptible population $\sigma_{S}$.
        %The oscillatory behavior of the system (\ref{ch1.sec0.eq8})-(\ref{ch1.sec0.eq11}) relative to $E_{0}$ obtained in Theorem~\ref{ch1.sec2.thm0}[1.][2.] under the influence of various intensity levels of the white noise processes in the system is discussed further in  Section~\ref{ch1.sec2-2}.
    \end{rem}
 %%%%%%
 %%%%%%
 %%%%%%
  %%%%%%%%%%%%%%%%%%%%%
 %%%%%%%%%%%%%%%%%%%%%%%%
 %%%%%%%%%%%%%%%%%%%%%%%%%%
\section{Existence and Asymptotic properties of the Endemic Equilibrium\label{ch1.sec3}}
In this section the existence and asymptotic behavior of the nontrivial steady state of the generalized systems (\ref{ch1.sec0.eq3})-(\ref{ch1.sec0.eq6}) and  (\ref{ch1.sec0.eq8})-(\ref{ch1.sec0.eq11})  are discussed. It is easy to see that the stochastic dynamic system  (\ref{ch1.sec0.eq8})-(\ref{ch1.sec0.eq11}) does not have a non-zero steady state, whenever the system is perturbed by the  white noise from at least one of the sources:  the disease transmission or the natural death rates, that is, when at least one of  $\sigma_{i}> 0,i=S,E,I,R,\beta$. The following result provides  sufficient conditions for the existence of the endemic equilibrium of the deterministic system (\ref{ch1.sec0.eq3})-(\ref{ch1.sec0.eq6}). That is, when the system is unperturbed by random fluctuations in the disease dynamics.
%%%
\begin{thm}\label{ch1.sec3.thm1}
 Let the threshold condition $R_{0}>1$ be satisfied, where $R_{0}$ is  defined in (\ref{ch1.sec2.theorem1.corollary1.eq3}). It follows that the deterministic system (\ref{ch1.sec0.eq3})-(\ref{ch1.sec0.eq6}) has a unique positive equilibrium state denoted by $E_{1}=(S^{*}_{1}, E^{*}_{1}, I^{*}_{1})$,  whenever
\begin{equation}\label{ch1.sec3.thm1.eq1}
E(e^{-\mu(T_{1}+T_{2})})\geq \frac{\hat{K}_{0}+\frac{\alpha}{\beta \frac{B}{\mu}}}{G'(0)},
\end{equation}
where $\hat{K}_{0}$ is defined in  Theorem~\ref{ch1.sec2.theorem1.corollary1}.
\end{thm}
Proof:\\
The nonzero steady state solution $E_{1}=(S^{*}_{1}, E^{*}_{1}, I^{*}_{1})$ of the decoupled deterministic system associated with (\ref{ch1.sec0.eq3})-(\ref{ch1.sec0.eq6})  is a solution to the following system:
\begin{eqnarray}
 &&B-\beta S\int^{h_{1}}_{t_{0}}f_{T_{1}}(s) e^{-\mu s}G(I))ds - \mu S+ \alpha \int_{t_{0}}^{\infty}f_{T_{3}}(r)Ie^{-\mu r}dr =0,\label{ch1.sec3.thm1.proof.eq1}\\
%%%
 &&\beta S\int^{h_{1}}_{t_{0}}f_{T_{1}}(s) e^{-\mu s}G(I)ds - \mu E-\beta \int_{t_{0}}^{h_{2}}f_{T_{2}}(u)S\int^{h_{1}}_{t_{0}}f_{T_{1}}(s) e^{-\mu s-\mu u}G(I)dsdu =0,\nonumber\\
 &&\label{ch1.sec3.thm1.proof.eq2}\\
%%%%
%%%
&&\beta \int_{t_{0}}^{h_{2}}f_{T_{2}}(u)S\int^{h_{1}}_{t_{0}}f_{T_{1}}(s) e^{-\mu s-\mu u}G(I)dsdu- (\mu +d+ \alpha) I=0.\label{ch1.sec3.thm1.proof.eq3}
%%%
\end{eqnarray}
%%%
%%%u
%%
Solving for $E$ and $S$ from  (\ref{ch1.sec3.thm1.proof.eq2}) and (\ref{ch1.sec3.thm1.proof.eq3}) respectively and substituting the result into (\ref{ch1.sec3.thm1.proof.eq1}), gives the following equation:
\begin{equation}\label{ch1.sec3.thm1.proof.eq3b}
H(I)=0
\end{equation}
 where,
\begin{equation}\label{ch1.sec3.thm1.proof.eq4}
H(I)= B-\frac{1}{E(e^{-\mu(T_{1}+T_{2})})}I\left[\frac{(\mu+d+\alpha)\mu}{\beta G(I)}+(\mu+d)E(e^{-\mu T_{1}})+\alpha E(e^{-\mu T_{1}})\left(1-E(e^{-\mu T_{2}})E(e^{-\mu T_{3}})\right)\right].
\end{equation}
But $0<E(e^{-\mu T_{i}})\leq 1, i=1,2,3$, and $\lim_{I\rightarrow \infty} G(I)=C<\infty$,  hence for sufficiently large positive value of $I$, it is easy to see that $H(I)$ is negative. Furthermore, the derivative of $H(I)$ is given by
\begin{eqnarray}
H'(I)&=&-\frac{(\mu+d+\alpha)\mu}{\beta  E(e^{-\mu (T_{1}+T_{2})})} \frac{(G(I)-IG'(I))}{G^{2}(I)}\nonumber\\
&&-\frac{1}{E(e^{-\mu (T_{1}+T_{2})})}\left((\mu+d)E(e^{-\mu T_{1}})+\alpha E(e^{-\mu T_{1}})(1-E(e^{-\mu T_{2}})E(e^{-\mu T_{3}}))\right).\label{ch1.sec3.thm1.proof.eq5}
\end{eqnarray}
Without loss of generality assume that $G'(I)>0$. It follows from the other properties of $G$ in Assumption~\ref{ch1.sec0.assum1}, that is, $G(0)=0$, $G''(I)<0$,  that $(G(I)-IG'(I))>0$ and this further implies that $H'(I)<0$ for all $I> 0$. That is, $H(I)$ is a decreasing function  for all $I> 0$. Therefore, a positive root of the equation (\ref{ch1.sec3.thm1.proof.eq3b}) suffices that $H(0)>0$. But, from (\ref{ch1.sec3.thm1.proof.eq4}),
\begin{equation}\label{ch1.sec3.thm1.proof.eq6}
H(0)=B\left(1- \frac{(\mu+d+\alpha)}{\beta \frac{B}{\mu} G'(0)E(e^{-\mu (T_{1}+T_{2})})}\right)\geq B\left(1- \frac{1}{R_{0}}\right).
\end{equation}
Under the assumptions in the hypothesis of Theorem~\ref{ch1.sec3.thm1}, it is easy to see that  $H(0)>0$.
%%%\begin{equation}\label{ch2.sec2.thm1.eq5aaa}
%R^{*}_{0}=\frac{\beta S^{*}_{0} +\frac{1}{2}\sigma^{2}_{I}}{(\mu+d+\alpha)},
%\begin{equation} \label{ch1.sec2.lemma2a.corrolary1.eq4}
%\hat{R}^{*}_{0}=\frac{\beta S^{*}_{0} }{(\mu+d+\alpha)},
%\end{equation}
%\end{equation}
%%%%%%%%%%%%%%%%%%%%%%%To be completed

To completely exhibit the influence of the delays in the system on the existence of the endemic or nontrivial steady state of the system $E_{1}$ in the absence of any random fluctuations in the disease dynamics, the following result which follows immediately from Theorem~\ref{ch1.sec3.thm1} and Theorem~\ref{ch1.sec2.lemma2a.corrolary1} for constant delays $T_{i}, i=1, 2,3$ in the system is stated as a theorem.
\begin{thm}\label{ch1.sec3.thm1.corrolary1}
Suppose the incubation periods of the malaria plasmodium inside the mosquito and human hosts $T_{1}$ and $T_{2}$, and also the period of effective natural immunity against malaria inside the human being $T_{3}$ are constant.  Let the threshold condition $R^{*}_{0}>1$ be satisfied, where $R^{*}_{0}$ is  defined in (\ref{ch2.sec2.thm1.eq5aaa}). It follows that the deterministic system (\ref{ch1.sec0.eq3})-(\ref{ch1.sec0.eq6}) has a unique positive equilibrium state denoted by $E_{1}=(S^{*}_{1}, E^{*}_{1}, I^{*}_{1})$,  whenever
\begin{equation}\label{ch1.sec3.thm1.corrolary1.eq1}
 T_{1}+T_{2}\leq\frac{1}{\mu}\log{(G^{'}(0))}.%E(e^{-\mu(T_{1}+T_{2})})\geq \frac{\hat{K}_{0}+\frac{\alpha}{\beta \frac{B}{\mu}}}{G'(0)},
\end{equation}
%where $\hat{K}_{0}$ is defined in  Theorem~\ref{ch1.sec2.theorem1.corollary1}.
\end{thm}
Proof:\\
The results in Theorem~\ref{ch1.sec3.thm1.corrolary1} follow easily from the proof of Theorem~\ref{ch1.sec3.thm1} by  letting the probability density functions $f_{T_{i}}, i=1,2,3$ be the dirac-delata function (\ref{ch1.sec2.eq4}), and further applying the translation properties of the dirac-delta function. It can also be easily seen from  (\ref{ch1.sec3.thm1.proof.eq6}) that
%%%%%%%%%%%%%%%%% to be completed.
\begin{equation}\label{ch1.sec3.thm1.corrolary1.proof.eq1}
H(0)=B\left(1- \frac{1}{R^{*}_{0}}\frac{1}{G^{'}(0)e^{-\mu(T_{1}+T_{2})}}\right)\geq B\left(1- \frac{1}{R_{0}}\right).
\end{equation}
Hence, $H(0)>0$, whenever (\ref{ch1.sec3.thm1.corrolary1.eq1}) holds and $R^{*}_{0}>1$, and the existence result follows immediately.
%%%%%%%%%%%%%%%%%%%
\begin{rem}\label{ch1.sec3.rem1}
\item[1.] It is noted that (\ref{ch1.sec3.thm1.eq1}) provides estimates for the expected survival rate $E(e^{-\mu(T_{1}+T_{2})})$ of the parasites or infectious agent inside the vectors and the human body over the full incubation period of the disease. Therefore, Theorem~\ref{ch1.sec3.thm1} provides threshold conditions for the basic reproduction number $R_{0}$, and the expected survival rate of the parasites of the disease described by the system (\ref{ch1.sec0.eq3})-(\ref{ch1.sec0.eq6}) that are sufficient for the dynamics to  establish an endemic steady state of the infection in the population. Thus, malaria control policy decisions should embark on not only reducing the basic reproduction number of the disease $R_{0}$, but also reduce the survival rate of the parasites over the total incubation period in the mosquito and human body.
    \item[2.] The importance of the incubation periods of the malaria plasmodium in the human and mosquito hosts are more highlighted in Theorem~\ref{ch1.sec3.thm1.corrolary1}, where it is assumed that the random variables $T_{i}, i=1,2,3$ are constant. Observe that the assumption in (\ref{ch1.sec3.thm1.corrolary1.eq1}) is also equivalent to condition on the survival probability rate  $e^{-\mu(T_{1}+T_{2})}\geq \frac{1}{G^{'}(0)}$. The term $\frac{1}{\mu}\log{(G^{'}(0))}$ signifies a fraction of the average lifespan $\frac{1}{\mu}$ of an individual in the population, where the fraction is determined by the magnitude of $\log{(G^{'}(0))}$.  Therefore, the assumption in  (\ref{ch1.sec3.thm1.corrolary1.eq1}) signifies that the existence of the endemic equilibrium  in the human population necessitates the total incubation period of the malaria plasmodium inside the mosquito and human being to be at most a fraction of the average lifespan of an individual in the population, whenever the basic reproduction number  satisfies $R^{*}_{0}>1$.

         For example, when the nonlinear incidence function $G$ defined in Assumption~\ref{ch1.sec0.assum1} takes one of the following forms: (1) $G(I)=\frac{I}{1+aI}, a>0$, or (2) $G(I)=\frac{I}{1+aI^{2}}, a>0$ etc., it can be easily seen that $\frac{1}{\mu}\log{(G^{'}(0))}=0$. This result implies from Theorem~\ref{ch1.sec3.thm1.corrolary1} and in particular (\ref{ch1.sec3.thm1.corrolary1.eq1}), that  when  $T_{1}+T_{2}=0$ and the basic reproduction number $R^{*}_{0}$ satisfies $R^{*}_{0}>1$, then an endemic equilibrium state $E_{1}$ exists. That is, in the case where the disease transmission is instantaneous ($T_{1}+T_{2}=0$) between humans and mosquitoes, an outbreak of the disease will rapidly establish a steady endemic population state, whenever  $R^{*}_{0}>1$.

         More generally, let the logarithmic function in  (\ref{ch1.sec3.thm1.corrolary1.eq1}) be written with the base of $b\geq 1$, that is, $\log(.)\equiv \log_{b}(.), b\geq 1$, and also denote $g_{b}(G^{'})=\frac{1}{\mu}\log{(G^{'}(0))}$. It is easy to see that for values of $G^{'}(0)\in [1, b)$, it follows that $0\leq g_{b}(G^{'})<\frac{1}{\mu}$, and for values of $G^{'}(0)\geq b$, it is seen that $g_{b}(G^{'})\geq \frac{1}{\mu}$. Therefore, the values of $G^{'}(0)\geq b$, lead to larger values of $g_{b}(G^{'})$, that further expand the length of the interval $[0,g_{b}(G^{'})]$, which from (\ref{ch1.sec3.thm1.corrolary1.eq1}), gives more leeway for the condition $T_{1}+T_{2}\in [0,g_{b}(G^{'})]$ to hold, and consequently lead to the existence of an endemic equilibrium $E_{1}$, whenever all other conditions in the hypothesis of Theorem~\ref{ch1.sec3.thm1.corrolary1} are satisfed. Conversely, the values of $G^{'}(0)\in [1, b)$, lead to smaller values of $g_{b}(G^{'})$, which further constrict the length of the interval $[0,g_{b}(G^{'})]$, and as a result from (\ref{ch1.sec3.thm1.corrolary1.eq1}), does not give much leeway for the requirement $T_{1}+T_{2}\in [0,g_{b}(G^{'})]$ to hold. This observation about the relationship between the properties of the nonlinear incidence function $G$, and the existence of the endemic equilibrium allows more insight about the properties of the incidence function controlling the persistence of malaria in the human population.
\end{rem}
%%%
%%%
%%%
%%%
The following lemma will be utilized to prove the results that characterize the asymptotic behavior of the stochastic system (\ref{ch1.sec0.eq8})-(\ref{ch1.sec0.eq11}) in the neighborhood of the nontrivial steady state $E_{1}=(S^{*}_{1}, E^{*}_{1}, I^{*}_{1})$,  whenever at least one of $\sigma_{i} \neq 0,i=S,E,I,R,\beta$. It should be noted from Assumption~\ref{ch1.sec0.assum1} that the nonlinear function $G$ is  bounded.  Therefore, suppose
\begin{equation}\label{ch1.sec3.rem1.eqn1}
G^{*}=\sup_{x>0}{G(x)},
\end{equation}
 then it is easy to see that $0\leq  G(x)\leq G^{*}$. It follows further from Assumption~\ref{ch1.sec0.assum1} that given $\lim _{I\rightarrow \infty}{G(I)}=C$, if $G$ is strictly monotonic increasing then $G^{*}\leq C$. Also, if $G$ is strictly monotonic decreasing then $G^{*}\geq C$.
\begin{lemma}\label{ch1.sec3.lemma1}
Let the hypothesis of Theorem~\ref{ch1.sec3.thm1} (or Theorem~\ref{ch1.sec3.thm1.corrolary1}) be satisfied and define the $\mathcal{C}^{2,1}-$ function $V:\mathbb{R}^{3}_{+}\times \mathbb{R}_{+}\rightarrow \mathbb{R}_{+}$ where
\begin{equation}\label{ch1.sec3.lemma1.eq1}
  V(t)=V_{1}(t)+V_{2}(t)+V_{3}(t) +V_{4}(t),
\end{equation}
where,
\begin{equation}\label{ch1.sec3.lemma1.eq2}
  V_{1}(t)=\frac{1}{2}\left(S(t)-S^{*}_{1}+E(t)-E^{*}_{1}+I(t)-I^{*}_{1}\right)^{2},
\end{equation}
\begin{equation}\label{ch1.sec3.lemma1.eq3}
  V_{2}(t)=\frac{1}{2}\left(S(t)-S^{*}_{1}\right)^{2}
\end{equation}
\begin{equation}\label{ch1.sec3.lemma1.eq4}
  V_{3}(t)=\frac{1}{2}\left(S(t)-S^{*}_{1}+E(t)-E^{*}_{1}\right)^{2}.
\end{equation}
 and
 \begin{eqnarray}
  V_{4}(t)&=&\frac{3}{2}\frac{\alpha}{\lambda(\mu)}\int_{t_{0}}^{\infty}f_{T_{3}}(r)e^{-2\mu r}dr(I(\theta)-I^{*})^{2}d\theta dr\nonumber\\
  &&+\frac{\beta S^{*}_{1}}{\lambda(\mu)}(G'(I^{*}_{1}))^{2} \int_{t_{0}}^{h_{2}}\int_{t_{0}}^{h_{1}}f_{T_{2}}(u)f_{T_{1}}(s)e^{-2\mu (s+u)}\int^{t}_{t-(s+u)}(I(\theta)-I^{*}_{1})^{2}d\theta dsdu\nonumber\\
  %%%%%
  %&&+\frac{\beta }{\lambda(\mu)}(G'(I^{*}_{1}))^{2} \int_{t_{0}}^{h_{2}}\int_{t_{0}}^{h_{1}}f_{T_{2}}(u)f_{T_{1}}(s)e^{-2\mu (s+u)}\int^{t}_{t-(s+u)}(I(\theta)-I^{*}_{1})^{2}d\theta %dsdu\nonumber\\
  %%%%%
  &&+[\frac{\beta \lambda(\mu)}{2} \int_{t_{0}}^{h_{2}}\int_{t_{0}}^{h_{1}}f_{T_{2}}(u)f_{T_{1}}(s)e^{-2\mu (s+u)}\int^{t}_{t-(s+u)}G^{2}(I(\theta))(S(\theta)-S^{*}_{1})^{2}d\theta dsdu\nonumber\\
  %%%%%
  %%%%%
  &&+\sigma^{2}_{\beta} \int_{t_{0}}^{h_{2}}\int_{t_{0}}^{h_{1}}f_{T_{2}}(u)f_{T_{1}}(s)e^{-2\mu (s+u)}\int^{t}_{t-(s+u)}G^{2}(I(\theta))(S(\theta)-S^{*}_{1})^{2}d\theta dsdu,\nonumber\\\label{ch1.sec3.lemma1.eq4b}
\end{eqnarray}
 where $\lambda(\mu)>0$ is a real valued function of $\mu$.
Suppose  $\tilde{\phi}_{1}$, $\tilde{\psi}_{1}$ and $\tilde{\varphi}_{1}$ are defined as follows
\begin{eqnarray}
\tilde{\phi}_{1}&=&3\mu-\left[2\mu\lambda{(\mu)}+(2\mu+d+\alpha)\frac{\lambda{(\mu)}}{2}+\alpha\lambda{(\mu)}+\frac{\beta S^{*}_{1}\lambda{(\mu)}}{2}+3\sigma ^{2}_{S}+\left(\frac{\beta (G^{*})^{2}}{2\lambda{(\mu)}}+\sigma^{2}_{\beta}(G^{*})^{2}\right)E(e^{-2\mu T_{1}})\right.\nonumber\\
&&\left.+\left(\frac{\beta \lambda{(\mu)}(G^{*})^{2}}{2}+\sigma^{2}_{\beta}(G^{*})^{2}\right)E(e^{-2\mu (T_{1}+T_{2})})\right]\label{ch1.sec3.lemma1.eq5a}\\
  \tilde{\psi}_{1} &=& 2\mu-\left[\frac{\beta }{2\lambda{(\mu)}}+\frac{\beta S^{*}_{1}\lambda{(\mu)}}{2}+ \frac{2\mu}{\lambda{(\mu)}}+(2\mu+d+\alpha)\frac{\lambda{(\mu)}}{2}+\alpha \lambda{(\mu)}+ 2 \sigma^{2}_{E} \right]\label{ch1.sec3.lemma1.eq5b}\\
    \tilde{\varphi}_{1}&=& (\mu + d+\alpha)-\left[(2\mu+d+\alpha)\frac{1}{\lambda{(\mu)}}+ \frac{\alpha\lambda{(\mu)}}{2} + \sigma^{2}_{I}+\frac{3\alpha}{2\lambda{(\mu)}}E(e^{-2\mu T_{3}})\right.\nonumber\\
    &&\left. +\left(\frac{\beta S^{*}_{1}(G'(I^{*}_{1}))^{2}}{\lambda{(\mu)}}
\right)E(e^{-2\mu (T_{1}+T_{2})})\right].\label{ch1.sec3.lemma1.eq5c}
   \end{eqnarray}%+\frac{\beta (G'(I^{*}_{1}))^{2}}{2\lambda{(\mu)}}
  The differential operator $dV$ applied to $V(t)$ with respect to the stochastic system (\ref{ch1.sec0.eq8})-(\ref{ch1.sec0.eq11}) can be written as follows:
 \begin{equation}\label{ch1.sec3.lemma1.eq5}
   dV=LV(t)dt + \overrightarrow{g}(S(t), E(t), I(t))d\overrightarrow{w(t)},
 \end{equation}
 where for $\overrightarrow{w(t)}=(w_{S},w_{E}, w_{I}, w_{\beta})^{T}$ and the  function $(S(t), E(t), I(t))\mapsto g(S(t), E(t), I(t))$, is defined as follows:
 \begin{eqnarray}
   &&\overrightarrow{g}(S(t), E(t), I(t))d\overrightarrow{w(t)}= -\sigma_{S}(3(S(t)-S^{*}_{1})+2(E(t)-E^{*}_{1})+I(t)-I^{*}_{1})S(t)dw_{S}(t)\nonumber\\
   &&-\sigma_{E}(2(S(t)-S^{*}_{1})+2(E(t)-E^{*}_{1})+I(t)-I^{*}_{1})E(t)dw_{E}(t)\nonumber\\
   &&-\sigma_{I}((S(t)-S^{*}_{1})+(E(t)-E^{*}_{1})+I(t)-I^{*}_{1})I(t)dw_{I}(t)\nonumber\\
   &&-\sigma_{\beta}(S(t)-S^{*}_{1})S(t)\int_{t_{0}}^{h_{1}}f_{T_{1}}(s)e^{-\mu s}G(I(t-s))dsdw_{\beta}(t)\nonumber\\
   &&-\sigma_{\beta}((S(t)-S^{*}_{1})+(E(t)-E^{*}_{1}))\int_{t_{0}}^{h_{2}}\int_{t_{0}}^{h_{1}}f_{T_{2}}(u)f_{T_{1}}(s)e^{-\mu (s+u)}S(t-u)G(I(t-s-u))dsdudw_{\beta}(t),\nonumber\\\label{ch1.sec3.lemma1.eq6}
 \end{eqnarray}
  and $LV$ satisfies the following inequality
  \begin{eqnarray}
    LV(t)&\leq& -\left\{\tilde{\phi}_{1}(S(t)-S^{*}_{1})^{2}+\tilde{\psi}_{1}(E(t)-E^{*}_{1})^{2}+\tilde{\varphi}_{1}(I(t)-I^{*}_{1})^{2}\right\}\nonumber\\
    &&+3\sigma^{2}_{S}(S^{*}_{1})^{2}+ 2\sigma^{2}_{E}(E^{*}_{1})^{2}+\sigma^{2}_{I}(I^{*}_{1})^{2}+\sigma^{2}_{\beta}(S^{*}_{1})^{2}(G^{*})^{2}E(e^{-2\mu (T_{1}+T_{2})})+\sigma^{2}_{\beta}(S^{*}_{1})^{2}(G^{*})^{2}E(e^{-\mu T_{1}}).\nonumber\\\label{ch1.sec3.lemma1.eq7}
  \end{eqnarray}
\end{lemma}
Proof\\
From (\ref{ch1.sec3.lemma1.eq2})-(\ref{ch1.sec3.lemma1.eq4}) the derivative of $V_{1}$, $V_{2}$ and $V_{3}$ with respect to the system (\ref{ch1.sec0.eq8})-(\ref{ch1.sec0.eq11}) can be written in the form:
\begin{eqnarray}
dV_{1}&=&LV_{1}dt -\sigma_{S}((S(t)-S^{*}_{1})+(E(t)-E^{*}_{1})+I(t)-I^{*}_{1})S(t)dw_{S}(t)\nonumber\\
   &&-\sigma_{E}((S(t)-S^{*}_{1})+(E(t)-E^{*}_{1})+I(t)-I^{*}_{1})E(t)dw_{E}(t)\nonumber\\
   &&-\sigma_{I}((S(t)-S^{*}_{1})+(E(t)-E^{*}_{1})+I(t)-I^{*}_{1})I(t)dw_{I}(t),\nonumber\\\label{ch1.sec3.lemma1.proof.eq1}
\end{eqnarray}
\begin{eqnarray}
dV_{2}&=&LV_{2}dt -\sigma_{S}((S(t)-S^{*}_{1}))S(t)dw_{S}(t)\nonumber\\
   &&-\sigma_{\beta}(S(t)-S^{*}_{1})S(t)\int_{t_{0}}^{h_{1}}f_{T_{1}}(s)e^{-\mu s}G(I(t-s))dsdw_{\beta}(t)\nonumber\\\label{ch1.sec3.lemma1.proof.eq2}
\end{eqnarray}
and
\begin{eqnarray}
dV_{3}&=&LV_{3}dt -\sigma_{S}((S(t)-S^{*}_{1})+(E(t)-E^{*}_{1}))S(t)dw_{S}(t)-\sigma_{S}((S(t)-S^{*}_{1})+(E(t)-E^{*}_{1}))E(t)dw_{E}(t)\nonumber\\
   &&-\sigma_{\beta}((S(t)-S^{*}_{1})+(E(t)-E^{*}_{1}))\int_{t_{0}}^{h_{2}}\int_{t_{0}}^{h_{1}}f_{T_{2}}(u)f_{T_{1}}(s)e^{-\mu (s+u)}S(t-u)G(I(t-s-u))dsdudw_{\beta}(t),\nonumber\\\label{ch1.sec3.lemma1.proof.eq3}
\end{eqnarray}
where utilizing (\ref{ch1.sec3.thm1.proof.eq1})-(\ref{ch1.sec3.thm1.proof.eq3}),  $LV_{1}$, $LV_{2}$ and $LV_{3}$ can be written as follows:
\begin{eqnarray}
LV_{1}(t)&=&-\mu (S(t)-S^{*}_{1})^{2}-\mu (E(t)-E^{*}_{1})^{2}-(\mu + d+ \alpha) (I(t)-I^{*}_{1})^{2}\nonumber\\
&&-2\mu (S(t)-S^{*}_{1})(E(t)-E^{*}_{1})-(2\mu + d+ \alpha) (S(t)-S^{*}_{1})(I(t)-I^{*}_{1})\nonumber\\
&&-(2\mu + d+ \alpha) (E(t)-E^{*}_{1})(I(t)-I^{*}_{1})\nonumber\\
&&+\alpha((S(t)-S^{*}_{1})+(E(t)-E^{*}_{1})+(I(t)-I^{*}_{1}))\int_{t_{0}}^{\infty}f_{T_{3}}(r)e^{-\mu r}(I(t-r)-I^{*}_{1})dr\nonumber\\
&&+\frac{1}{2}\sigma^{2}_{S}S^{2}(t)+\frac{1}{2}\sigma^{2}_{E}E^{2}(t)+\frac{1}{2}\sigma^{2}_{I}I^{2}(t),\label{ch1.sec3.lemma1.proof.eq4}
\end{eqnarray}
%%%
%%%%
%%%%
\begin{eqnarray}
LV_{2}(t)&=&-\mu (S(t)-S^{*}_{1})^{2}+\alpha(S(t)-S^{*}_{1})\int_{t_{0}}^{\infty}f_{T_{3}}(r)e^{-\mu r}(I(t-r)-I^{*}_{1})dr\nonumber\\
&&-\beta(S(t)-S^{*}_{1})^{2}\int_{t_{0}}^{h_{1}}f_{T_{1}}(s)e^{-\mu s}G(I(t-s))ds\nonumber\\
&&-\beta S^{*}_{1}(S(t)-S^{*}_{1})\int_{t_{0}}^{h_{1}}f_{T_{1}}(s)e^{-\mu s}(G(I(t-s))-G(I^{*}_{1}))ds\nonumber\\
&&+\frac{1}{2}\sigma^{2}_{S}S^{2}(t)+\frac{1}{2}\sigma^{2}_{\beta}S^{2}(t)\left(\int_{t_{0}}^{h_{1}}f_{T_{1}}(s)e^{-\mu s}G(I(t-s))ds\right)^{2},\label{ch1.sec3.lemma1.proof.eq5}
\end{eqnarray}
and
%%%%
%%%%
\begin{eqnarray}
LV_{3}(t)&=&-\mu (S(t)-S^{*}_{1})^{2}-\mu (E(t)-E^{*}_{1})^{2}-2\mu (S(t)-S^{*}_{1})(E(t)-E^{*}_{1})\nonumber\\
&&+\alpha((S(t)-S^{*}_{1})+(E(t)-E^{*}_{1}))\int_{t_{0}}^{\infty}f_{T_{3}}(r)e^{-\mu r}(I(t-r)-I^{*}_{1})dr\nonumber\\
&&-\beta(S(t)-S^{*}_{1})\int_{t_{0}}^{h_{2}}\int_{t_{0}}^{h_{1}}f_{T_{2}}(u)f_{T_{1}}(s)e^{-\mu (s+u)}(S(t-u)-S^{*}_{1})G(I(t-s-u))dsdu\nonumber\\
&&-\beta(E(t)-E^{*}_{1})\int_{t_{0}}^{h_{2}}\int_{t_{0}}^{h_{1}}f_{T_{2}}(u)f_{T_{1}}(s)e^{-\mu (s+u)}(S(t-u)-S^{*}_{1})G(I(t-s-u))dsdu\nonumber\\
&&-\beta S^{*}_{1}(S(t)-S^{*}_{1})\int_{t_{0}}^{h_{2}}\int_{t_{0}}^{h_{1}}f_{T_{2}}(u)f_{T_{1}}(s)e^{-\mu (s+u)}(G(I(t-s-u))-G(I^{*}_{1}))dsdu\nonumber\\
&&-\beta S^{*}_{1}(E(t)-E^{*}_{1})\int_{t_{0}}^{h_{2}}\int_{t_{0}}^{h_{1}}f_{T_{2}}(u)f_{T_{1}}(s)e^{-\mu (s+u)}(G(I(t-s-u))-G(I^{*}_{1}))dsdu\nonumber\\
%%%%%%%%%&&
&&+\frac{1}{2}\sigma^{2}_{S}S^{2}(t)+\frac{1}{2}\sigma^{2}_{E}E^{2}(t)\nonumber\\
%%%%%%%%%&&
&&+\frac{1}{2}\sigma^{2}_{\beta}\left(\int_{t_{0}}^{h_{2}}\int_{t_{0}}^{h_{1}}f_{T_{2}}(u)f_{T_{1}}(s)e^{-\mu (s+u)}S(t-u)G(I(t-s-u))dsdu\right)^{2}.\label{ch1.sec3.lemma1.proof.eq6}
\end{eqnarray}
%%%
%%%%
 From (\ref{ch1.sec3.lemma1.proof.eq4})-(\ref{ch1.sec3.lemma1.proof.eq6}), the set of inequalities that follow will be used to estimate the sum $LV_{1}(t)+LV_{2}(t)+LV_{3}(t)$:-
  Applying  $Cauchy-Swartz$ and  $H\ddot{o}lder$ inequalities,  and also applying the algebraic inequality (\ref{ch2.sec2.thm2.proof.eq2a}),
%\begin{equation}\label{ch2.sec2.thm2.proof.eq2}
%2ab\leq \frac{a^{2}}{g(c)}+b^{2}g(c)
%\end{equation}
%where $a,b,c\in \mathbb{R}$,  and the function $g$ is such that $g(c)> 0$
 the terms associated with the integral term (sign)    $\int_{t_{0}}^{\infty}f_{T_{3}}(r)e^{-\mu r}(I(t-r)-I^{*}_{1})dr$ are estimated as follows:
 \begin{equation}\label{ch1.sec3.lemma1.proof.eq7}
  (a(t)-a^{*})\int_{t_{0}}^{\infty}f_{T_{3}}(r)e^{-\mu r}(I(t-r)-I^{*}_{1})dr\leq  \frac{\lambda(\mu)}{2}(a-a^{*})^{2} + \frac{1}{2\lambda(\mu)}\int_{t_{0}}^{\infty}f_{T_{3}}(r)e^{-2\mu r}(I(t-r)-I^{*}_{1})^{2}dr,
 \end{equation}
 where $a(t)\in \{S(t), E(t), I(t)\}$ and $a^{*}\in \{S^{*}_{1}, E^{*}_{1}, I^{*}_{1}\}$. Furthermore,  the terms with the integral sign that depend on $G(I(t-s))$ and $G(I(t-s-u))$ are estimated as follows:
 \begin{eqnarray}
 &&-\beta(S(t)-S^{*}_{1})^{2}\int_{t_{0}}^{h_{1}}f_{T_{1}}(s)e^{-\mu s}G(I(t-s))ds \leq  \frac{\beta\lambda(\mu)}{2}(S(t)-S^{*}_{1})^{2}\nonumber\\
&&+\frac{\beta}{2\lambda(\mu)}(S(t)-S^{*}_{1})^{2}\int_{t_{0}}^{h_{1}}f_{T_{1}}(s)e^{-2\mu s}G^{2}(I(t-s))ds.\nonumber\\
&& -\beta(E(t)-E^{*}_{1})\int_{t_{0}}^{h_{2}}\int_{t_{0}}^{h_{1}}f_{T_{2}}(u)f_{T_{1}}(s)e^{-\mu (s+u)}(S(t-u)-S^{*}_{1})G(I(t-s-u))dsdu\leq  \frac{\beta}{2\lambda(\mu)}(E(t)-E^{*}_{1})^{2} \nonumber\\
&&+\frac{\beta\lambda(\mu)}{2}\int_{t_{0}}^{h_{2}}\int_{t_{0}}^{h_{1}}f_{T_{2}}(u)f_{T_{1}}(s)e^{-2\mu (s+u)}(S(t-u)-S^{*}_{1})^{2}G^{2}(I(t-s-u))dsdu. \label{ch1.sec3.lemma1.proof.eq8}
 \end{eqnarray}
%%%%
The terms with the integral sign that depend on $G(I(t-s))-G(I^{*}_{1})$ and $G(I(t-s-u))-G(I^{*}_{1})$ are estimated as follows:
\begin{eqnarray}
&&-\beta S^{*}_{1}(S(t)-S^{*}_{1})\int_{t_{0}}^{h_{1}}f_{T_{1}}(s)e^{-\mu s}(G(I(t-s))-G(I^{*}_{1}))ds\leq \frac{\beta S^{*}_{1}\lambda(\mu)}{2}(S(t)-S^{*}_{1})^{2}\nonumber\\
&& +\frac{\beta S^{*}_{1}}{2\lambda(\mu)}\int_{t_{0}}^{h_{1}}f_{T_{1}}(s)e^{-2\mu s}(I(t-s)-I^{*}_{1})^{2}\left(\frac{G(I(t-s))-G(I^{*}_{1})}{I(t-s)-I^{*}_{1}}\right)^{2}ds\nonumber\\
&&\leq \frac{\beta S^{*}_{1}\lambda(\mu)}{2}(S(t)-S^{*}_{1})^{2}\nonumber\\
&& +\frac{\beta S^{*}_{1}}{2\lambda(\mu)}\int_{t_{0}}^{h_{1}}f_{T_{1}}(s)e^{-2\mu s}(I(t-s)-I^{*}_{1})^{2}\left(G'(I^{*}_{1})\right)^{2}ds.\nonumber\\
&&-\beta S^{*}_{1}(E(t)-E^{*}_{1})\int_{t_{0}}^{h_{2}}\int_{t_{0}}^{h_{1}}f_{T_{2}}(u)f_{T_{1}}(s)e^{-\mu (s+u)}(G(I(t-s-u))-G(I^{*}_{1}))dsdu\leq \frac{\beta S^{*}_{1}\lambda(\mu)}{2}(E(t)-E^{*}_{1})^{2}\nonumber\\
&& +\frac{\beta S^{*}_{1}}{2\lambda(\mu)}\int_{t_{0}}^{h_{2}}\int_{t_{0}}^{h_{1}}f_{T_{2}}(u)f_{T_{1}}(s)e^{-2\mu s}(I(t-s-u)-I^{*}_{1})^{2}\left(\frac{G(I(t-s-u))-G(I^{*}_{1})}{I(t-s-u)-I^{*}_{1}}\right)^{2}ds\nonumber\\
&&\leq \frac{\beta S^{*}_{1}\lambda(\mu)}{2}(E(t)-E^{*}_{1})^{2}\nonumber\\
&& +\frac{\beta S^{*}_{1}}{2\lambda(\mu)}\int_{t_{0}}^{h_{2}}\int_{t_{0}}^{h_{1}}f_{T_{2}}(u)f_{T_{1}}(s)e^{-2\mu s}(I(t-s-u)-I^{*}_{1})^{2}\left(G'(I^{*}_{1})\right)^{2}ds,\nonumber\\\label{ch1.sec3.lemma1.proof.eq9}
\end{eqnarray}
where the inequality in (\ref{ch1.sec3.lemma1.proof.eq9}) follows from Assumption~\ref{ch1.sec0.assum1}. That is, $G$ is a differentiable monotonic function with $G''(I)<0$, and consequently,  $0< \frac{G(I)-G(I^{*}_{1})}{(I-I^{*}_{1})}\leq G'(I^{*}_{1}), \forall I>0$.
%%%

By also employing the  $Cauchy-Swartz$ and $H\ddot{o}lder$ inequalities, and also applying the following algebraic inequality $(a+b)^{2}\leq 2a^{2}+ 2b^{2}$, the last set of terms with integral signs on (\ref{ch1.sec3.lemma1.proof.eq5})-(\ref{ch1.sec3.lemma1.proof.eq6}) are estimated as follows:
\begin{eqnarray}
&&\frac{1}{2}\sigma^{2}_{\beta}S^{2}(t)\left(\int_{t_{0}}^{h_{1}}f_{T_{1}}(s)e^{-\mu s}G(I(t-s))ds\right)^{2}\leq \sigma^{2}_{\beta}\left((S(t)-S^{*}_{1})^{2}+(S^{*}_{1})^{2}\right)\times\nonumber\\
&&\times\int_{t_{0}}^{h_{1}}f_{T_{1}}(s)e^{-2\mu s}G^{2}(I(t-s))ds.\nonumber\\
&&\frac{1}{2}\sigma^{2}_{\beta}\left(\int_{t_{0}}^{h_{2}}\int_{t_{0}}^{h_{1}}f_{T_{2}}(u)f_{T_{1}}(s)e^{-\mu (s+u)}S(t-u)G(I(t-s-u))dsdu\right)^{2}\leq \sigma^{2}_{\beta}\times\nonumber\\
&&\times\int_{t_{0}}^{h_{2}}\int_{t_{0}}^{h_{1}}f_{T_{2}}(u)f_{T_{1}}(s)e^{-2\mu (s+u)}\left((S(t-u)-S^{*}_{1})^{2}+(S^{*}_{1})^{2}\right)G^{2}(I(t-s-u))dsdu.\nonumber\\\label{ch1.sec3.lemma1.proof.eq10}
\end{eqnarray}
By further applying the algebraic inequality (\ref{ch2.sec2.thm2.proof.eq2a}) and the inequalities (\ref{ch1.sec3.lemma1.proof.eq7})-(\ref{ch1.sec3.lemma1.proof.eq10}) on the sum $LV_{1}(t)+LV_{2}(t)+LV_{3}(t)$, it is easy to see from (\ref{ch1.sec3.lemma1.proof.eq4})-(\ref{ch1.sec3.lemma1.proof.eq6}) that
\begin{eqnarray}
&&LV_{1}(t)+LV_{2}(t)+LV_{3}(t)\leq (S(t)-S^{*}_{1})^{2}\left[-3\mu +2\mu \lambda(\mu) +(2\mu+ d+\alpha)\frac{\lambda(\mu)}{2}+\alpha\lambda(\mu)\right.\nonumber\\
&&\left.+ \frac{\beta\lambda(\mu)}{2}+\frac{\beta S^{*}_{1}\lambda(\mu)}{2}+ \frac{\beta}{2\lambda(\mu)}(G^{*})^{2}E(e^{-2\mu T_{1}})+ 3\sigma^{2}_{S}+\sigma^{2}_{S} (G^{*})^{2}E(e^{-2\mu T_{1}})\right]\nonumber\\
&&(E(t)-E^{*}_{1})^{2}\left[-2\mu +\frac{2\mu }{\lambda(\mu)} +(2\mu+ d+\alpha)\frac{\lambda(\mu)}{2}+\alpha\lambda(\mu)+ \frac{\beta}{2\lambda(\mu)}+\frac{\beta S^{*}_{1}\lambda(\mu)}{2}+ 2\sigma^{2}_{E}\right]\nonumber\\
%%%%%\right.\nonumber\\&&\left.
&&+(I(t)-I^{*}_{1})^{2}\left[-(\mu+d+\alpha)  +(2\mu+ d+\alpha)\frac{1}{\lambda(\mu)}+\frac{\alpha\lambda(\mu)}{2}+ \frac{\beta}{2\lambda(\mu)}+\frac{\beta S^{*}_{1}\lambda(\mu)}{2}+ \sigma^{2}_{I}\right]\nonumber\\
%\right.\nonumber\\
%&&\left.
&&+\frac{3\alpha}{2\lambda(\mu)}\int_{t_{0}}^{\infty}f_{T_{3}}(r)e^{-2\mu r}(I(t-r)-I^{*}_{1})^{2}dr\nonumber\\
&&+\frac{\beta S^{*}_{1}}{\lambda(\mu)}(G'(I^{*}_{1}))^{2}\int_{t_{0}}^{h_{2}}\int_{t_{0}}^{h_{1}}f_{T_{2}}(u)f_{T_{1}}(s)e^{-2\mu (s+u)}(I(t-s-u)-I^{*}_{1})^{2}dsdu\nonumber\\
&&+\frac{\beta \lambda(\mu)}{2}\int_{t_{0}}^{h_{2}}\int_{t_{0}}^{h_{1}}f_{T_{2}}(u)f_{T_{1}}(s)e^{-2\mu (s+u)}G^{2}(I(t-s-u))(S(t-s)-S^{*}_{1})^{2}dsdu\nonumber\\
&&+3\sigma^{2}_{S}(S^{*}_{1})^{2}+2\sigma^{2}_{E}(E^{*}_{1})^{2}+ \sigma^{2}_{I}(I^{*}_{1})^{2}\nonumber\\
&&+\sigma^{2}_{\beta}(S^{*}_{1})^{2}\int_{t_{0}}^{h_{1}}f_{T_{1}}(r)e^{-2\mu }f_{T_{1}}(s)e^{-2\mu s}G^{2}(I(t-s))ds\nonumber\\
&&+\sigma^{2}_{\beta}\int_{t_{0}}^{h_{2}}\int_{t_{0}}^{h_{1}}f_{T_{2}}(u)f_{T_{1}}(s)e^{-2\mu (s+u)}G^{2}(I(t-s-u))(S(t-s-u)-S^{*}_{1})^{2}dsdu\nonumber\\\label{ch1.sec3.lemma1.proof.eq11}
\end{eqnarray}
But $V(t)=V_{1}(t)+V_{2}(t)+V_{3}(t)+V_{4}(t)$, therefore from (\ref{ch1.sec3.lemma1.proof.eq11}),  (\ref{ch1.sec3.lemma1.eq4b}) and (\ref{ch1.sec3.lemma1.proof.eq4})-(\ref{ch1.sec3.lemma1.proof.eq6}), the results in (\ref{ch1.sec3.lemma1.eq5})-(\ref{ch1.sec3.lemma1.eq7}) follow directly.
%%%%%
%%%%%%

Theorem~\ref{ch1.sec3.thm1} asserts that the deterministic system (\ref{ch1.sec0.eq3})-(\ref{ch1.sec0.eq6}) has an endemic equilibrium denoted $E_{1}=(S^{*}_{1}, E^{*}_{1}, I^{*}_{1})$. To obtain insight about the endemic asymptotic properties of the disease dynamics in the absence of any random fluctuations in the system, it is necessary to study the global stability of the endemic equilibrium of the deterministic system.
For convenience, the following notations are introduced and used in the rest of the results that follow in this section.  Let $a_{1}(\mu, d, \alpha, \beta, B, \sigma ^{2}_{S}, \sigma^{2}_{\beta})$, $a_{2}(\mu, d, \alpha, \beta, B, \sigma ^{2}_{I})$,  $a_{3}(\mu, d, \alpha, \beta, B)$, and $a_{3}(\mu, d, \alpha, \beta, B, \sigma ^{2}_{E})$ represent the following set of parameters
\begin{eqnarray}
a_{1}(\mu, d, \alpha, \beta, B, \sigma ^{2}_{S}, \sigma^{2}_{\beta})&=&2\mu\lambda{(\mu)}+(2\mu+d+\alpha)\frac{\lambda{(\mu)}}{2}+\alpha\lambda{(\mu)}+\frac{\beta S^{*}_{1}\lambda{(\mu)}}{2}+3\sigma ^{2}_{S}\nonumber\\
&&+\left(\frac{\beta \lambda{(\mu)}(G^{*})^{2}}{2}+\sigma^{2}_{\beta}(G^{*})^{2}\right)\left(\frac{1}{G'(0)}\right)^{2}\label{ch1.sec3.lemma1.proof.eq13a}\\
a_{1}(\mu, d, \alpha, \beta, B)&=&2\mu\lambda{(\mu)}+(2\mu+d+\alpha)\frac{\lambda{(\mu)}}{2}+\alpha\lambda{(\mu)}+\frac{\beta S^{*}_{1}\lambda{(\mu)}}{2}\nonumber\\
&&+\left(\frac{\beta \lambda{(\mu)}(G^{*})^{2}}{2}\right)\left(\frac{1}{G'(0)}\right)^{2}\label{ch1.sec3.lemma1.proof.eq13a1}\\
%\hat{K}_{0}+\frac{\alpha}{\beta \frac{B}{\mu}}
%%%%%%%%%%%%%%%%%%%%%%%%%%%%%%%%%%%%%%%%%%%%%
a_{2}(\mu, d, \alpha, \beta, B, \sigma ^{2}_{I})&=&(2\mu+d+\alpha)\frac{1}{\lambda{(\mu)}}+ \frac{\alpha\lambda{(\mu)}}{2} + \sigma^{2}_{I}\nonumber\\
&&+\left(\frac{\beta S^{*}_{1}(G'(I^{*}_{1}))^{2}}{\lambda{(\mu)}}
\right)\left(\frac{1}{G'(0)}\right)^{2}\label{ch1.sec3.lemma1.proof.eq13b}\\
a_{2}(\mu, d, \alpha, \beta, B)&=&(2\mu+d+\alpha)\frac{1}{\lambda{(\mu)}}+ \frac{\alpha\lambda{(\mu)}}{2} \nonumber\\
&&+\left(\frac{\beta S^{*}_{1}(G'(I^{*}_{1}))^{2}}{\lambda{(\mu)}}
\right)\left(\frac{1}{G'(0)}\right)^{2}\label{ch1.sec3.lemma1.proof.eq13b1}\\
%\hat{K}_{0}+\frac{\alpha}{\beta \frac{B}{\mu}}%
%%%%%%%%%%%%%%%%%%%%%%%%%%%%%
a_{3}(\mu, d, \alpha, \beta, B, \sigma ^{2}_{E})&=&\frac{\beta }{2\lambda{(\mu)}}+\frac{\beta S^{*}_{1}\lambda{(\mu)}}{2}+ \frac{2\mu}{\lambda{(\mu)}}+(2\mu+d+\alpha)\frac{\lambda{(\mu)}}{2}+\alpha \lambda{(\mu)}+ 2 \sigma^{2}_{E}\nonumber\\\label{ch1.sec3.lemma1.proof.eq13c}
a_{3}(\mu, d, \alpha, \beta, B)&=&\frac{\beta }{2\lambda{(\mu)}}+\frac{\beta S^{*}_{1}\lambda{(\mu)}}{2}+ \frac{2\mu}{\lambda{(\mu)}}+(2\mu+d+\alpha)\frac{\lambda{(\mu)}}{2}+\alpha \lambda{(\mu)}\nonumber\\\label{ch1.sec3.lemma1.proof.eq13c1}
\end{eqnarray}

Also let $\tilde{a}_{1}(\mu, d, \alpha, \beta, B, \sigma ^{2}_{S}, \sigma^{2}_{\beta})$, $\tilde{a}_{1}(\mu, d, \alpha, \beta, B)$, $\tilde{a}_{2}(\mu, d, \alpha, \beta, B, \sigma ^{2}_{I})$, $\tilde{a}_{2}(\mu, d, \alpha, \beta, B)$  represent the following set of parameters
\begin{eqnarray}
\tilde{a}_{1}(\mu, d, \alpha, \beta, B,\sigma ^{2}_{S}, \sigma^{2}_{\beta})&=&2\mu\lambda{(\mu)}+(2\mu+d+\alpha)\frac{\lambda{(\mu)}}{2}+\alpha\lambda{(\mu)}+\frac{\beta S^{*}_{1}\lambda{(\mu)}}{2}+3\sigma ^{2}_{S}\nonumber\\
&&+\left(\frac{\beta \lambda{(\mu)}(G^{*})^{2}}{2}+\sigma^{2}_{\beta}(G^{*})^{2}\right)+\left(\frac{\beta (G^{*})^{2}}{2\lambda{(\mu)}}+\sigma^{2}_{\beta}(G^{*})^{2}\right)\label{ch1.sec3.thm2.proof.eq4a}\\
\tilde{a}_{1}(\mu, d, \alpha, \beta, B)&=&2\mu\lambda{(\mu)}+(2\mu+d+\alpha)\frac{\lambda{(\mu)}}{2}+\alpha\lambda{(\mu)}+\frac{\beta S^{*}_{1}\lambda{(\mu)}}{2}\nonumber\\
&&+\left(\frac{\beta \lambda{(\mu)}(G^{*})^{2}}{2}\right)+\left(\frac{\beta (G^{*})^{2}}{2\lambda{(\mu)}}\right)\label{ch1.sec3.thm2.proof.eq4a1}\\
%%%%%%%%%%%%%%%%%%%%%%%%%%%%%
%%%%%%%%%%%%%%%%%%%%%%%%%%%%
\tilde{a}_{2}(\mu, d, \alpha, \beta, B, \sigma^{2}_{I})&=&(2\mu+d+\alpha)\frac{1}{\lambda{(\mu)}}+ \frac{\alpha\lambda{(\mu)}}{2} + \sigma^{2}_{I}\nonumber\\
&&+\left(\frac{\beta S^{*}_{1}(G'(I^{*}_{1}))^{2}}{\lambda{(\mu)}}
\right)+\frac{3\alpha}{2\lambda(\mu)},\label{ch1.sec3.thm2.proof.eq4b}\\
\tilde{a}_{2}(\mu, d, \alpha, \beta, B)&=&(2\mu+d+\alpha)\frac{1}{\lambda{(\mu)}}+ \frac{\alpha\lambda{(\mu)}}{2} +\left(\frac{\beta S^{*}_{1}(G'(I^{*}_{1}))^{2}}{\lambda{(\mu)}}
\right)+\frac{3\alpha}{2\lambda(\mu)}.\label{ch1.sec3.thm2.proof.eq4b1}
%a_{3}(\mu, d, \alpha, \beta, B)&=&\frac{\beta }{2\lambda{(\mu)}}+\frac{\beta S^{*}_{1}\lambda{(\mu)}}{2}+ \frac{2\mu}{\lambda{(\mu)}}+(2\mu+d+\alpha)\frac{\lambda{(\mu)}}{2}+\alpha %\lambda{(\mu)}+ 2 \sigma^{2}_{E}\nonumber\\\label{ch1.sec3.lemma1.proof.eq13c}
\end{eqnarray}
%and $a_{3}(\mu, d, \alpha, \beta, B)$  is defined in (\ref{ch1.sec3.lemma1.proof.eq13c}).
%%%%%%%%%%%%%%%%%%%%%%%%%%%%%%%%%%%%%%%%%%%%%%%%%%%%%%%%%%%%%%%%%%%%%%%%%%%%%%%%%%%%%%%%%%%%%%%%%%reformulated later on called theorem3a
\begin{thm}\label{ch2.sec3.thm3a}
Let the hypotheses of  Theorem~\ref{ch1.sec3.thm1}  and  Lemma~\ref{ch1.sec3.lemma1} be satisfied and let
 \begin{equation}\label{ch1.sec3.thm3a.eq1}
 \mu> \max\left\{\frac{1}{3}\tilde{a}_{1}(\mu, d, \alpha, \beta, B),\frac{1}{2}a_{3}(\mu, d, \alpha, \beta, B)\right\}\quad and\quad(\mu+d+\alpha)>\tilde{a}_{2}(\mu, d, \alpha, \beta, B).
 \end{equation}
  For any probability distribution of the delay times $T_{1}, T_{2}$ and $T_{3}$, the following are true:-
  \item[1.]
There exists a positive real number $\mathfrak{m}_{2}>0$, such that the differential operator $\dot{V}$ applied to the Lyapunov functional $V$ defined in (\ref{ch1.sec3.lemma1.eq1}) with respect to the deterministic system (\ref{ch1.sec0.eq3})-(\ref{ch1.sec0.eq6}), satisfies the following inequality
\begin{eqnarray}
  \dot{V}(t)&\leq &-\mathfrak{m}_{2}||X(t)-E_{1}||^{2},\nonumber\\%\mathfrak{m}_{2}\left[ (S(v)-S^{*}_{1})^{2}+ (E(v)-E^{*}_{1})^{2}+  (I(v)-I^{*}_{1})^{2}\right]=
    \label{ch2.sec3.thm3a.eq2}
\end{eqnarray}%\l
where $X(t)$ is defined in  (\ref{ch1.sec0.eq13b}) and $||.||$ is the natural norm defined on $\mathbb{R}^{2}$.
\item[2.] The endemic equilibrium  $E_{1}$ is globally uniformly and asymptotically stable.
\end{thm}
%%%%
Proof:\\
From Lemma~\ref{ch1.sec3.lemma1},  (\ref{ch1.sec3.lemma1.eq5a})-(\ref{ch1.sec3.lemma1.eq5c}), and  letting the intensity values $\sigma_{i}=0, i=S, E, I, \beta$, it follows that
 \begin{eqnarray}
-\tilde{\phi}_{1}&=&-3\mu+\left[2\mu\lambda{(\mu)}+(2\mu+d+\alpha)\frac{\lambda{(\mu)}}{2}+\alpha\lambda{(\mu)}+\frac{\beta S^{*}_{1}\lambda{(\mu)}}{2}+\left(\frac{\beta (G^{*})^{2}}{2\lambda{(\mu)}}\right)E(e^{-2\mu T_{1}})\right.\nonumber\\
&&\left.+\left(\frac{\beta \lambda{(\mu)}(G^{*})^{2}}{2}\right)E(e^{-2\mu (T_{1}+T_{2})})\right]\nonumber\\
&&\leq -3\mu+\left[2\mu\lambda{(\mu)}+(2\mu+d+\alpha)\frac{\lambda{(\mu)}}{2}+\alpha\lambda{(\mu)}+\frac{\beta S^{*}_{1}\lambda{(\mu)}}{2}+\left(\frac{\beta (G^{*})^{2}}{2\lambda{(\mu)}}\right)\right.\nonumber\\
&&\left.+\left(\frac{\beta \lambda{(\mu)}(G^{*})^{2}}{2}\right)\right]\nonumber\\
&&=-\left(3\mu-\tilde{a}_{1}(\mu, d, \alpha, \beta, B)\right)\label{ch1.sec3.thm3a.proof.eq1}\\
%%%%%%
  -\tilde{\psi}_{1} &=& -2\mu+\left[\frac{\beta }{2\lambda{(\mu)}}+\frac{\beta S^{*}_{1}\lambda{(\mu)}}{2}+ \frac{2\mu}{\lambda{(\mu)}}+(2\mu+d+\alpha)\frac{\lambda{(\mu)}}{2}+\alpha \lambda{(\mu)} \right]\nonumber\\
  &&=-\left(2\mu-a_{3}(\mu, d, \alpha, \beta, B)\right)\label{ch1.sec3.thm3a.proof.eq2}\\
    -\tilde{\varphi}_{1}&=& -(\mu + d+\alpha)+\left[(2\mu+d+\alpha)\frac{1}{\lambda{(\mu)}}+ \frac{\alpha\lambda{(\mu)}}{2} +\frac{3\alpha}{2\lambda{(\mu)}}E(e^{-2\mu T_{3}})\right.\nonumber\\
    &&\left. +\left(\frac{\beta S^{*}_{1}(G'(I^{*}_{1}))^{2}}{\lambda{(\mu)}}
\right)E(e^{-2\mu (T_{1}+T_{2})})\right]\nonumber\\
&\leq& -(\mu + d+\alpha)+\left[(2\mu+d+\alpha)\frac{1}{\lambda{(\mu)}}+ \frac{\alpha\lambda{(\mu)}}{2} +\frac{3\alpha}{2\lambda{(\mu)}}\right.\nonumber\\
    &&\left. +\left(\frac{\beta S^{*}_{1}(G'(I^{*}_{1}))^{2}}{\lambda{(\mu)}}
\right)\right]\nonumber\\
&&=-\left((\mu + d+\alpha)-\tilde{a}_{2}(\mu, d, \alpha, \beta, B)\right),\label{ch1.sec3.thm3a.proof.eq3}
   \end{eqnarray}
since $0<E(e^{-2\mu (T_{i})})\leq 1, i=1, 2,3$. It follows from (\ref{ch1.sec3.lemma1.eq7}) and (\ref{ch1.sec3.thm3a.proof.eq1})-(\ref{ch1.sec3.thm3a.proof.eq3}) that
\begin{eqnarray}
   LV&\leq &-\min\{\left(3\mu-\tilde{a}_{1}(\mu, d, \alpha, \beta, B)\right), \left(2\mu-a_{3}(\mu, d, \alpha, \beta, B)\right), \left((\mu + d+\alpha)-\tilde{a}_{2}(\mu, d, \alpha, \beta, B)\right)\}\times\nonumber\\
   &&\times \left[ (S(t)-S^{*}_{1})^{2}+ (E(t)-E^{*}_{1})^{2}+  (I(t)-I^{*}_{1})^{2}\right].\label{ch1.sec3.thm3a.proof.eq4}
 \end{eqnarray}
 Define
\begin{equation}\label{ch2.sec3.thm3a.proof.eq5}
\mathfrak{m}_{2}=\min\{\left(3\mu-\tilde{a}_{1}(\mu, d, \alpha, \beta, B)\right), \left(2\mu-a_{3}(\mu, d, \alpha, \beta, B)\right), \left((\mu + d+\alpha)-\tilde{a}_{2}(\mu, d, \alpha, \beta, B)\right)\}.
\end{equation}
 Then under the assumption (\ref{ch1.sec3.thm3a.eq1}) in the hypothesis, $\mathfrak{m}_{2}>0$. Thus, the result in  (\ref{ch2.sec3.thm3a.eq2}) follows directly from (\ref{ch1.sec3.thm3a.proof.eq4}). Moreover, the global uniform asymptotic stability of the steady state $E_{1}$ follows easily by applying the comparison stability results in \cite{ladde,wanduku-determ}.
 %%%%%

 %%%%%%%%%%%%%%%%%%%%%%%%%%%
 The following result presence a clearer picture of the influence of the delays in the system on the stability of the endemic equilibrium $E_{1}$, whenever the system is unperturbed by the random fluctuations in the disease dynamics.
%%%%%%%%%%%%%%%%%%%%%edit
%%%%
\begin{thm}\label{ch2.sec3.thm3a.corrolary1}
Let the hypotheses of Theorem~\ref{ch1.sec3.thm1.corrolary1} and  Lemma~\ref{ch1.sec3.lemma1} be satisfied and let
 \begin{equation}\label{ch2.sec3.thm3a.corrolary1.eq1}
 \mu>\max{ \left(\frac{1}{3}a_{1}(\mu, d, \alpha, \beta, B),\frac{1}{2}a_{3}(\mu, d, \alpha, \beta, B)\right)} ,\quad and\quad(\mu+d+\alpha)>a_{2}(\mu, d, \alpha, \beta, B ).%\quad and\quad 2\mu> a_{3}(\mu, d, \alpha, \beta, B).
 \end{equation}
 Also let the delay times $T_{1}, T_{2}$ and $T_{3}$ be constant, that is, the probability density functions of $T_{1}, T_{2}$ and $T_{3}$ respectively denoted by $f_{T_{i}}, i=1, 2, 3$ are the  dirac-delta functions defined in (\ref{ch1.sec2.eq4}). Furthermore, let the constants $T_{1}, T_{2}$ and $T_{3}$ satisfy the following set of inequalities:
\begin{equation}\label{ch2.sec3.thm3a.corrolary1.eq2}
T_{1}>\frac{1}{2\mu}\log{\left(\frac{\left(\frac{\beta (G^{*})^{2}}{2\lambda{(\mu)}}\right)}{(3\mu-a_{1}(\mu, d, \alpha, \beta, B))}\right)},
\end{equation}
\begin{equation}\label{ch2.sec3.thm3a.corrolary1.eq3}
T_{2}<\frac{1}{2\mu}\log{\left(\frac{(3\mu-a_{1}(\mu, d, \alpha, \beta, B))}{\left(\frac{\beta (G^{*})^{2}}{2\lambda{(\mu)}}\right)\left(\frac{1}{G'(0)}\right)^{2}}\right)},
\end{equation}%\frac{\hat{K}_{0}+\frac{\alpha}{\beta \frac{B}{\mu}}
and
\begin{equation}\label{ch2.sec3.thm3a.corrolary1.eq4}
T_{3}>\frac{1}{2\mu}\log{\left(\frac{\frac{3\alpha}{2\lambda(\mu)}}{(\mu+d+\alpha)-a_{2}(\mu, d, \alpha, \beta, B)}\right)}.
\end{equation}
There exists a positive real number $\mathfrak{m}_{1}>0$, such that the differential operator $\dot{V}$ applied to the Lyapunov functional $V$ defined in (\ref{ch1.sec3.lemma1.eq1}) with respect to the deterministic system (\ref{ch1.sec0.eq3})-(\ref{ch1.sec0.eq6}), satisfies the following inequality
\begin{eqnarray}
  \dot{V}(t)&\leq &-\mathfrak{m}_{1}||X(t)-E_{1}||^{2},\nonumber\\%\mathfrak{m}_{2}\left[ (S(v)-S^{*}_{1})^{2}+ (E(v)-E^{*}_{1})^{2}+  (I(v)-I^{*}_{1})^{2}\right]=
    \label{ch2.sec3.thm3a.corrolary1.eq5}
\end{eqnarray}%\l
where $X(t)$ is defined in  (\ref{ch1.sec0.eq13b}) and $||.||$ is the natural norm defined on $\mathbb{R}^{2}$.

Furthermore, the endemic equilibrium  $E_{1}$ is globally uniformly and asymptotically stable. Moreover, it is exponentially stable.
\end{thm}
Proof:\\
From Lemma~\ref{ch1.sec3.lemma1}  and (\ref{ch1.sec3.lemma1.eq5a})-(\ref{ch1.sec3.lemma1.eq5c}), by setting $\sigma_{i}=0, i=S, E, I, R, \beta$, it is easy to see that under the assumptions in (\ref{ch1.sec3.thm1.corrolary1.eq1}), and (\ref{ch2.sec3.thm3a.corrolary1.eq1})-(\ref{ch2.sec3.thm3a.corrolary1.eq4}), then $\tilde{\phi}_{1}>0$, $\tilde{\psi}_{1}>0$ and $\tilde{\varphi}_{1}>0$. Therefore,  from (\ref{ch1.sec3.lemma1.eq5})-(\ref{ch1.sec3.lemma1.eq7}) it is also easy to see that
\begin{eqnarray}
   \dot{V}(t)&\leq&-\min{(\tilde{\phi}_{1}, \tilde{\psi}_{1}, \tilde{\varphi}_{1})}\left\{(S(t)-S^{*}_{1})^{2}+(E(t)-E^{*}_{1})^{2}+(I(t)-I^{*}_{1})^{2}\right\}.\label{ch2.sec3.thm3a.corrolary1.proof.eq1}
 \end{eqnarray}
 Letting $\mathfrak{m}_{1}=\min{(\tilde{\phi}_{1}, \tilde{\psi}_{1}, \tilde{\varphi}_{1})}$, the result in (\ref{ch2.sec3.thm3a.corrolary1.eq5}) follows immediately.

Moreover, the global uniform asymptotic stability of the steady state $E_{1}$ follows easily by applying the comparison stability results in \cite{divine-proceeding1,divine-proceeding2,wanduku-determ}.
 %%%%
\begin{rem}\label{ch1.sec3.rem2a}
  Theorem~\ref{ch2.sec3.thm3a} presents minimum sufficient conditions independent of the probability distribution of the delay time random variables $T_{1}, T_{2}$ and $T_{3}$ that lead to the global uniform stability of the endemic equilibrium $E_{1}$. That is, when the disease dynamics is unperturbed by random fluctuations in the disease, then all solutions of the system (\ref{ch1.sec0.eq3})-(\ref{ch1.sec0.eq6}) that start  near the nontrivial steady state $E_{1}$ stay in the neighborhood of the steady state, and converge asymptotically at $E_{1}$, whenever the conditions in hypothesis of Theorem~\ref{ch2.sec3.thm3} are satisfied. In other words, the disease is persistent near the nontrivial steady state, whenever the conditions in the hypothesis of Theorem~\ref{ch2.sec3.thm3} are satisfied.

     A clearer picture of the impacts of the delays in the system $T_{i}, i=1,2,3$ on the stability of the endemic equilibrium is presented in Theorem~\ref{ch2.sec3.thm3a.corrolary1}, where it is assumed that the incubation delay period of malaria inside the vector and human hosts, and also the period of effective natural immunity are all constant for all individuals in the population. The conditions in (\ref{ch2.sec3.thm3a.corrolary1.eq1}) and (\ref{ch2.sec3.thm3a.corrolary1.eq2})-(\ref{ch2.sec3.thm3a.corrolary1.eq4}) are sufficient for the disease to  be persistent in the human population.  The threshold bounds for the delay times in (\ref{ch2.sec3.thm3a.corrolary1.eq2})-(\ref{ch2.sec3.thm3a.corrolary1.eq4}) are all fractions of the average lifespan $\frac{1}{\mu}$ of individuals in the population. Moreover, the threshold bounds depend on the parameters of the system and are computationally attractive, whenever a specific set of parameter values are given. Therefore, in a disease scenario where the incubation delays and natural immunity delay period are constant, an estimate for the values  $T_{1}, T_{2}$ and $T_{3}$ can be determine which would lead to malaria persistence in the population.
 \end{rem}%%%%
%%%%%%%%
%%%%%%%
The smooth behavior of the disease dynamics near the endemic equilibrium $E_{1}$ depicted in Theorem~\ref{ch2.sec3.thm3a} and Theorem~\ref{ch2.sec3.thm3a.corrolary1} is complicated by the presence of noise due to random fluctuations in the disease dynamics as it is shown by the subsequent results. The following theorem characterizes the behavior of the stochastic system (\ref{ch1.sec0.eq8})-(\ref{ch1.sec0.eq11}) in the neighborhood of the nontrivial steady states $E_{1}=(S^{*}_{1}, E^{*}_{1}, I^{*}_{1})$ defined in Theorem~\ref{ch1.sec3.thm1} and Theorem~\ref{ch1.sec3.thm1.corrolary1}, whenever the incubation and natural immunity delay times of the disease denoted by $T_{1}$,  $T_{2}$, and  $T_{3}$ are constant for all individuals in the population. The assumption that $T_{1}, T_{2}$ and $T_{3}$ are constant, is also equivalent to the special case of letting the probability density functions of $T_{1}, T_{2}$ and $T_{3}$ to be the dirac-delta function defined in (\ref{ch1.sec2.eq4}).
Moreover, under the assumption that $T_{1}\geq 0, T_{2}\geq 0$ and $T_{3}\geq 0$ are constant, it follows  from (\ref{ch1.sec3.lemma1.eq5a})-(\ref{ch1.sec3.lemma1.eq5c}),  that  $E(e^{-2\mu (T_{1}+T_{2})})=e^{-2\mu (T_{1}+T_{2})} $, $E(e^{-2\mu T_{1}})=e^{-2\mu T_{1}} $ and $E(e^{-2\mu T_{3}})=e^{-2\mu T_{3}} $.
%%%%%%%%%%%%%%%%%%%%%%%%%%%%%
%%%%%%
%%%%%
%%%%
%%%%
\begin{thm}\label{ch1.sec3.thm2}
Let the hypotheses of Theorem~\ref{ch1.sec3.thm1.corrolary1} and  Lemma~\ref{ch1.sec3.lemma1} be satisfied and let
 \begin{equation}\label{ch1.sec3.thm2.eq1}
 \mu>\max{ \left(\frac{1}{3}a_{1}(\mu, d, \alpha, \beta, B, \sigma^{2}_{S},\sigma^{2}_{\beta}),\frac{1}{2}a_{3}(\mu, d, \alpha, \beta, B, \sigma^{2}_{E})\right)} ,\quad and\quad(\mu+d+\alpha)>a_{2}(\mu, d, \alpha, \beta, B,\sigma^{2}_{I} ).%\quad and\quad 2\mu> a_{3}(\mu, d, \alpha, \beta, B).
 \end{equation}
 Also let the delay times $T_{1}, T_{2}$ and $T_{3}$ be constant, that is, the probability density functions of $T_{1}, T_{2}$ and $T_{3}$ respectively denoted by $f_{T_{i}}, i=1, 2, 3$ are the  dirac-delta functions defined in (\ref{ch1.sec2.eq4}). Furthermore, let the constants $T_{1}, T_{2}$ and $T_{3}$ satisfy the following set of inequalities:
\begin{equation}\label{ch1.sec3.thm2.eq2}
T_{1}>\frac{1}{2\mu}\log{\left(\frac{\left(\frac{\beta (G^{*})^{2}}{2\lambda{(\mu)}}+\sigma^{2}_{\beta}(G^{*})^{2}\right)}{(3\mu-a_{1}(\mu, d, \alpha, \beta, B, \sigma^{2}_{S},\sigma^{2}_{\beta}))}\right)},
\end{equation}
\begin{equation}\label{ch1.sec3.thm2.eq3}
T_{2}<\frac{1}{2\mu}\log{\left(\frac{(3\mu-a_{1}(\mu, d, \alpha, \beta, B, \sigma^{2}_{S},\sigma^{2}_{\beta}))}{\left(\frac{\beta (G^{*})^{2}}{2\lambda{(\mu)}}+\sigma^{2}_{\beta}(G^{*})^{2}\right)\left(\frac{1}{G'(0)}\right)^{2}}\right)},
\end{equation}%\frac{\hat{K}_{0}+\frac{\alpha}{\beta \frac{B}{\mu}}
and
\begin{equation}\label{ch1.sec3.thm2.eq3b}
T_{3}>\frac{1}{2\mu}\log{\left(\frac{\frac{3\alpha}{2\lambda(\mu)}}{(\mu+d+\alpha)-a_{2}(\mu, d, \alpha, \beta, B, \sigma^{2}_{I})}\right)}.
\end{equation}
There exists a positive real number $\mathfrak{m}_{1}>0$, such that
\begin{eqnarray}
  &&\limsup_{t\rightarrow \infty}\frac{1}{t}E\int^{t}_{0}\left[ (S(v)-S^{*}_{1})^{2}+ (E(v)-E^{*}_{1})^{2}+  (I(v)-I^{*}_{1})^{2}\right]dv\nonumber\\
  &&\leq \frac{3\sigma^{2}_{S}(S^{*}_{1})^{2}+ 2\sigma^{2}_{E}(E^{*}_{1})^{2}+\sigma^{2}_{I}(I^{*}_{1})^{2}+\sigma^{2}_{\beta}(S^{*}_{1})^{2}(G^{*})^{2}e^{-2\mu (T_{1}+T_{2})}+\sigma^{2}_{\beta}(S^{*}_{1})^{2}(G^{*})^{2}e^{-\mu T_{1}}}{\mathfrak{m}_{1}}.\nonumber\\
  \label{ch2.sec3.thm2.eq4}
\end{eqnarray}%\l
\end{thm}
Proof:\\
From Lemma~\ref{ch1.sec3.lemma1},  (\ref{ch1.sec3.lemma1.eq5a})-(\ref{ch1.sec3.lemma1.eq5c}), it is easy to see that under the assumptions in (\ref{ch1.sec3.thm1.corrolary1.eq1}) and (\ref{ch1.sec3.thm2.eq1})-(\ref{ch1.sec3.thm2.eq3b}), then $\tilde{\phi}_{1}>0$, $\tilde{\psi}_{1}>0$ and $\tilde{\varphi}_{1}>0$. Therefore,  from (\ref{ch1.sec3.lemma1.eq5})-(\ref{ch1.sec3.lemma1.eq7}) it is also easy to see that
\begin{eqnarray}
   dV&=&LV(t)dt + \overrightarrow{g}(S(t), E(t), I(t))d\overrightarrow{w(t)},\nonumber\\
   &\leq&-\min\{\tilde{\phi}_{1}, \tilde{\psi}_{1}, \tilde{\varphi}_{1}\}\left[ (S(t)-S^{*}_{1})^{2}+ (E(t)-E^{*}_{1})^{2}+  (I(t)-I^{*}_{1})^{2}\right]\nonumber\\
   && +3\sigma^{2}_{S}(S^{*}_{1})^{2}+ 2\sigma^{2}_{E}(E^{*}_{1})^{2}+\sigma^{2}_{I}(I^{*}_{1})^{2}+\sigma^{2}_{\beta}(S^{*}_{1})^{2}(G^{*})^{2}e^{-2\mu (T_{1}+T_{2})}+\sigma^{2}_{\beta}(S^{*}_{1})^{2}(G^{*})^{2}e^{-\mu T_{1}}\nonumber\\
   &&+ \overrightarrow{g}(S(t), E(t), I(t))d\overrightarrow{w(t)},\label{ch1.sec3.thm2.proof.eq1}
 \end{eqnarray}
 Integrating both sides of (\ref{ch1.sec3.thm2.proof.eq1}) from 0 to $t$ and taking the expectation, it follows that
 %%%%%%%%%%%%%%%%%%%%%%%%%%%%%
 \begin{eqnarray}
 &&E(V(t)-V(0))\leq -\mathfrak{m_{1}}E\int^{t}_{0}\left[ (S(v)-S^{*}_{1})^{2}+ (E(v)-E^{*}_{1})^{2}+  (I(v)-I^{*}_{1})^{2}\right]dv\nonumber\\
   && +\left(3\sigma^{2}_{S}(S^{*}_{1})^{2}+ 2\sigma^{2}_{E}(E^{*}_{1})^{2}+\sigma^{2}_{I}(I^{*}_{1})^{2}+\sigma^{2}_{\beta}(S^{*}_{1})^{2}(G^{*})^{2}e^{-2\mu (T_{1}+T_{2})}+\sigma^{2}_{\beta}(S^{*}_{1})^{2}(G^{*})^{2}e^{-\mu T_{1}}\right)t,\nonumber\\
   \label{ch2.sec3.thm2.proof.eq2}
\end{eqnarray}
where $V(0)$ is constant and
\begin{equation}\label{ch2.sec3.thm2.proof.eq3}
\mathfrak{m}_{1}=min(\tilde{\phi},\tilde{ \psi},\tilde{\varphi})>0.
\end{equation}
 Hence, diving both sides of (\ref{ch2.sec3.thm2.proof.eq2}) by $t$ and $\mathfrak{m}_{1}$, and taking the $\limsup_{t\rightarrow \infty}$, then (\ref{ch2.sec3.thm2.eq4}) follows directly.
 %%%%
%%%%
%%%%
\begin{rem}\label{ch2.sec3.thm2.rem1}
When the disease dynamics is perturbed by random fluctuations in the disease transmission or natural death rates, that is,  when at least one of the  intensities $\sigma^{2}_{i}\neq 0, i= S, E, I, \beta$, it has been noted earlier that the stochastic system (\ref{ch1.sec0.eq8})-(\ref{ch1.sec0.eq11}) does not have an endemic equilibrium state. Nevertheless, the conditions in Theorem~\ref{ch1.sec3.thm2} provide estimates for the constant delay times $T_{1}, T_{2}$ and $T_{3}$ in (\ref{ch1.sec3.thm2.eq2})-(\ref{ch1.sec3.thm2.eq3b}) in addition to other parametric restrictions in (\ref{ch1.sec3.thm2.eq1}) that are sufficient for the solutions of the perturbed stochastic system (\ref{ch1.sec0.eq8})-(\ref{ch1.sec0.eq11}) to oscillate near the nontrivial steady state, $E_{1}$, of the deterministic system (\ref{ch1.sec0.eq3})-(\ref{ch1.sec0.eq6}) found in Theorem~\ref{ch1.sec3.thm1}. The result in (\ref{ch2.sec3.thm2.eq4}) that characterizes the average distance between the trajectories of the stochastic system and the nontrivial steady state $E_{1}$ also signifies that the size of the oscillations of the trajectories relative to $E_{1}$ depends on the size of the intensity values, $\sigma^{2}_{i}\neq 0, i= S, E, I, \beta$,  of the random fluctuations.

As a physical interpretation and deduction, the results in Theorem~\ref{ch1.sec3.thm2} suggest that the presence of random fluctuations in the malaria dynamics stemming from the disease transmission or natural death rates completely perturbs the stability of the endemic equilibrium $E_{1}$ shown in Theorem~\ref{ch2.sec3.thm3a.corrolary1}. This means that the presence of noise in the malaria epidemic promotes the persistence of malaria in the human population as described in Remark~\ref{ch1.sec3.rem2a},  where the human population continues to oscillates in character near the nonzero steady state $E_{1}$ as depicted in (\ref{ch2.sec3.thm2.eq4}). Furthermore, for smaller intensity values of the random fluctuations, the  human population also oscillates closely to the nonzero steady state $E_{1}$, and vice versa. While (\ref{ch2.sec3.thm2.eq4}) indicates that larger intensity values of the random fluctuations in the natural death and disease transmission rates lead to the human population oscillating further away from the nonzero steady state $E_{1}$ asymptotically, it can be propositioned that the human population gets extinct overtime due to the high intensity of natural death rate of human beings asymptotically. The numerical simulation results in Section~\ref{ch1.sec4} confirm the proposition that the human population becomes extinct asymptotically for sufficiently large intensities of the random fluctuations.

 These facts suggest that malaria control policies in the event where the disease is persistent, should focus on reducing the intensities of the fluctuations in the disease transmission and natural death rates, perhaps through vector control and better care of the people in the population, in order to reduce the number of deaths that may lead to human extinction by the disease.
\end{rem}
The subsequent result provides more general conditions irrespective of the probability distribution of the random variables $T_{1}, T_{2}$ and $T_{3}$, that are sufficient for the trajectories of the stochastic system (\ref{ch1.sec0.eq8})-(\ref{ch1.sec0.eq11}) to oscillate near the nontrivial steady state $E_{1}$,  of the deterministic system (\ref{ch1.sec0.eq3})-(\ref{ch1.sec0.eq6}).
% when the system is perturbed by at least one of the intensities of the system $\sigma^{2}_{i}=0, i= S, E, I, \beta$ irrespective of the distribution of the random variables  $T_{1}, T_{2}$ and $T_{3}$.
%%%%%%%
%For simplicity, the following notations are introduced.  Let $\tilde{a}_{1}(\mu, d, \alpha, \beta, B)$ , $\tilde{a}_{2}(\mu, d, \alpha, \beta, B)$ and $a_{3}(\mu, d, \alpha, \beta, B)$ represent the following set of parameters
%\begin{eqnarray}
%\tilde{a}_{1}(\mu, d, \alpha, \beta, B)&=&2\mu\lambda{(\mu)}+(2\mu+d+\alpha)\frac{\lambda{(\mu)}}{2}+\alpha\lambda{(\mu)}+\frac{\beta S^{*}_{1}\lambda{(\mu)}}{2}+3\sigma ^{2}_{S}\nonumber\\
%&&+\left(\frac{\beta \lambda{(\mu)}(G^{*})^{2}}{2}+\sigma^{2}_{\beta}(G^{*})^{2}\right)+\left(\frac{\beta (G^{*})^{2}}{2\lambda{(\mu)}}+\sigma^{2}_{\beta}(G^{*})^{2}\right)\label{ch1.sec3.thm2.proof.eq4a}\\
%\tilde{a}_{2}(\mu, d, \alpha, \beta, B)&=&(2\mu+d+\alpha)\frac{1}{\lambda{(\mu)}}+ \frac{\alpha\lambda{(\mu)}}{2} + \sigma^{2}_{I}\nonumber\\
%&&+\left(\frac{\beta S^{*}_{1}(G'(I^{*}_{1}))^{2}}{\lambda{(\mu)}}
%\right)+\frac{3\alpha}{2\lambda(\mu)}\label{ch1.sec3.thm2.proof.eq4b}
%a_{3}(\mu, d, \alpha, \beta, B)&=&\frac{\beta }{2\lambda{(\mu)}}+\frac{\beta S^{*}_{1}\lambda{(\mu)}}{2}+ \frac{2\mu}{\lambda{(\mu)}}+(2\mu+d+\alpha)\frac{\lambda{(\mu)}}{2}+\alpha %\lambda{(\mu)}+ 2 \sigma^{2}_{E}\nonumber\\\label{ch1.sec3.lemma1.proof.eq13c}
%\end{eqnarray}
%and $a_{3}(\mu, d, \alpha, \beta, B)$  is defined in (\ref{ch1.sec3.lemma1.proof.eq13c}).
\begin{thm}\label{ch2.sec3.thm3}
Let the hypotheses of Theorem~\ref{ch1.sec3.thm1} and  Lemma~\ref{ch1.sec3.lemma1} be satisfied, and let
 \begin{equation}\label{ch1.sec3.thm3.eq1}
 \mu> \max\left\{\frac{1}{3}\tilde{a}_{1}(\mu, d, \alpha, \beta, B, \sigma ^{2}_{S}, \sigma^{2}_{\beta}),\frac{1}{2}a_{3}(\mu, d, \alpha, \beta, B, \sigma ^{2}_{E})\right\}\quad and\quad(\mu+d+\alpha)>\tilde{a}_{2}(\mu, d, \alpha, \beta, B, \sigma ^{2}_{I}).
 \end{equation}
  It follows that for any arbitrary probability distribution of the delay times: $T_{1}, T_{2}$ and $T_{3}$,
there exists a positive real number $\mathfrak{m}_{2}>0$ such that
\begin{eqnarray}
  &&\limsup_{t\rightarrow \infty}\frac{1}{t}E\int^{t}_{0}\left[ (S(v)-S^{*}_{1})^{2}+ (E(v)-E^{*}_{1})^{2}+  (I(v)-I^{*}_{1})^{2}\right]dv\nonumber\\
  &&\leq \frac{3\sigma^{2}_{S}(S^{*}_{1})^{2}+ 2\sigma^{2}_{E}(E^{*}_{1})^{2}+\sigma^{2}_{I}(I^{*}_{1})^{2}+\sigma^{2}_{\beta}(S^{*}_{1})^{2}(G^{*})^{2}E(e^{-2\mu (T_{1}+T_{2})})+\sigma^{2}_{\beta}(S^{*}_{1})^{2}(G^{*})^{2}E(e^{-\mu T_{1}})}{\mathfrak{m}_{2}}.\nonumber\\
  \label{ch2.sec3.thm3.eq2}
\end{eqnarray}%\l
\end{thm}
%%%%
Proof:\\
From Lemma~\ref{ch1.sec3.lemma1},  (\ref{ch1.sec3.lemma1.eq5a})-(\ref{ch1.sec3.lemma1.eq5c}),
 \begin{eqnarray}
-\tilde{\phi}_{1}&=&-3\mu+\left[2\mu\lambda{(\mu)}+(2\mu+d+\alpha)\frac{\lambda{(\mu)}}{2}+\alpha\lambda{(\mu)}+\frac{\beta S^{*}_{1}\lambda{(\mu)}}{2}+3\sigma ^{2}_{S}+\left(\frac{\beta (G^{*})^{2}}{2\lambda{(\mu)}}+\sigma^{2}_{\beta}(G^{*})^{2}\right)E(e^{-2\mu T_{1}})\right.\nonumber\\
&&\left.+\left(\frac{\beta \lambda{(\mu)}(G^{*})^{2}}{2}+\sigma^{2}_{\beta}(G^{*})^{2}\right)E(e^{-2\mu (T_{1}+T_{2})})\right]\nonumber\\
&&\leq -3\mu+\left[2\mu\lambda{(\mu)}+(2\mu+d+\alpha)\frac{\lambda{(\mu)}}{2}+\alpha\lambda{(\mu)}+\frac{\beta S^{*}_{1}\lambda{(\mu)}}{2}+3\sigma ^{2}_{S}+\left(\frac{\beta (G^{*})^{2}}{2\lambda{(\mu)}}+\sigma^{2}_{\beta}(G^{*})^{2}\right)\right.\nonumber\\
&&\left.+\left(\frac{\beta \lambda{(\mu)}(G^{*})^{2}}{2}+\sigma^{2}_{\beta}(G^{*})^{2}\right)\right]\nonumber\\
&&=-\left(3\mu-\tilde{a}_{1}(\mu, d, \alpha, \beta, B, \sigma ^{2}_{S}, \sigma^{2}_{\beta})\right)\label{ch1.sec3.thm3.proof.eq1}\\
%%%%%
%%%%%%
  -\tilde{\psi}_{1} &=& -2\mu+\left[\frac{\beta }{2\lambda{(\mu)}}+\frac{\beta S^{*}_{1}\lambda{(\mu)}}{2}+ \frac{2\mu}{\lambda{(\mu)}}+(2\mu+d+\alpha)\frac{\lambda{(\mu)}}{2}+\alpha \lambda{(\mu)}+ 2 \sigma^{2}_{E} \right]\nonumber\\
  &&=-\left(2\mu-a_{3}(\mu, d, \alpha, \beta, B, \sigma ^{2}_{E})\right)\label{ch1.sec3.thm3.proof.eq2}\\
    -\tilde{\varphi}_{1}&=& -(\mu + d+\alpha)+\left[(2\mu+d+\alpha)\frac{1}{\lambda{(\mu)}}+ \frac{\alpha\lambda{(\mu)}}{2} + \sigma^{2}_{I}+\frac{3\alpha}{2\lambda{(\mu)}}E(e^{-2\mu T_{3}})\right.\nonumber\\
    &&\left. +\left(\frac{\beta S^{*}_{1}(G'(I^{*}_{1}))^{2}}{\lambda{(\mu)}}
\right)E(e^{-2\mu (T_{1}+T_{2})})\right]\nonumber\\
&\leq& -(\mu + d+\alpha)+\left[(2\mu+d+\alpha)\frac{1}{\lambda{(\mu)}}+ \frac{\alpha\lambda{(\mu)}}{2} + \sigma^{2}_{I}+\frac{3\alpha}{2\lambda{(\mu)}}\right.\nonumber\\
    &&\left. +\left(\frac{\beta S^{*}_{1}(G'(I^{*}_{1}))^{2}}{\lambda{(\mu)}}
\right)\right]\nonumber\\
&&=-\left((\mu + d+\alpha)-\tilde{a}_{2}(\mu, d, \alpha, \beta, B, \sigma ^{2}_{I})\right),\label{ch1.sec3.thm3.proof.eq3}
   \end{eqnarray}%+\frac{\beta (G'(I^{*}_{1}))^{2}}{2\lambda{(\mu)}it is easy to see that under the assumptions in (\ref{ch1.sec3.thm1.eq1}) and (\ref{ch1.sec3.thm2.eq1})-(\ref{ch1.sec3.thm2.eq3}),
%%%%
since $0<E(e^{-2\mu (T_{i})})\leq 1, i=1, 2,3$. It follows from (\ref {ch1.sec3.lemma1.eq5})-(\ref{ch1.sec3.lemma1.eq7}) and (\ref{ch1.sec3.thm3.proof.eq1})-(\ref{ch1.sec3.thm3.proof.eq3}) that
\begin{eqnarray}
   dV&=&LV(t)dt + \overrightarrow{g}(S(t), E(t), I(t))d\overrightarrow{w(t)},\nonumber\\
   &\leq&-\min\{\left(3\mu-\tilde{a}_{1}(\mu, d, \alpha, \beta, B)\right), \left(2\mu-a_{3}(\mu, d, \alpha, \beta, B)\right), \left((\mu + d+\alpha)-\tilde{a}_{2}(\mu, d, \alpha, \beta, B)\right)\}\times\nonumber\\
   &&\times \left[ (S(t)-S^{*}_{1})^{2}+ (E(t)-E^{*}_{1})^{2}+  (I(t)-I^{*}_{1})^{2}\right]\nonumber\\
   && +3\sigma^{2}_{S}(S^{*}_{1})^{2}+ 2\sigma^{2}_{E}(E^{*}_{1})^{2}+\sigma^{2}_{I}(I^{*}_{1})^{2}+\sigma^{2}_{\beta}(S^{*}_{1})^{2}(G^{*})^{2}e^{-2\mu (T_{1}+T_{2})}+\sigma^{2}_{\beta}(S^{*}_{1})^{2}(G^{*})^{2}e^{-\mu T_{1}}\nonumber\\
   &&+ \overrightarrow{g}(S(t), E(t), I(t))d\overrightarrow{w(t)},\label{ch1.sec3.thm3.proof.eq4}
 \end{eqnarray}
 where under the assumptions (\ref{ch1.sec3.thm3.eq1}) in the hypothesis,
\begin{equation}\label{ch2.sec3.thm3.proof.eq5}
\mathfrak{m}_{2}=\min\{\left(3\mu-\tilde{a}_{1}(\mu, d, \alpha, \beta, B)\right), \left(2\mu-a_{3}(\mu, d, \alpha, \beta, B)\right), \left((\mu + d+\alpha)-\tilde{a}_{2}(\mu, d, \alpha, \beta, B)\right)\}>0.
\end{equation}
Integrating both sides of (\ref{ch1.sec3.thm3.proof.eq4}) from 0 to $t$ and taking the expectation, it follows that
 %%%%%%%%%%%%%%%%%%%%%%%%%%%%%
 \begin{eqnarray}
 &&E(V(t)-V(0))\leq -\mathfrak{m_{2}}E\int^{t}_{0}\left[ (S(v)-S^{*}_{1})^{2}+ (E(v)-E^{*}_{1})^{2}+  (I(v)-I^{*}_{1})^{2}\right]dv\nonumber\\
   && +\left(3\sigma^{2}_{S}(S^{*}_{1})^{2}+ 2\sigma^{2}_{E}(E^{*}_{1})^{2}+\sigma^{2}_{I}(I^{*}_{1})^{2}+\sigma^{2}_{\beta}(S^{*}_{1})^{2}(G^{*})^{2}E(e^{-2\mu (T_{1}+T_{2})})+\sigma^{2}_{\beta}(S^{*}_{1})^{2}(G^{*})^{2}E(e^{-\mu T_{1}})\right)t,\nonumber\\
   \label{ch2.sec3.thm3.proof.eq6}
\end{eqnarray}
where $V(0)$ is constant.  Hence, diving both sides of (\ref{ch2.sec3.thm3.proof.eq6}) by $t$ and $\mathfrak{m}_{2}$, and taking the $\limsup_{t\rightarrow \infty}$, then (\ref{ch2.sec3.thm3.eq2}) follows directly.
 %%%%
 \begin{rem}\label{ch1.sec3.rem2}
  When the disease dynamics is perturbed by random fluctuations in the disease transmission or natural death rates, that is,  when at least one of the  intensities $\sigma^{2}_{i}\neq 0, i= S, E, I, \beta$, it has been noted earlier that the stochastic system (\ref{ch1.sec0.eq8})-(\ref{ch1.sec0.eq11}) does not have an endemic equilibrium state. Nevertheless, the conditions in Theorem~\ref{ch2.sec3.thm3} provide minimum general parametric restrictions in (\ref{ch1.sec3.thm3.eq1}) irrespective of the probability distribution of the random variable delay times $T_{1}, T_{2}$ and $T_{3}$  that are sufficient for the solutions of the perturbed stochastic system (\ref{ch1.sec0.eq8})-(\ref{ch1.sec0.eq11}) to oscillate near the nontrivial steady state, $E_{1}$, of the deterministic system (\ref{ch1.sec0.eq3})-(\ref{ch1.sec0.eq6}) found in Theorem~\ref{ch1.sec3.thm1}. The result in (\ref{ch2.sec3.thm3.eq2}) that characterizes the average distance between the trajectories of the stochastic system and the nontrivial steady state $E_{1}$ also signifies that the size of the oscillations of the trajectories relative to $E_{1}$ depends on the intensity values, $\sigma^{2}_{i}\neq 0, i= S, E, I, \beta$,  of the random fluctuations.

  In addition, comparing the conditions in (\ref{ch1.sec3.thm3.eq1}) with the deterministic case in (\ref{ch1.sec3.thm3a.eq1}), it is easy to see that
  \begin{eqnarray}
 &&\mu> \max\left\{\frac{1}{3}\tilde{a}_{1}(\mu, d, \alpha, \beta, B, \sigma ^{2}_{S}, \sigma^{2}_{\beta}),\frac{1}{2}a_{3}(\mu, d, \alpha, \beta, B, \sigma ^{2}_{E})\right\}\nonumber\\
 &&>\max\left\{\frac{1}{3}\tilde{a}_{1}(\mu, d, \alpha, \beta, B),\frac{1}{2}a_{3}(\mu, d, \alpha, \beta, B)\right\}\nonumber\\
  &&and\quad(\mu+d+\alpha)>\tilde{a}_{2}(\mu, d, \alpha, \beta, B, \sigma ^{2}_{I})>\tilde{a}_{2}(\mu, d, \alpha, \beta, B),\label{ch1.sec3.rem2.eq1}
 \end{eqnarray}
 whenever $\sigma_{i}> 0, i= S, E, I, \beta$. It can be  deduced from the observation in (\ref{ch1.sec3.rem2.eq1}) and conclusions of Theorem~\ref{ch2.sec3.thm3} and Theorem~\ref{ch2.sec3.thm3a} that in the absence of any random fluctuations in the disease dynamics from natural death and disease transmission rates, a solution that starts in the neighborhood of the nontrivial steady state $E_{1}$ stays in the neighborhood and converges asymptotically to the steady state $E_{1}$. However, the occurrence of random fluctuations in the disease dynamics from any of the sources-disease transmission or natural death rates with intensity values $\sigma_{i}> 0, i= S, E, I, \beta$,   destabilize the system from the equilibrium state $E_{1}$. Nevertheless, the solutions of the perturbed stochastic system (\ref{ch1.sec0.eq8})-(\ref{ch1.sec0.eq11}) continue to oscillate near the nontrivial steady state $E_{1}$, provided that the natural death rate $\mu$ and the total removal by recovery, natural and disease related death rates $(\mu+d+\alpha)$ satisfy the conditions in (\ref{ch1.sec3.thm3.eq1}). But, sufficiently large values of $\sigma_{i}> 0, i= S, E, I, \beta$, that is,  $\sigma_{i}\rightarrow \infty , i= S, E, I, \beta$,  imply from (\ref{ch1.sec3.thm3.eq1}) and (\ref{ch2.sec3.thm3.eq2}) that $\mu\rightarrow \infty$, $(\mu+d+\alpha)\rightarrow \infty$, and further imply that the solutions of (\ref{ch1.sec0.eq8})-(\ref{ch1.sec0.eq11}) oscillate at much further distance away from $E_{1}$.
%%%% \begin{equation}\label{ch1.sec3.thm3a.eq1}
%% \mu> \max\left\{\frac{1}{3}\tilde{a}_{1}(\mu, d, \alpha, \beta, B),\frac{1}{2}a_{3}(\mu, d, \alpha, \beta, B)\right\}\quad and\quad(\mu+d+\alpha)>\tilde{a}_{2}(\mu, d, \alpha, \beta, B).
%% \end{equation}

  As a physical interpretation and deduction similar to Remark~\ref{ch2.sec3.thm2.rem1}, these observations above suggest that the presence of random fluctuations in the malaria dynamics stemming from the disease transmission or natural death rates leads to a persistent state of the disease in the human population, where the population oscillates in character near the nonzero steady state $E_{1}$. Furthermore, for smaller intensity values of the random fluctuations, the  human population also oscillates close to the nonzero steady state $E_{1}$, and vice versa. But sufficiently large intensity values lead to very high natural and disease related death rates, that is, $\mu\rightarrow \infty$, and $(\mu+d+\alpha)\rightarrow \infty$, which postulate the fact that the human population goes extinct asymptotically. Indeed, while (\ref{ch2.sec3.thm3.eq2}) indicates that larger intensity values of the random fluctuations lead to the human population oscillating further away from the nonzero steady state $E_{1}$ asymptotically, the numerical simulation results in Section~\ref{ch1.sec4} confirm the proposition that the human population becomes extinct asymptotically for sufficiently large intensities of the random fluctuations.

 These facts suggest that malaria control policies in the event where the disease is persistent, should focus on reducing the intensities of the fluctuations in the disease transmission and natural death rates, perhaps through vector control and better care of the people in the population, in order to reduce the number of deaths that may lead to human extinction by the disease.
   \end{rem}
 %%%%%%%%%%%%%%%%%%%%%%%%%%%
 %%%%%%%%%%%%%%%%%%%%%%%%%%%%
  \section{Example}\label{ch1.sec4}
  It should be noted that some of the numerical examples discussed in this section are utilized in various capacities elsewhere to address different sub-objectives of the current on going project. In this study, the examples presented below are used to facilitate understanding about the  influence of the noise in the system  on the trajectories of  the stochastic system (\ref{ch1.sec0.eq8})-(\ref{ch1.sec0.eq11}) relative to the equilibria of the stochastic/deterministic systems in Theorem~\ref{ch1.sec2.thm0} and Theorem~\ref{ch1.sec3.thm1}.
 \subsection{Example 1: The effect of the intensity of the white noise process on disease eradication: }
 %random fluctuations in the system on threshold parameters ($R_{1}$, $U_{0}$, and $V_{0}$)
 This example illustrates the results in Theorem~\ref{ch1.sec2.theorem1}, Theorem~\ref{ch1.sec2.theorem1.corollary1} and Lemma~\ref{ch1.sec2.lemma2}, and also provides numerical evidence in support of the results in Section~\ref{ch1.sec2} that characterize the effects of the intensity of the white noise processes in the system originating from the random fluctuations in the disease dynamics ( that is, from disease transmission and natural death rates), on the stochastic asymptotic stability of the  system in relation to the disease free equilibrium $E_{0}=(S^{*}_{0},0,0), S^{*}_{0}=\frac{B}{\mu} $. Recall, Theorem~\ref{ch1.sec2.theorem1} and Lemma~\ref{ch1.sec2.lemma2} provide conditions for the threshold values $R_{1}$, $U_{0}$, and $V_{0}$ defined in (\ref{ch2.sec2.thm1.eq5a})-(\ref{ch2.sec2.thm1.eq5c}) which are sufficient for the stochastic stability of $E_{0}$ and consequently for disease eradication.  For simplicity in this example, the following assumptions are considered:- ($a_{1}$) there are no random fluctuations in the disease dynamics due to the natural death of susceptible individuals, that is, the intensity of the white noise due to the random fluctuations in the natural death of susceptible individuals $\sigma_{S}=0$. Indeed, from Theorem~\ref{ch1.sec2.theorem1} and Lemma~\ref{ch1.sec2.lemma2}, there exists a stable disease free equilibrium $E_{0}$, whenever $\sigma_{S}=0$ and the threshold values satisfy $R_{1}\leq 1$, $U_{0}\leq 1$, and $V_{0}\leq 1$. ($a_{2}$) It is also assumed that the intensities of the white noise processes in the system due to the random fluctuations in the natural death  and disease transmission rates for the other disease classes-exposed, infectious and removal are equal,  that is, $\sigma_{E}=\sigma_{I}=\sigma_{R}=\sigma_{\beta}=\sigma$.

\begin{figure}[H]
  \centering
  \includegraphics[height=8cm]{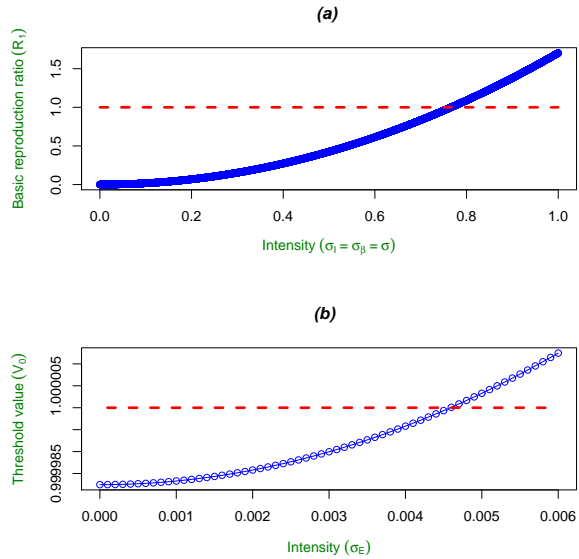}
  \caption{(a) and (b) Show the values of the noise modified basic reproduction number, $R_{1}$, (defined in (\ref{ch2.sec2.thm1.eq5a})) and the threshold parameter $V_{0}$ (defined in (\ref{ch2.sec2.thm1.eq5c})) over continuous changes in the values of the intensities of white noise processes due to random fluctuations in natural death and disease transmission processes of exposed, infectious and removal individuals, that is, $\sigma=\sigma_{E}=\sigma_{\beta}=\sigma_{I}=\sigma_{R}$.  The curves in (a) and (b) show the values of $R_{1}$ and $V_{0}$ respectively. In addition, the broken horizontal lines depict the threshold mark, $1$, for the threshold values  $R_{1}$ and $V_{0}$, where  for the values of $R_{1}$ and $V_{0}$ below the threshold mark $1$, the disease free equilibrium $E_{0}$ is stochastically asymptotically stable, and the  disease can consequently be eradicated.   It is easy to see that low values of $\sigma\in[0, 0.7661]$ lead to $R_{1}\leq 1$, and $R_{1}> 1$ other wise. For $V_{0}$, the low values of $\sigma\in[0, 0.0045]$ lead to $V_{0}\leq 1$, and $V_{0}> 1$ other wise. Therefore, values for $R_{1}$, $U_{0}$, and $V_{0}$ that satisfy $R_{1}\leq 1$, $U_{0}\leq 1$, and $V_{0}\leq 1$ are achieved for very low values of $\sigma$. This observation signifies that for a disease scenario where the physical processes lead to the specific parameter values defined in Table~\ref{ch1.sec4.table1}, the disease can only be eradicated when the random fluctuations in the disease dynamics exhibit very low intensity values of $\sigma\in[0, 0.0045]$. For any intensity values higher than $0.0045$, the disease-free equilibrium $E_{0}$ is unstable, and this signifies that the disease outbreak becomes naturally uncontrollable and establishes either a stable endemic population $E_{1}$, whenever the intensity value $\sigma$  is small (see Theorem~\ref{ch2.sec3.thm3a}) or the disease oscillates near the endemic population $E_{1}$ as shown in Theorems~[\ref{ch1.sec3.thm2} \& \ref{ch2.sec3.thm3}].}\label{ch1.sec4.figure1}
\end{figure}

  The convenient list of system parameter values in Table~\ref{ch1.sec4.table1} are used to generate different values for $R_{1}$, $U_{0}$, and $V_{0}$ under continuous changes in the values of $\sigma=\sigma_{E}=\sigma_{I}=\sigma_{R}=\sigma_{\beta}$. The Figure~\ref{ch1.sec4.figure1} depicts the results for $R_{1}$  and $V_{0}$. For $U_{0}$, it follows from Table~\ref{ch1.sec4.table1} and (\ref{ch2.sec2.thm1.eq5b}) that $U_{0}\approx 1$, where $\tilde{K}(\mu)=0.999991$.
%%%%%%
 \begin{table}[h]
  \centering
  \caption{A list of specific values chosen for the system parameters for Example 1. }\label{ch1.sec4.table1}
  \begin{tabular}{l l l}
  Disease transmission rate&$\beta$& $6.277E-66$\\\hline
  Constant Birth rate&$B$&$ \frac{22.39}{1000}$\\\hline
  Recovery rate& $\alpha$& $5.5067E-07$\\\hline
  Disease death rate& $d$& 0.11838\\\hline
  Natural death rate& $\mu$& $0.6$\\\hline
  %Intensity of fluctuations& $\sigma_{i}, i=S, E, I, R, \beta$& 0.05\\\hline
  %Incubation delay in vector& $T_{1}$& 2 units of time\\\hline
  %Incubation delay in host& $T_{2}$& 1 unit of time\\\hline
  %Immunity delay time& $T_{3}$& 4 units of time\\\hline
  \end{tabular}
\end{table}
  %%%%%%%%%%%%%Note that the observations of this example are consistent with the results in Theorems~[\ref{ch1.sec2-2.thm1}-\ref{ch1.sec2-2.thm3},$\&$ \ref{ch1.sec2-2.thm5}-\ref{ch1.sec2-2.thm6}].
%%%%%
%\newpage
 \subsection{Example 2: Effect of the intensity of white noise on the trajectories of the system  }%random fluctuations in the system on sample population density over time
\begin{figure}[H]
\begin{center}
\includegraphics[height=8cm]{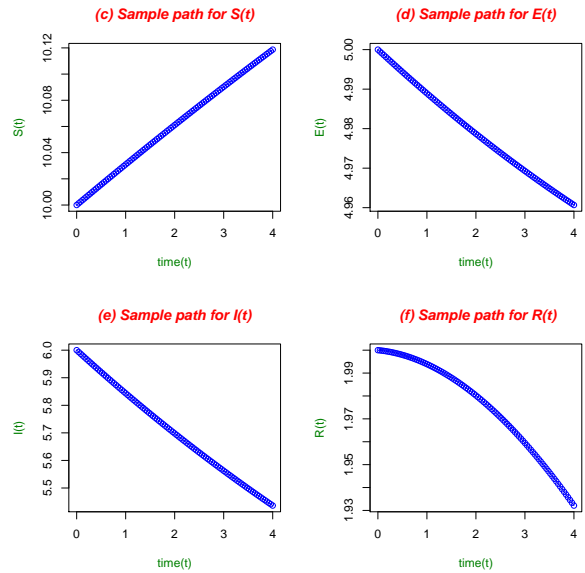}
\caption{(c), (d), (e) and (f) show the trajectories of the disease classes $(S,E,I,R)$ respectively, when there are no or only infinitesimally small random fluctuations in the disease dynamics, that is,  when the intensities of the white noise processes in the system due to random fluctuations in the natural death  and disease transmission processes in all the classes $(S,E,I,R)$ are described as follows:   $\sigma_{S}=\sigma_{E}=\sigma_{\beta}=\sigma_{I}=\sigma_{R}=0$.}\label{ch1.sec4.figure 2}
\end{center}
\end{figure}

\begin{figure}[H]
\begin{center}
\includegraphics[height=8cm]{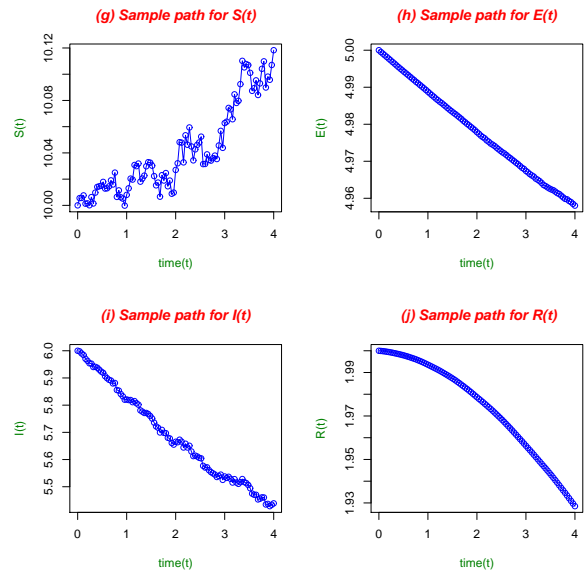}
\caption{(g), (h), (i) and (j) show the trajectories of the disease classes $(S,E,I,R)$ respectively, when there are no or only infinitesimally small random fluctuations in the disease dynamics from the natural death rate of the classes $(S,E,I,R)$, that is, when  $\sigma_{S}=\sigma_{E}=\sigma_{I}=\sigma_{R}=0$, but there are random fluctuations in the disease transmission process with low intensity value of $\sigma_{\beta}=0.5$.}\label{ch1.sec4.figure 3}
\end{center}
\end{figure}

\begin{figure}[H]
\begin{center}
\includegraphics[height=8cm]{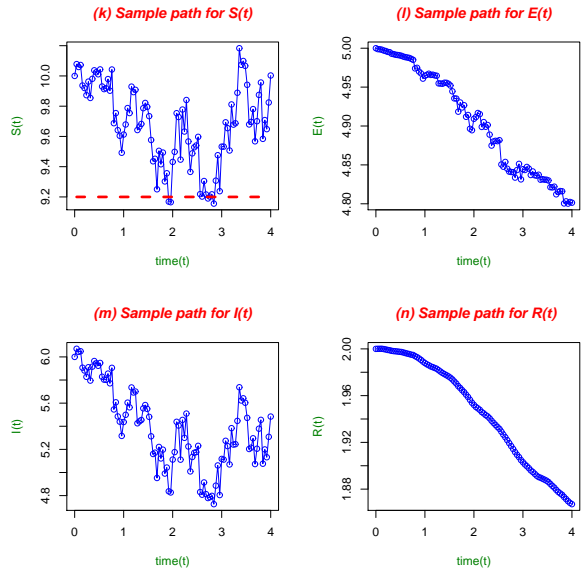}
\caption{(k), (l), (m) and (n) show the trajectories of the disease classes $(S,E,I,R)$ respectively, when there are no or only infinitesimally small random fluctuations in the disease dynamics from the natural death of the classes (S,E,I,R), that is, when  $\sigma_{S}=\sigma_{E}=\sigma_{I}=\sigma_{R}=0$, but there are random fluctuations in the disease transmission process with a high intensity value of $\sigma_{\beta}=9$. In addition, the broken line on the sample path for $S(t)$ in (k) depicts the $S$-coordinate $S^{*}_{0}=\frac{B}{\mu}=9.199999$ for the disease free steady state $E_{0}=(S^{*}_{0},0,0,0), S^{*}_{0}=\frac{B}{\mu}, E^{*}_{0}=0, I^{*}_{0}=0, R^{*}_{0}=0$. }\label{ch1.sec4.figure 4}
\end{center}
\end{figure}

\begin{figure}[H]
\begin{center}
\includegraphics[height=8cm]{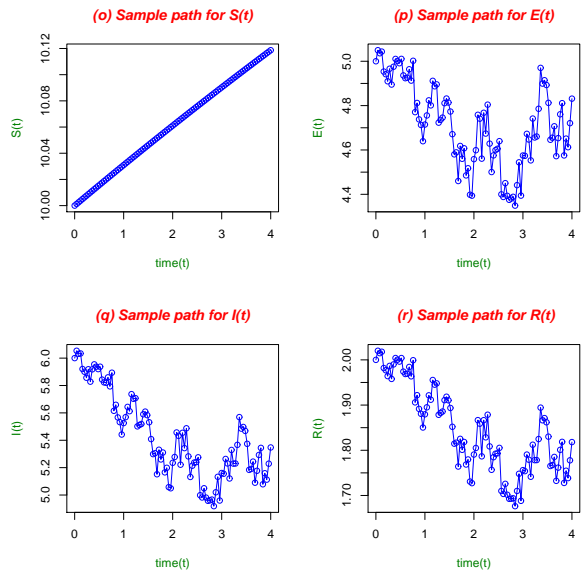}
\caption{(o), (p), (q) and (r) show the trajectories of the disease classes $(S,E,I,R)$ respectively, when there are significant small random fluctuations in the disease dynamics from the natural death process of exposed, infectious and removal classes, with intensity  value $\sigma_{E}=\sigma_{I}=\sigma_{R}=0.5$, but there are no or only infinitesimally small fluctuations in the disease dynamics due to the  disease transmission and natural death processes of susceptible individuals, that is,  $\sigma_{S}=\sigma_{\beta}=0$.}\label{ch1.sec4.figure 5}
\end{center}
\end{figure}

\begin{figure}[H]
\begin{center}
\includegraphics[height=8cm]{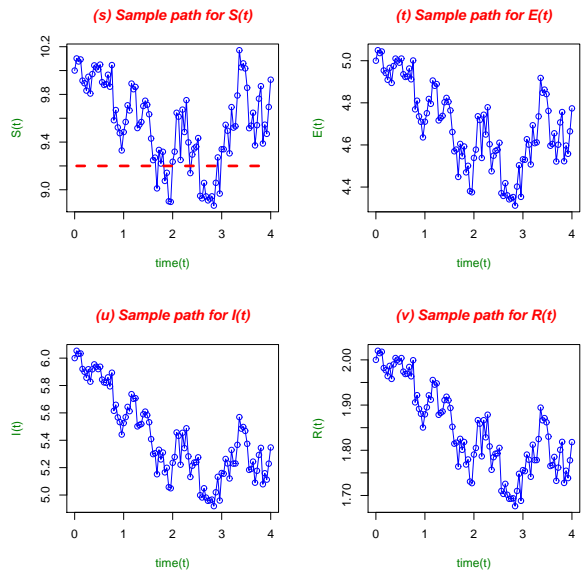}
\caption{(s), (t), (u) and (v) show the trajectories of the disease classes $(S,E,I,R)$ respectively, when there are significant but small random fluctuations in the disease dynamics from the natural death process in all the disease classes- susceptible, exposed, infectious and removal classes with low intensity value of  $\sigma_{S}=\sigma_{E}=\sigma_{I}=\sigma_{R}=0.5$, but there are no or only infinitesimally small fluctuations in the disease dynamics from the disease transmission process,  that is, $\sigma_{\beta}=0$. In addition, the broken line on the sample path for $S(t)$ in (s) depicts the $S$-coordinate $S^{*}_{0}=\frac{B}{\mu}=9.199999$ for the disease free steady state $E_{0}=(S^{*}_{0},0,0,0), S^{*}_{0}=\frac{B}{\mu}, E^{*}_{0}=0, I^{*}_{0}=0, R^{*}_{0}=0$.}\label{ch1.sec4.figure 6}
\end{center}
\end{figure}

\begin{figure}[H]
\begin{center}
\includegraphics[height=8cm]{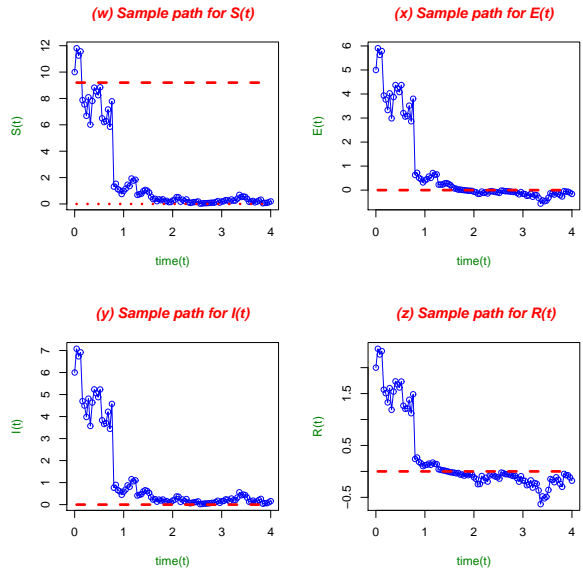}
\caption{(w), (x), (y) and (z) show the trajectories of the disease classes $(S,E,I,R)$ respectively, when there are significant and large random fluctuations in the disease dynamics from the natural death process in all the disease classes- susceptible, exposed, infectious and removal classes with sufficiently high intensity value of  $\sigma_{S}=\sigma_{E}=\sigma_{I}=\sigma_{R}=9$, but there are no or only infinitesimally small fluctuations in the disease dynamics from the disease transmission process,  that is, $\sigma_{\beta}=0$. In addition, the broken line on the sample paths for $S(t)$, $E(t)$, $I(t)$ and $R(t)$  depict the $S, E, I, R$-coordinates $S^{*}_{0}=\frac{B}{\mu}=9.199999, E^{*}_{0}=0, I^{*}_{0}=0, R^{*}_{0}=0$ for the disease free steady state $E_{0}=(S^{*}_{0},0,0,0), S^{*}_{0}=\frac{B}{\mu}, E^{*}_{0}=0, I^{*}_{0}=0, R^{*}_{0}=0$. (w), (x), (y) and (z) also show that the population goes extinct over time due to the high intensity of the white noise.}\label{ch1.sec4.figure 7}
\end{center}
\end{figure}

\begin{figure}[H]
\begin{center}
\includegraphics[height=8cm]{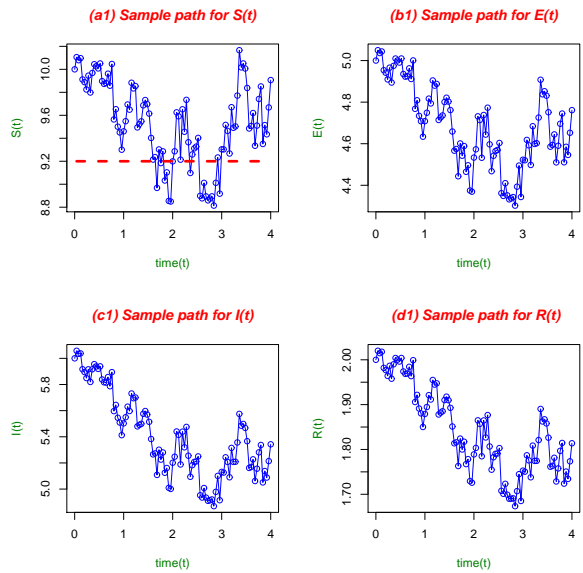}
\caption{(a1), (b1), (c1) and (d1) show the trajectories of the disease classes $(S,E,I,R)$ respectively, when there are significant but small random fluctuations in the disease dynamics from the natural death process in all the disease classes- susceptible, exposed, infectious and removal classes with low intensity value of  $\sigma_{S}=\sigma_{E}=\sigma_{I}=\sigma_{R}=0.5$,  and there are also significant fluctuations in the disease dynamics from the disease transmission with a low intensity value of $\sigma_{\beta}=0.5$. In addition, the broken line on the sample path for $S(t)$ in (a1) depicts the $S$-coordinate $S^{*}_{0}=\frac{B}{\mu}=9.199999$ for the disease free steady state $E_{0}=(S^{*}_{0},0,0,0), S^{*}_{0}=\frac{B}{\mu}, E^{*}_{0}=0, I^{*}_{0}=0, R^{*}_{0}=0$.}\label{ch1.sec4.figure 8}
\end{center}
\end{figure}

\begin{figure}[H]
\begin{center}
\includegraphics[height=8cm]{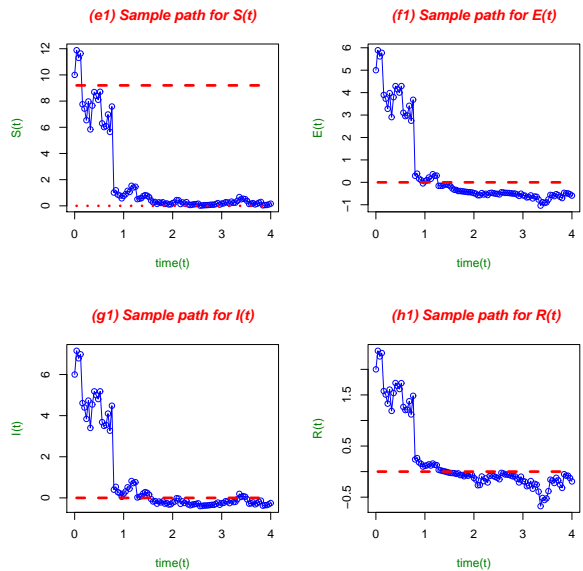}
\caption{(e1), (f1), (g1) and (h1) show the trajectories of the disease classes $(S,E,I,R)$ respectively, when there are significant and large random fluctuations in the disease dynamics from the natural death process in all the disease classes- susceptible, exposed, infectious and removal classes with a sufficiently high intensity value of  $\sigma_{S}=\sigma_{E}=\sigma_{I}=\sigma_{R}=9$,  and there are also significant fluctuations in the disease dynamics from the disease transmission process with a sufficiently high intensity value of $\sigma_{\beta}=9$. In addition, the broken line on the sample paths for $S(t)$, $E(t)$, $I(t)$ and $R(t)$  depict the $S, E, I, R$-coordinates $S^{*}_{0}=\frac{B}{\mu}=9.199999, E^{*}_{0}=0, I^{*}_{0}=0, R^{*}_{0}=0$ for the disease free steady state $E_{0}=(S^{*}_{0},0,0,0), S^{*}_{0}=\frac{B}{\mu}, E^{*}_{0}=0, I^{*}_{0}=0, R^{*}_{0}=0$. Furthermore,  (e1), (f1), (g1) and (h1) show that the population goes extinct over time due to the high intensity of the white noise.}\label{ch1.sec4.figure 9}
\end{center}
\end{figure}

%%%%
The list of convenient choice of parameter values in Table~\ref{ch1.sec4.table2} are used to generate the trajectories of the stochastic system (\ref{ch1.sec0.eq8})-(\ref{ch1.sec0.eq11}) in order to (1.) illustrate the impact of the source of the white noise processes in the system (owing to the random fluctuations in  the natural death or disease transmission rates) on the disease dynamics, and also to (2.) illustrate the effect of the intensity of the  white noise processes in the system on the trajectories of the different disease classes $(S, E, I, R)$ in the system. These illustrations also uncover the overall behavior of the stochastic system over time.
%%%%%%%%%%
 %Figure~\ref{ch1.sec4.figure2} depicts the behavior of the stochastic system (\ref{ch1.sec0.eq8})-(\ref{ch1.sec0.eq11}) when the system has no fluctuations or whenever the intensity of the %fluctuations are infinitesimally small, that is $\sigma_{S}=\sigma_{E}=\sigma_{\beta}=\sigma_{I}=\sigma_{R}=0(\epsilon)$.
 \begin{table}[h]
  \centering
  \caption{A list of specific values chosen for the system parameters for Example 2.}\label{ch1.sec4.table2}
  \begin{tabular}{l l l}
  Disease transmission rate&$\beta$& 0.6277\\\hline
  Constant Birth rate&$B$&$ \frac{22.39}{1000}$\\\hline
  Recovery rate& $\alpha$& 0.05067\\\hline
  Disease death rate& $d$& 0.01838\\\hline
  Natural death rate& $\mu$& $0.002433696$\\\hline
  %Intensity of fluctuations& $\sigma_{i}, i=S, E, I, R, \beta$& 0.05\\\hline
  Incubation delay time in vector& $T_{1}$& 2 units \\\hline
  Incubation delay time in host& $T_{2}$& 1 unit \\\hline
  Immunity delay time& $T_{3}$& 4 units\\\hline
  \end{tabular}
\end{table}
The Euer-Maruyama stochastic approximation scheme\footnote{A seed is set on the random number generator to reproduce  the same sequence of random numbers for the Brownian motion in order to generate reliable graphs for the trajectories of the system under different intensity values for the white noise processes, so that comparison can be made to identify differences that reflect the effect of intensity values.} is used to generate trajectories for the different states $S(t), E(t), I(t), R(t)$ over the time interval $[0,T]$, where $T=\max(T_{1}+T_{2}, T_{3})=4$. The special nonlinear incidence  functions $G(I)=\frac{aI}{1+I}, a=0.05$ in \cite{gumel} is utilized to generate the numeric results. Furthermore, the following initial conditions are used
\begin{equation}\label{ch1.sec4.eq1}
\left\{
\begin{array}{l l}
S(t)= 10,\\
E(t)= 5,\\
I(t)= 6,\\
R(t)= 2,
\end{array}
\right.
\forall t\in [-T,0], T=\max(T_{1}+T_{2}, T_{3})=4.
\end{equation}
%%%%%%%%%%%%%%%%%
%%%%%%%%%%%%%%%
The sample means for the sample paths of the $S, E, I, R$ states generated over time $t\in [0, T]$  are summarized in Table~\ref{ch1.sec4.table3}, and will be used to compare the effect of the intensity values of the white noise processes in the system on the trajectories of the system over time.
\begin{table}[h]
  \centering
  \caption{ Shows the intensity values of the white noise processes in the system and the corresponding sample means for the trajectories of the $S, E, I, R$ states generated over time $t\in [0, 4]$ in Example 2. The sample means for $S, E, I, R$ are denoted $\bar{S}, \bar{E}, \bar{I}, \bar{R}$ respectively. }\label{ch1.sec4.table3}
  \begin{tabular}{l l l l l l}
  $\sigma_{i}, i= S, E, I, R, \beta$&Figure \# &$\bar{S}$& $\bar{E}$&$\bar{I}$&$\bar{R}$\\\hline
  $\sigma_{i}=0, i= S, E, I, R, \beta$&Figure~\ref{ch1.sec4.figure 2}&10.06048&4.979256&5.704827&1.975407\\\hline
  $\sigma_{i}=0, i= S, E, I, R$, and $\sigma_{\beta}=0.5$&Figure~\ref{ch1.sec4.figure 3}&10.04129&4.978257&5.687113&1.973783\\\hline
  $\sigma_{i}=0, i= S, E, I, R$, and $\sigma_{\beta}=9$&Figure~\ref{ch1.sec4.figure 4}&9.681482&4.906452&5.385973&1.94617\\\hline
  $\sigma_{i}=0.5, i=  E, I, R$, and $\sigma_{S}=\sigma_{\beta}=0$&Figure~\ref{ch1.sec4.figure 5}&10.06048&4.715779&5.42661&1.845652\\\hline
  $\sigma_{i}=0.5, i= S, E, I, R$, and $\sigma_{\beta}=0$&Figure~\ref{ch1.sec4.figure 6}&9.553725&4.692877&5.42661&1.845652\\\hline
  $\sigma_{i}=9, i= S, E, I, R$, and $\sigma_{\beta}=0$&Figure~\ref{ch1.sec4.figure 7}&1.980488&0.8066963&1.200498&0.240599\\\hline
  $\sigma_{i}=0.5, i= S, E, I, R$, and $\sigma_{\beta}=0.5$&Figure~\ref{ch1.sec4.figure 8}&9.529665&4.687529&5.406493&1.843888\\\hline
  $\sigma_{i}=9, i= S, E, I, R$, and $\sigma_{\beta}=9$&Figure~\ref{ch1.sec4.figure 9}&1.88787&0.4633994&0.8659143&0.2315721\\\hline
    \end{tabular}
\end{table}
The following observations can be made from Table~\ref{ch1.sec4.table3}:
\begin{rem}\label{ch1.sec4.rem1}
\item[1.] When  $\sigma_{i}=0, i= S, E, I, R$, there is   moderate decrease in the average values $\bar{S}, \bar{E}, \bar{I}, \bar{R}$ of $S, E, I, R$ from the trajectories in Figures~\ref{ch1.sec4.figure 2}-\ref{ch1.sec4.figure 4} as $\sigma_{\beta}$ increases from $\sigma_{\beta}=0$ to $\sigma_{\beta}=9$.
    \item[2.] When  $\sigma_{\beta}=0$, there is  a sharp decrease in the average values $\bar{S}, \bar{E}, \bar{I}, \bar{R}$ of $S, E, I, R$ from the trajectories in Figure~\ref{ch1.sec4.figure 2}, and Figures~\ref{ch1.sec4.figure 6}-\ref{ch1.sec4.figure 7} as $\sigma_{i}, i= S, E, I, R$ increases from $\sigma_{i}=0, i= S, E, I, R$ to $\sigma_{i}=9, i= S, E, I, R$.
        \item[3.] When all the $\sigma_{i}$'s , that is,  $\sigma_{i}, i= S, E, I, R, \beta$ equally increase together from $\sigma_{i}=0, i= S, E, I, R, \beta$ to $\sigma_{i}=9, i= S, E, I, R, \beta$, there is  a sharper decrease in the average values $\bar{S}, \bar{E}, \bar{I}, \bar{R}$ of $S, E, I, R$ from the trajectories in Figure~\ref{ch1.sec4.figure 2},  and Figures~\ref{ch1.sec4.figure 8}-\ref{ch1.sec4.figure 9}.
            \item[4.] When  $\sigma_{S}=\sigma_{\beta}=0 $, there is no change in the average value $\bar{S}$ of $S$  and there is  moderate decrease in the average values $ \bar{E}, \bar{I}, \bar{R}$ of $ E, I, R$  from the trajectories in Figure~\ref{ch1.sec4.figure 2} and Figure~\ref{ch1.sec4.figure 5} as $\sigma_{i}, i=  E, I, R$ increases from $\sigma_{i}=0, i=  E, I, R$ to $\sigma_{i}=0.5, i=  E, I, R$.
    \end{rem}
%%%%%%%%%%%%%%%%
%%%%%%%%%%%%%%%%\footnote{No random fluctuations in the system signifies that the system is deterministic in character.}
%%%%SEIRS compartmental framework  noise-in-disease-transmission-only--point-5
%%%%%%%%%%%%%%%%%%%%%%
%%%%%%%%%%%%%%%%%%%%%%suceptible-naturaldeath-0-disease-transmission-0-othernatural-death-point-5
%%%%%%%%%%%%%%%%%%%%%%%%%%%%%%%%%%%%%%%%%%%%%%%%%%%%%%%%%%%%%%%
The Figures~\ref{ch1.sec4.figure 2}-\ref{ch1.sec4.figure 4} can be used to examine the effect of increasing the intensity value of the white noise process, $\sigma_{\beta}$, originating from the random fluctuations in the disease transmission process on the trajectories for $(S,E,I,R)$ in the absence of any significant random fluctuations in the disease dynamics due to the natural death process for all the disease classes $(S,E,I,R)$, that is, $\sigma_{i}=0, i=S,E,I,R$ .
%For intensity values for the white noise process, it follows from
It can be observed from Figure~\ref{ch1.sec4.figure 2} that when the intensity value  $\sigma_{\beta}$ is  infinitesimally small, that is, $\sigma_{\beta}=0$,  no significant oscillations occur over time on the trajectories for $S, E, I, R$ in (a), (b), (c) and (d) respectively. Furthermore, for significant but low intensity values for $\sigma_{\beta}$, that is, $\sigma_{\beta}=0.5$, Figure~\ref{ch1.sec4.figure 3} shows that some significant oscillations  occur on the trajectories for the susceptible (g) and infectious (i) populations. Moreover, the size of the oscillations observed on the trajectories for the susceptible (g) and infectious population (i) seem to be small in value over time compared to Figure~\ref{ch1.sec4.figure 4}. In addition, no significant oscillations are observed on the trajectories for the exposed (h) and removal (j) populations. In Figure~\ref{ch1.sec4.figure 4}, with an increase in the intensity value for $\sigma_{\beta}$ to $\sigma_{\beta}=9$, more disease classes exhibit significant oscillations on their trajectories, for instance, more significant sized oscillations are observed on the trajectory of one additional class- exposed population (l) than is observed in Figure~\ref{ch1.sec4.figure 3} (h). Moreover, it appears that the high intensity value $\sigma_{\beta}=9$  has increased  the size of the oscillations in the susceptible (k) and infectious  (m) populations,  and  further deviating  the trajectories of the system away from the noise-free state in Figure~\ref{ch1.sec4.figure 2}. In addition, the trajectories for the states- $(S,E,I)$ in Figure~\ref{ch1.sec4.figure 4} (k), (l), (m) respectively,  oscillate near the disease free state $E_{0}=(S^{*}_{0},0,0,0)$, where $ S^{*}_{0}=\frac{B}{\mu}=9.199999, E^{*}_{0}=0, I^{*}_{0}=0, R^{*}_{0}=0$.

 One can also observe from Table~\ref{ch1.sec4.table3} and Remark~\ref{ch1.sec4.rem1} that for small values $\sigma_{i}=0, i= S, E, I, R$, the average values of  $S,E,I,R$ over time on the trajectories in Figures~\ref{ch1.sec4.figure 2}-\ref{ch1.sec4.figure 4} decrease continuously with increase in the intensity value of $\sigma_{\beta}$ from $\sigma_{\beta}=0$ to $\sigma_{\beta}=9$.
 These observations related to the oscillatory behavior of the system, for example, comparing the trajectory of $S$ in  Figure~\ref{ch1.sec4.figure 2}(c), Figure~\ref{ch1.sec4.figure 3}(g) and Figure~\ref{ch1.sec4.figure 4}(k), and also comparing the trajectory for $I$ in Figure~\ref{ch1.sec4.figure 2}(e), Figure~\ref{ch1.sec4.figure 3}(i) and Figure~\ref{ch1.sec4.figure 4}(m) suggest that continuously increasing the intensity value for $\sigma_{\beta}$ tends to increase the oscillatory behavior of the trajectories of the system that results in an average decrease in the size of the susceptible, exposed, infectious and removal populations over time. Furthermore, the size of the oscillations in the system is proportional to the size of the intensity values of the white noise process as remarked in Remark~\ref{ch1.sec2.rem2}.
%%%%%%%%%%%%%%%%%%%%%%%%%%%%%%%%%%%%%%%%%%%%%%%%%%%%%%%%%%%%%%%%%
%%%%%%%%%%%%%%%%%%%%%%noise-in-disease-transmissio-0-all-natural-death-point5
Figure~\ref{ch1.sec4.figure 2} , Figure~\ref{ch1.sec4.figure 5}, and Figure~\ref{ch1.sec4.figure 6} can be used as an example to examine the effect of the intensity of the white noise process, $\sigma_{i}, i= S, E, I, R$, originating from the random fluctuations in the natural death process of each class-$S, E, I, R$, on the trajectories of the system, in the absence of any significant fluctuation in the disease dynamics owing to the disease transmission process, that is, $\sigma_{\beta}=0$. For example, to examine the effect of $\sigma_{\beta}$ for the susceptible class, $S$, on the trajectories of the stochastic stochastic system, observe that in Figure~\ref{ch1.sec4.figure 5}, when $\sigma_{S}=\sigma_{\beta}=0$ and $\sigma_{i}=0.5, i= E, I, R$, no significant oscillations occur on the trajectories of $S$ in Figure~\ref{ch1.sec4.figure 5}(o) and also on  Figure~\ref{ch1.sec4.figure 2}(c). Furthermore, when $\sigma_{S}$ is increased to $\sigma_{S}=0.5$,  Figure~\ref{ch1.sec4.figure 6}(s) depicts significant sized oscillations on the trajectory of $S$. Moreover, the trajectory for $S$ oscillates near the disease free steady state $S^{*}_{0}=9.199999$. It can be further observed using Table~\ref{ch1.sec4.table3} and Remark~\ref{ch1.sec4.rem1} that no major differences have occurred on the trajectories of the other states $E, I, R$ in both  Figure~\ref{ch1.sec4.figure 5}(p),(q),(r) and Figure~\ref{ch1.sec4.figure 6}(t),(u),(v) respectively. In addition, it can be seen from Table~\ref{ch1.sec4.table3} and Remark~\ref{ch1.sec4.rem1} that when $\sigma_{\beta}=0$,  the increase in the intensity value of $\sigma_{S}$ from $\sigma_{S}=0$ to $\sigma_{S}=0.5$ on average leads to a decrease in the susceptible population size over time in Figure~\ref{ch1.sec4.figure 6}(s) than it is observed in Figure~\ref{ch1.sec4.figure 5}(o) and  Figure~\ref{ch1.sec4.figure 2}(c).   These observations suggest that in the absence of random fluctuations in the disease dynamics from the  disease transmission process, that is, $\sigma_{\beta}=0$, the intensity of the white noise process, $\sigma_{S}$, owing to the natural death of the susceptible class $S$, (1.) exhibits a significant effect primarily on its trajectory, and (2.) the effect of increasing the intensity value of $\sigma_{S}$ leads to  an oscillatory behavior on the trajectory of $S$ that  decreases the susceptible population averagely over time. Note that a similar numerical and graphical diagnostic approach can be used to examine the effects of the other classes $E, I, R$, whenever $\sigma_{\beta}=0$.
%%%%%%%%%%%%%%%%%%%%%%suceptible-naturaldeath-0-disease-transmission-0-othernatural-death-point-5
%%%%%%%%%%%%%%%%%%%%%%noise-in-disease-transmissio-0-all-natural-death-point9
%%%%%%%%%%%%%%%%%%%%%%%%%%%%%%%%%%%%%%%%
%%%%%%%%%%%%%%%%%%%%%%s
%%%%%%%%%%%%%%%%%%%%%%
Figure~\ref{ch1.sec4.figure 2}, Figure~\ref{ch1.sec4.figure 6} and Figure~\ref{ch1.sec4.figure 7} can be used to examine the effect of increasing the intensity value of the  white noise process originating from the natural death, $\sigma_{i}, i= S, E, I, R$, in the absence of any significant random fluctuations in the disease dynamics from the disease transmission process, that is, when $\sigma_{\beta}=0$. Figure~\ref{ch1.sec4.figure 6} (s),(t),(u),(v) show that the trajectories for  $S, E, I, R$ respectively, oscillate near the disease free equilibrium $E_{0}=(9.199999, 0, 0, 0)$ over time when the intensity value is increased from $\sigma_{i}=0, i= S, E, I, R$ to $\sigma_{i}=0.5, i= S, E, I, R$ than is observed in the Figure~\ref{ch1.sec4.figure 2} (c), (d), (e), (f). Furthermore, the oscillations on the trajectories seem to be  small in size over time. When the intensity value,  $\sigma_{i}, i= S, E, I, R$, is  further increased to $\sigma_{i}=9, i= S, E, I, R$, the oscillations on the trajectories in Figure~\ref{ch1.sec4.figure 7} (w),(x),(y),(z), appear to have increased in size. Furthermore, Table~\ref{ch1.sec4.table3} and Remark~\ref{ch1.sec4.rem1} show that the oscillations lead to a decrease in the average values of $S, E, I, R$ over time, and  further away from the disease free state of $S^{*}_{0}=9.199999$. Moreover, the population rapidly becomes extinct over time. These observations suggest that the increase in the intensity value of the white noise due to natural death in all classes, $\sigma_{i}, i= S, E, I, R$,  in the population (1.)leads to an increase in the oscillatory behavior of the system which decreases the population size averagely over time and also (2.)leads to population  extinction over time. Note that this observation is consistent with the results of Theorem~\ref{ch1.sec2.theorem2}, whenever $\sigma_{S}$ is large magnitude.
%%%%%%%%%%%%%%%%%%%%%%%%%%%%%%%%%%%%%%%%
%%%%%%%%%%%%%%%%%%%%%%s
%%%%%%%%%%%%%%%%%%%%%%%%%%%%%%%%%%%%%%%%%%%%%
%%%%%%%%%%%%%%%%%%%%%%%%%%%%%%%%%%%%%%%%%%%%%%
%%%%%%%%%%%%%%%%%%%%%%%%%%%%%%%%%%%%%%%%%%%%%
%%%%%%%%%%%%%%%%%%%%%%%%%%%%%%%%%%%%%%%%%%%%
Figure~\ref{ch1.sec4.figure 2}, Figure~\ref{ch1.sec4.figure 8} and Figure~\ref{ch1.sec4.figure 9} can be used to examine the effect of increasing the intensity values, $\sigma_{i}, i= S, E, I, R,\beta$, of all the white noise processes in the system on the trajectories of the system. Figure~\ref{ch1.sec4.figure 8} (a1),(b1),(c1),(d1) show that the trajectories for $S, E, I, R$ respectively oscillate near the disease free steady state $E_{0}=(9.199999, 0, 0, 0)$ over time when the intensity value is increased  from $\sigma_{i}=0, i= S, E, I, R,\beta$ to $\sigma_{i}=0.5, i= S, E, I, R,\beta$ than it is observed in Figure~\ref{ch1.sec4.figure 2}(c),(d),(e),(f). Furthermore, the oscillations of the trajectories seem to be  small in size compared to the corresponding trajectories in Figure~\ref{ch1.sec4.figure 9}.  When the intensity values of $\sigma_{i}, i= S, E, I, R, \beta$ are  further increased to $\sigma_{i}=9, i= S, E, I, R$, it can be seen from  Figure~\ref{ch1.sec4.figure 9}(e1),(f1),(g1),(h1), Table~\ref{ch1.sec4.table3} and Remark~\ref{ch1.sec4.rem1} that the oscillations increase in size and lead to a sharp decrease in the average values  of $S, E, I, R$ on their trajectories over time, and also further deviating the average susceptible population size away from the disease free state of $S^{*}_{0}=9.199999$. Moreover, the population rapidly becomes extinct over time. These observations suggests that the increase in the intensity value of the white noise processes in the system due to the random fluctuations in the disease dynamics originating from the disease transmission  and natural death processes for all disease classes in the population leads to (1.) an increase in the oscillatory behavior of the system which decreases the average total population size over time, and also leads to (2.) the rapid extinction of the population over time.

It can also be observed by comparing Figure~\ref{ch1.sec4.figure 7}(w),(x),(y),(z), and  Figure~\ref{ch1.sec4.figure 9}(e1),(f1),(g1),(h1), that for a fixed value of $\sigma_{i}=9, i= S, E, I, R$,  if $\sigma_{\beta}$  increases from  $\sigma_{\beta}=0$ in Figure~\ref{ch1.sec4.figure 7}(w),(x),(y),(z) to $\sigma_{\beta}=9$  in Figure~\ref{ch1.sec4.figure 9}(e1),(f1),(g1),(h1), then the population more rapidly becomes extinct than it is observed in Figure~\ref{ch1.sec4.figure 7}(w),(x),(y),(z). Indeed, in Figure~\ref{ch1.sec4.figure 7}(w),(x),(y),(z), the trajectories for the  susceptible $S$, exposed $E$, infectious $I$ and Removal $R$  states go extinct at approximately the following times $t=2, t=1.8, t=2$ and $t=1.8$ respectively. Meanwhile,  in Figure~\ref{ch1.sec4.figure 9}(e1),(f1),(g1),(h1), the trajectories for  susceptible $S$, exposed $E$, infectious $I$ and Removal $R$ go extinct earlier at the approximate times $t=1.5, t=1, t=1$ and $t=1.4$ respectively.  Note that this observation is consistent with the results of Theorem~\ref{ch1.sec2.theorem2}, whenever $\sigma_{S}$ is large in magnitude.

 The following  pairs of figures:- (Figure~\ref{ch1.sec4.figure 3} (g),(h),(i),(j) \& Figure~\ref{ch1.sec4.figure 4} (k),(l),(m),(n))  and (  Figure~\ref{ch1.sec4.figure 6} (s),(t),(u),(v) \& Figure~\ref{ch1.sec4.figure 7} (w),(x),(y),(z)), can be used with reference to Figure~\ref{ch1.sec4.figure 2}, to examine and compare the two major sources of random fluctuations in the disease dynamics namely-natural death and disease transmission processes, in order to determine the source which has stronger effect on the trajectories of the system, whenever the intensity values of the white noise processes increase in value. In the absence of random fluctuations in the natural death process, that is, $\sigma_{i}=0, i= S, E, I, R$, as the intensity value of $\sigma_{\beta}$ is increased from $\sigma_{\beta}=0.5$ to $\sigma_{\beta}=9$,   the pair of figures (Figure~\ref{ch1.sec4.figure 3} (g),(h),(i),(j) \& Figure~\ref{ch1.sec4.figure 4} (k),(l), (m),(n)) show an increase in the oscillatory behavior on the trajectories of the system  which is more significant in size for  the $S$ and $I$ classes  over time. Furthermore, the oscillatory behavior leads to a decrease in the average susceptible and infectious populations over time than it is observed in Figure~\ref{ch1.sec4.figure 2} (c) and  Figure~\ref{ch1.sec4.figure 2}(e) respectively, as shown in Table~\ref{ch1.sec4.table3} and Remark~\ref{ch1.sec4.rem1}.   Moreover, the general disease population does not go extinct over time.

 Meanwhile, in the absence of random fluctuations in the disease transmission process, that is, $\sigma_{\beta}=0$,  the increase in the intensity value of $\sigma_{i}, i= S, E, I, R$  from $\sigma_{i}=0.5, i= S, E, I, R$ to $\sigma_{i}=9, i= S, E, I, R$,   the pair of figures (  Figure~\ref{ch1.sec4.figure 6}(s),(t),(u),(v) \& Figure~\ref{ch1.sec4.figure 7}(w),(x),(y),(z)) show very strong increase in the oscillatory behavior on the trajectories of the system which is significant in all the states-  $S$, $E$, $I$ and $R$ . Furthermore, from Table~\ref{ch1.sec4.table3} and Remark~\ref{ch1.sec4.rem1}, it can be seen that the oscillatory behavior of the system leads to a rapid decrease in the average values of all the states-$S$, $E$, $I$ and $R$ over time, with the mean susceptible population size deviating much further away from the disease free steady state $S^{*}_{0}=9.199999$,  than it is observed in Figure~\ref{ch1.sec4.figure 2}.  Moreover, the disease population goes extinct over time with the increase in the intensity value of $\sigma_{i}, i= S, E, I, R$.
 \section{Conclusion}
 The presented class of stochastic and deterministic SEIRS epidemic dynamic models with nonlinear incidence rates, three distributed delays and random perturbations characterizes the general dynamics of vector-borne diseases such as malaria and dengue fever, that are influenced by random environmental fluctuations from (1.) the disease transmission rate between susceptible humans and infectious vectors mainly mosquitoes, and also from (2.)  the natural death rates in the sub-categories - susceptible, exposed, infectious and removal individuals of the human population. Moreover, the random fluctuations in the disease dynamics are incorporated into the epidemic dynamic models via  white noise or  Wiener processes.  Furthermore, the three delays are random variables. Whereas, two of the delays represent the incubation periods of the infectious agent mainly the malaria parasites or dengue fever virus in the vector and human being, the third delay represents the period of effective naturally acquired immunity  against the disease, which is conferred to individuals  after recovery from infection. This study presents two classes of epidemic dynamic models, one represented as a system of Ito-Doob type stochastic differential equations, and the other represented as a system of ordinary differential equations, where the class type is determined by a general nonlinear incidence function $G$, satisfying a set of mathematical conditions. The nonlinear incidence function $G$ can be used to characterize the nonlinear character of disease transmission rates for disease scenarios that exhibit a striking initial increase or decrease in disease transmission rates which become steady or bounded as the infectious population size grows and becomes large.

Comparative conditions and threshold values pertaining to both stochastic and deterministic systems that are sufficient for the existence of unique global positive solutions, and other asymptotic properties  are presented. %,extinction and permanence of disease are in the human population
 For instance, detailed results  that characterize the  asymptotic behavior of the trajectories or solutions of the  stochastic and deterministic dynamic systems are presented  namely:- (1.) the existence and asymptotic stochastic stability of  feasible equilibria of the systems- disease free $E_{0}$ and endemic $E_{1}$ steady states, and (2.) the asymptotic oscillatory character of the solutions of the stochastic system near the  potential steady states- disease-free and endemic equilibria of the deterministic system.  In addition, comparative threshold values  for the stochastic  and deterministic stability of the disease free steady state, and consequently for disease eradication, such as the noise dependent and noise-free basic reproduction numbers for the disease dynamics are computed. Moreover, the threshold values are exhibited for special cases when the delay times in the system are both constant and random.

 The results from the comparative asymptotic analyses of the stochastic and deterministic  systems suggest that the source (disease transmission or natural death rates ) and size of the intensity values of the white noise processes in the  system exhibit direct consequences on the overall qualitative behavior of the class of epidemic dynamic models with respect to the equilibria of the systems. For example, (1.) the existence and stability of the disease free and endemic steady state populations of the epidemic dynamic models, and their consequences on disease eradication or persistence depend on the source and intensity of the white noise processes in the system, (2) the oscillatory character of the sample paths of the stochastic system near the disease free and endemic equilibria, and the consequences on the persistence of the disease in the population also depend on the source and intensity of the white noise processes in the system.

 Further thorough numerical examination of the asymptotic properties of the trajectories of the stochastic and deterministic systems under the influence of various intensity levels of the white noise processes in the system is conducted. The numerical simulation results suggest that higher intensity values of the white processes in the system drive the sample paths of the stochastic system further away from the common deterministic and stochastic disease free steady state, and also further away from the endemic equilibrium of the deterministic system. Moreover, the population seems to become extinct over time, whenever the intensity values of the white noise processes become large.


\begin{thebibliography}{300}
\bibitem{macdonald} Macdonald, G., The analysis of infection rates in diseases in which superinfection occurs. Trop. Dis. Bull. 47, 907-915
\end{thebibliography}
\end{document}